\documentclass[fleqn,usenatbib]{mnras}

\usepackage{newtxtext,newtxmath}
\usepackage{caption}
\usepackage{xcolor}

\usepackage[T1]{fontenc}

\DeclareRobustCommand{\VAN}[3]{#2}
\let\VANthebibliography\thebibliography
\def\thebibliography{\DeclareRobustCommand{\VAN}[3]{##3}\VANthebibliography}

\usepackage{graphicx}	
\usepackage{amsmath}	
\usepackage{subfigure}

\usepackage{booktabs}
\usepackage{multirow}

\title[PCA of real and synthetic spectra]{Evaluating quenching in cosmological simulations of galaxy formation with spectral covariance in the optical window}

\author[Sharbaf et al.]{\Large {Z. Sharbaf$^{1,2}$}\thanks{E-mail: \href{mailto:zsharbaf@iac.es}{zsharbaf@iac.es}}, {I. Ferreras$^{3,1,2}$\thanks{Corresponding author: \href{mailto:i.ferreras@ucl.ac.uk}{i.ferreras@ucl.ac.uk}}}, {A. Negri$^{1,2,4}$},
{J. Angthopo$^{5}$}, {C. Dalla Vecchia$^{1,2}$}, {O. Lahav$^{3}$}, {R.~S. Somerville$^{6}$}\\
$^{1}$Instituto de Astrofìsica de Canarias, C/ Vìa La ctea s/n, La Laguna, E-38200 La Laguna, Tenerife, Spain
\\
$^{2}$Departamento de Astrofsica, Universidad de La Laguna, E-38205 La Laguna, Tenerife, Spain
\\
$^{3}$Department of Physics and Astronomy, University College London, Gower Street, London WC1E 6BT, UK
\\
$^{4}$Facultad de F\'\i sica, Universidad de Sevilla, Avda. Reina Mercedes s/n, Campus de Reina Mercedes, E-41012 Sevilla, Spain\\
$^{5}$INAF - Osservatorio Astronomico di Brera, Via Brera 28, 20121, Milano, Italy\\
$^{6}$Center for Computational Astrophysics, Flatiron Institute, New York, NY 10010, USA\\
}

\date{Accepted for publication in MNRAS: 2025 April 01. Received 2025 April 01; in original form 2024 November 11}

\pubyear{2025}

\begin{document}
\label{firstpage}
\pagerange{\pageref{firstpage}--\pageref{lastpage}}
\maketitle

\begin{abstract}
Cosmological hydrodynamical simulations provide valuable insights on galaxy evolution when coupled with observational data. Comparisons with real galaxies are typically performed via scaling relations of the observables. Here we follow an alternative approach based on the spectral covariance in a model-independent way. We build upon previous work by Sharbaf et al. that studied the covariance of high quality SDSS continuum-subtracted spectra in a relatively narrow range of velocity dispersion ($\sigma\in [100,150]$\,km\,s$^{-1}$). Here the same analysis is applied to synthetic data from the EAGLE and Illustris TNG100 simulations, to assess the ability of these runs to mimic real galaxies. The real and simulated spectra are consistent regarding spectral covariance, although with subtle differences that can inform the implementation of subgrid physics. Spectral fitting done a posteriori on stacks segregated with respect to latent space reveals that the first principal component (PC1) is predominantly influenced by the stellar age distribution, with an underlying age-metallicity degeneracy. Good agreement is found regarding star formation prescriptions but there is disagreement with AGN feedback, that also affects the subset of quiescent galaxies. We show a substantial difference in the implementation of the AGN subgrid prescriptions, regarding central black hole seeding, that could lead to the mismatch.  Differences are manifest between these two simulations in the star formation histories stacked with respect to latent space.  We emphasise that this methodology only relies on the spectral covariance to assess whether simulations provide a true representation of galaxy formation.
\end{abstract}

\begin{keywords}
methods: data analysis – methods: statistical – techniques: spectroscopic – galaxies: evolution – galaxies: formation - galaxies: stellar content.
\end{keywords}


\section{Introduction}
Galaxy formation and evolution represent one of the most significant frontiers of astrophysics over the past decade. An exploration of the formation history of galaxies provides insights into the various physical processes involved in creating the stellar and gaseous components that we can observe through telescopes or investigate through simulations. Cosmological hydrodynamical simulations, such as EAGLE \citep{schaye:2015, Crain:15, McAlpine:16} and IllustrisTNG \citep{pillepich:2018, nelson:2018, springel:2018, Naiman:2018,marinacci:2018}, offer valuable insights when coupled with high-quality survey data, such as the Sloan Digital Sky Survey \citep{SDSS}, enhancing our understanding of galaxy evolution. There is a complementary role for both observation and simulation data. This is because observations are used to constrain various parameters in simulations, while simulations are used to interpret the observations with fundamental properties of galaxies.

Galaxies form and evolve as a result of the interaction between diverse physical processes that, in addition to gravity, influence baryonic matter. Given the inherent resolution limit of simulations that rely on a finite set of particles, or gridpoints, sub-grid physics is employed for the modeling of baryonic processes below the galactic scale, such as the formation of Black Holes (BHs), their growth, and feedback.  Incorporating these processes into simulations represents a significant challenge because these are complex physical processes and it is difficult to develop numerical algorithms that can accurately model their effects in a computationally efficient manner \citep[see, e.g,][]{SD:15,Naab:2017,CvdV23}. The simulations employ a variety of sub-grid models, including different criteria for the seeding of BHs, various models to compute BH accretion rates, various efficiency factors, and modeling techniques regarding the injection of energy from the AGN into the gas phase. In some models, AGN feedback channels are explicitly determined by the BH mass and the feedback receipt can vary, whereas uniform feedback is assumed in others, for which only one type of AGN feedback exists. Feedback from star formation is another important subgrid process that is expected to affect in a fundamental way the observed distribution of galaxies\citep[e.g.,][]{Schaye:2008}.
From a theoretical standpoint, state-of-the-art hydrodynamical simulations such as EAGLE \citep{schaye:2015} and IllustrisTNG \citep{pillepich:2018} reproduce the general fundamental properties of galaxies i.e. the evolution of the galaxy mass function \citep{Furlong:2015,Kaviraj:2017,pillepich:2018}, AGN luminosity \citep{rosas:2016, Volonteri:2016,McAlpine:2017} as well as the bimodality of galaxy color \citep{Trayford:2015, Trayford:2016, nelson:2018}, and the SFR and UVJ-based quenched fraction at $z \lesssim 2-3$ \citep{Donnari:2019,Donnari:2021}. Despite the good agreement between observational constraints and simulations, and the recent tremendous progress that has been made in these areas, there are still challenges to overcome. Non-trivial subgrid physics is therefore the major source of uncertainty in cosmological simulations, and adjusting these parameters can significantly alter results \citep{Okamoto:2005,Schaye:2010,Scannapieco:2012, Haas:2013, Haas:2013b,LeBrun:2014, Torrey:2014,NV:17}.

By comparing simulations and observations, sub-grid physics can be tested. Simulations and observations are frequently compared in papers \citep[e.g.,][]{Nelson:2015, pillepich:2018, vogelsberger:14, Habouzit:2021}, but the comparison with observational constraints can be challenging since these constraints often require physical modeling or assumptions. Using variance\footnote{In the strictest sense, we refer here to covariance, as we are dealing with multivariate analysis, but we use both terms with a similar meaning and prefer to use the term ``spectral variance''.} analysis, it is possible to obtain information from galaxy spectra in a model-independent manner, and without imposing physical constraints on the model \citep[e.g.,][]{Ferreras:06,Rogers:07,Rogers:10,variance}. Galaxy spectra encode the kinematics, age, and chemical composition of the stellar populations underlying them, thus representing one of the most reliable sources of information about galaxies. The spectral variance can be combined with stellar population synthesis models \citep[e.g.,][]{BC03,EMILES} to retrieve information from galaxy spectra. Following the methodology of \citet{Rogers:07}, we use a multivariate analysis method to explore the  retrieval of information from galaxy spectra on a model-independent basis. This method has been applied to a general sample of SDSS galaxies in \citet{variance}, hereafter referenced as PCA-SDSS. We performed principal component analysis on three separate groups of galaxy spectra: Star Forming (SF), AGN, and Quiescent (Q), based on the nebular emission properties. We emphasize that the variance of the input data in this work only relates to the absorption lines in the photospheres of stellar populations. The PCA-SDSS study analyses SDSS optical spectra using PCA to determine what physical phenomena contribute to the spectral variance, and suggested that galaxy structure may be controlled by a single (or a few) parameters, since only one component demonstrates a correlation with age, and plays a primary role as an evolutionary trend. In this study, we evaluate how the variance in the synthetic spectra created from the EAGLE \citep{schaye:2015} and IllustrisTNG \citep{pillepich:2018} simulations behaves in comparison with the optical spectra \citep{SDSS}, and evaluate the subgrid physics. A comparison of the different properties that are successfully reproduced and those that are not is made. We are interested in understanding how different sub-grid models can produce different spectral variance.

The structure of the paper is as follows: the sample and the simulations are presented in section \ref{sample}, followed by data pre-processing and an explanation of the restrictions in section \ref{pre-process}.
In section~\ref{resol-noise} we show how the synthetic spectra are produced from the simulations. The decomposition of optical spectra into principal components and projection of the synthetic and optical spectra to those principal components are explained in section \ref{method}, and the projections are explored in section \ref{sec-proj}, along with models of population synthesis. We discuss the results and present our conclusions in section \ref{sec-discussion}.

\section{Properties of the general sample}\label{sample}
This work adopts as observational reference and constraint a sample of 
optical spectra retrieved from the
Sloan Digital Sky Survey (SDSS, \citealt{SDSS}). The analysis is performed on a set of  spectroscopic synthetic data from the EAGLE (RefL0100N1504) simulation \citep{schaye:2015, Crain:15} and the IllustrisTNG (TNG100) simulation \citep{Genel:14, vogelsberger:14, Springel:10}. See the sections below for detailed descriptions of each sample and the applied restrictions to these samples regarding our analysis.

\subsection{Observational data (SDSS)}
The spectroscopic sample is taken from the SDSS archive\footnote{https://sdss.org}, in particular the Legacy dataset which contains single fiber spectroscopy at R=2,000 resolution \citep{Smee:13} from Data Release 16 \citep{DR16}. The spectra were de-reddened and de-redshifted using a linear interpolation algorithm with redshift and foreground dust estimates supplied by SDSS. The SEDs were normalized to the same average flux across the 6000-6500\AA\ rest-frame wavelength range.  The estimates of redshift, velocity dispersion, stellar mass (total) and star formation rate (total) are taken from catalogues {\tt galSpecInfo} and {\tt galSpecExtra} of the official SDSS database of the JHU-MPA group \citep{Brinchmann:04}. Accordingly, we optimize SDSS galaxy spectra constraints in terms of the variance analysis, particularly principal component analysis, based on PCA-SDSS. It is optimized in such a way as to minimize spurious signals in the variance unrelated to the physical phenomena underlying the presence of different galaxies. We restrict the stellar velocity dispersion to the range of 100--150\,km\,s$^{-1}$ and redshift to $z\in[0.05,0.1]$. We impose a threshold on the signal to noise ratio ($>$15 per $\Delta\log(\lambda/$\AA$)=10^{-4}$ pixel in the SDSS-r band ). With these thresholds applied, the final sample comprises 68,794 high quality spectra.

\subsection{Synthetic data (EAGLE and IllustrisTNG)} \label{synthetic-data}
Synthetic spectra are produced from simulated galaxies with the same instrumental signature as the SDSS observations, for a consistent comparison with the SDSS spectra. To produce spectra comparable to those from the 3-arcsec fibre-fed spectrograph, spectra and simulation parameters are extracted within a physical aperture with projected radius R=3\,kpc. At $z=0.1$, a standard $\Lambda$CDM cosmology maps the 3\,arcsec diameter fibre into a physical distance of 5.5\,kpc, so this choice is optimal for our sample. Choosing other areas in the range R$\sim$2-5\,kpc produces indistinguishable results. For each galaxy, the spectra corresponding to all stellar particles within this radius of the galactic centre are combined. Note that the stellar mass, which is used to characterize the overall properties of a galaxy, for both simulation and SDSS data, refers to the whole galaxy, not within a 3\,kpc aperture.  Given that the SDSS galaxy redshift range is from $z=0.05$ to $0.1$, we take the $z=0.1$ snapshot of the simulations. This is a valid approximation as the evolutionary differences between $z=0$ and $0.1$ are minimal for this study. 

We give below a very brief description of the two cosmological models used in this study, with emphasis on the modelling of the star formation and AGN activity, which are the fundamental parts of the subgrid physics that control the star formation histories that eventually determine the spectra. For more details, the reader should  check the references listed below, as well as the general description presented in Section~2 of \citet{Angthopo:2021} that follows a similar approach to select galaxies from the simulations.

\subsubsection{The EAGLE (RefL0100N1504) simulation}\label{EAG}
The EAGLE simulations \citep{schaye:2015,Crain:15,McAlpine:16} encompass a series of numerical hydrodynamical runs in a cosmological context, featuring various box sizes and resolutions. As part of the implementation, various theoretical considerations have been taken into account, such as radiative cooling, stellar feedback, star formation, and the seeding and feedback of black holes \citep{Schaye:2008, Wiersma:2009, Dalla:2012, rosas:2015, schaye:2015}. In each run of the EAGLE simulation, the stellar and black hole feedback has been calibrated differently to reproduce a set of observables. In this study, we focus on the fiducial EAGLE simulation RefL0100N1504, denoted hereafter EAGLE. The computational engine is a modified version of GADGET 3 \citep{springel:2005}. This simulation was executed using the ANARCHY code \citep{Schaller:2015}, based on the Smoothed Particle Hydrodynamics (SPH) technique. This simulation is characterized by a comoving box size of L$=$68$h^{-1}$\,Mpc (equivalent to $\sim$100\,Mpc), housing $1504^{3}$ dark matter (DM) particles, and an equal number of baryonic particles. The baryonic particle mass is $m_{\rm b}$$=$$1.81\times 10^{6}$\(M_\odot\), and the dark matter particle mass is $m_{\rm DM}$$=$$9.70\times 10^{6}$\(M_\odot\). A $\Lambda$CDM cosmological framework is assumed, adopting the \citet{planck:2014} parameters as a reference: 
$\Omega_{m}$$=$0.307, $\Omega_{\Lambda}$$=$0.693, $\Omega_{b}$$=$0.048, $h$$=$0.6777, 
$\sigma_{8}$$=$0.8288.

EAGLE simulates subgrid physics in a manner that reproduces many of the observed galaxy scaling relations -- more specifically the galaxy stellar mass function and the relation between galaxy mass and central BH mass --  and produces galaxies with the observed size (i.e. effective radius) distribution. Using the assumption that star-forming gas is self-gravitating, \cite{Schaye:2008}, the star formation rate is determined stochastically from gas pressure rather than gas density, providing a better match to the Kennicutt–Schmidt law. Moreover, a metallicity-dependent star formation threshold proposed by \citet{schaye:2004} is imposed, motivated by the fact that a certain density of cold, dense gas is required for star formation to occur. Gas cooling occurs at a lower density and pressure, in metal-rich gas, which allows a metallicity-dependent threshold for star formation.

As a result of feedback associated with either star formation or black hole accretion, a characteristic scale is produced in the stellar mass function, represented by M$^\ast$ that locates the 'knee' of the distribution. It is essential that both processes are efficient in order to reproduce the observed population of galaxies. As a result of the lack of resolution, simulations suffer from an 'overcooling' problem in terms of stellar feedback. When we are unable to model self-consistent outflows from feedback injected on the scale of individual clusters of stars, too much gas converts into stars too early, which is incompatible with the formation of high-mass galaxies.  This limitation is addressed by implementing a method \citep{Dalla:2012}, that makes stellar feedback a stochastic thermal process, thus enabling the control of energy obtainable per event of stellar feedback. 

An essential component of the EAGLE simulations is AGN feedback associated with the growth of BHs, which mostly quenches star formation in massive galaxies and shapes the gas profiles of the host haloes. A black hole seed of mass $m_{\rm seed} = 1.48 \times 10^{5} $ \(M_\odot\) is placed at the center of every halo more massive than $M_{h,{\rm thresh}} = 1.48 \times 10^{10}$ \(M_\odot\)  that does not already contain a black hole \citep{springel:2005}. This is done by converting the gas particle with the highest density into a BH that acts as a collisionless particle. AGN feedback is treated like star formation feedback – energy is injected thermally and stochastically. The two major modes of AGN feedback are quasar- and radio-modes. At present, the simulations do not have the resolution necessary to differentiate between the two \citep{Naab:2017}. Consequently, EAGLE simulations implement only one mode of AGN. It has been determined that the chosen method behaves similarly to quasar-mode feedback in so far as the input thermal energy rate is proportional to the gas accretion rate at the location of the SMBH.

\subsubsection{The IllustrisTNG (TNG100) simulation}\label{TNG}
The IllustrisTNG project comprises a set of simulations: TNG50, TNG100, and TNG300 \citep{marinacci:2018, Naiman:2018, nelson:2018, pillepich:2018, springel:2018}. They represent improved simulations over the original Illustris project \citep{Genel:14, vogelsberger:14} based on the moving-mesh AREPO code \citep{Springel:10}. In our analysis, we take the publicly available datasets from the TNG100 run \citep{nelson:2019}. It has a box size L=75$h^{-1}$\,Mpc$\sim$110\,Mpc, and  shares its initial conditions with the previous Illustris simulation. The initial conditions of the density field are determined at redshift z$=$127, and adopt the \citet{planck:2016} cosmological parameters, with matter density $\Omega_{m}$$=$0.3089, baryon density $\Omega_{b}$$=$0.0486, dark energy density $\Omega_{\Lambda}$$=$0.6911, Hubble constant $H_{0}$$=$100 h\,km\,s$^{-1}$\,Mpc$^{-1}$ with $h=0.6774$, power spectrum normalization $\sigma_{8} = 0.8159$ . This simulation has an equal number of initial gas cells and dark matter particles, N$_{\rm gas}$$=$N$_{DM}$$=$1820$^{3}$. In this simulation, the mass of a DM particle is $m_{\rm DM}=7.5 \times 10^{6} M_\odot$ and the typical mass of a baryonic resolution element is $m_{\rm b}=1.4\times 10^{6} M_\odot$.

As in EAGLE and the original Illustris simulation, subgrid physics includes radiative cooling, star formation, SN feedback, black hole formation and growth, and feedback from AGN. The approaches used in the IllustrisTNG simulations for stellar feedback, enrichment, and the low-mass end of the galaxy stellar mass function (GSMF) are presented in \citet{Pillepich2018TNG_SFH}.
The AGN feedback model and the high-mass end of the GSMF are described in detail by \citet{weinberger:2017}. The IllustrisTNG model significantly improved the original Illustris project in several areas: stellar evolution, gas chemical enrichment, growth, and feedback from SMBHs and galactic winds. TNG100 tracks star formation by gas stochastically converting into star particles if its density exceeds a critical threshold, $n_{H} \simeq 0.1 cm^{-3}$. This threshold is necessary to produce the observed Kennicutt-Schmidt relation.

Feedback associated with star formation drives galactic scale outflows. These outflows originate from star-forming gas isotropically and have a wind velocity that scales with local DM velocity dispersion. Kinetic wind schemes are considered. The wind particles are generated stochastically and hydrodynamically decoupled till they leave the local ISM. The wind mass loading for a given speed is given by the available SN energy. In addition, the wind metal content is assumed to be a constant fraction of the ISM value. A hydrodynamic recoupling occurs between wind particles outside the dense ISM, allowing them to deposit their mass, momentum, metals, and thermal energy.  A detailed description of these galactic-scale, star formation-driven, kinetic winds can be found in \citet{pillepich:2018}.

The implementation of feedback from BHs in TNG100 includes feedback injection at low accretion rates in the form of a kinetic, and supermassive BH-driven wind to minimize discrepancies in comparison to observational data at the massive end ($10^{13} - 10^{14}$ \(M_\odot\)) of the halo mass function. At high accretion rates, the TNG100 model invokes thermal feedback that heats gas surrounding the BH. The BH seeding is based on dark matter halo mass: when the halo mass surpasses a threshold,  $M_{h, {\rm thresh}} = 7.38 \times 10^{10}$ \(M_\odot\), a BH with a mass of $ M_{\rm seed} = 1.8 \times 10^{6}$ \(M_\odot\) is seeded \citep{weinberger:2017}. Following the observational trend of a mass-dependent change of the AGN mode -- quasar vs radio mode, with the radio mode occurring predominantly in evolved high-mass galaxies \citep{best:2005} -- the adopted  switch in AGN feedback mode in the simulation is based on a specific threshold, a stellar mass of roughly $10^{10.5}$ \(M_\odot\).

\section{Sample selection}\label{pre-process}
In this section we give details about the way our working samples have been extracted from the observational survey and the simulations. The sample is classified according to their activity into star forming, AGN or quiescent. This classification is typically done with nebular emission lines in observational data, but simulations offer us the benefit of focusing on the fundamental parameters that control this activity. We explore the classification process of the simulated data in some detail below. Splitting the sample into these three subsets is especially important to assess how well simulations trace star formation quenching, noting that any variance analysis is strongly dependent on the working sample. The separation allows us to focus on the most relevant spectral features in these three key evolutionary phases. Our analysis depends on choosing samples whose statistical distributions are compatible with respect to a chosen fundamental parameter that strongly correlates with stellar population properties. Following \citet{Angthopo:2021}, we choose total stellar mass as the fundamental parameter. While stellar velocity dispersion shows a stronger correlation, stellar mass is a more reliable parameter in simulations. In this section we also discuss the need for a sample homogenisation (mass matching) between samples. This process is necessary as the starting samples have different mass distributions. Given that stellar mass strongly correlates with stellar population content, differences in the subsamples would mostly be caused by a systematic in the mass distributions. In order to ensure a proper comparison of galaxy spectra in the three evolutionary phases, a careful mass homogenisation is needed.

\begin{figure}
    \centering
    \includegraphics[width=0.5\textwidth]{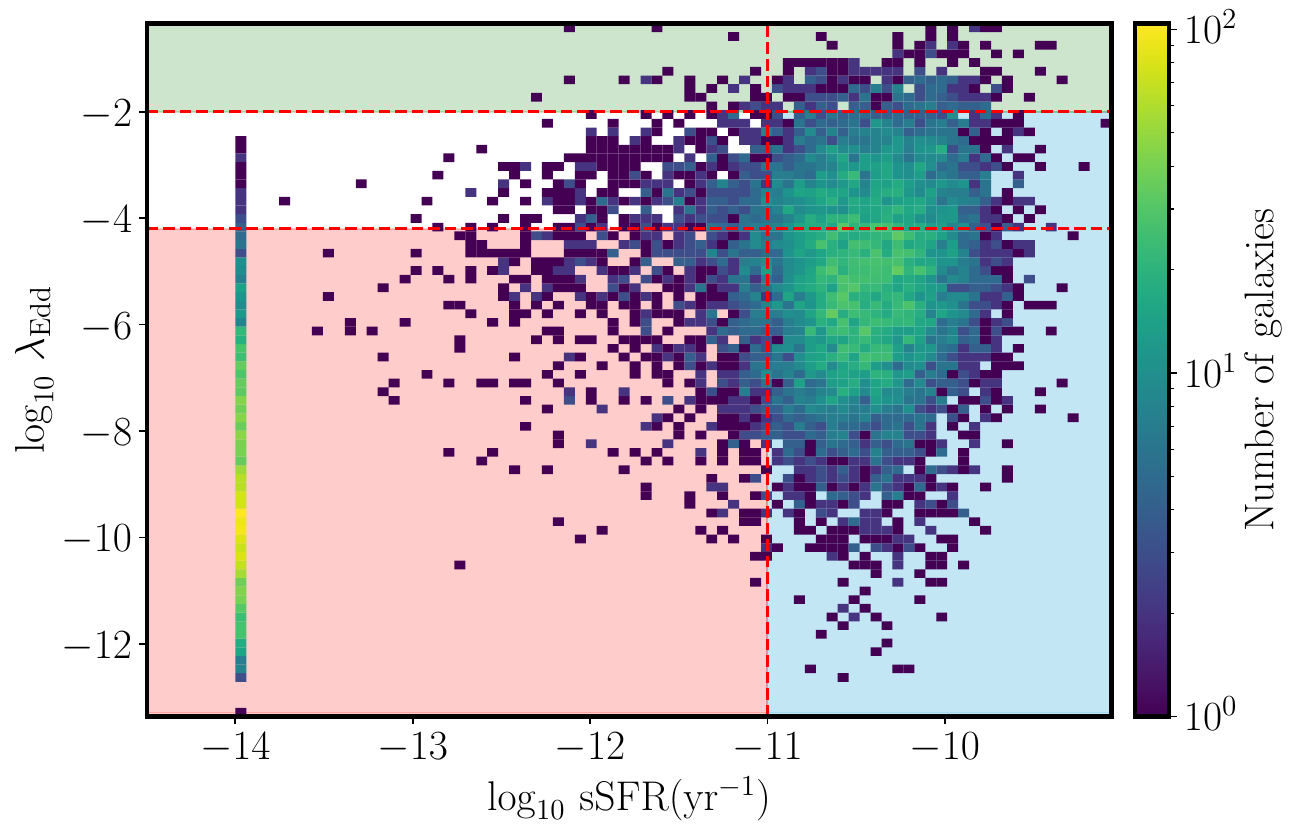}
    \includegraphics[width=0.5\textwidth]{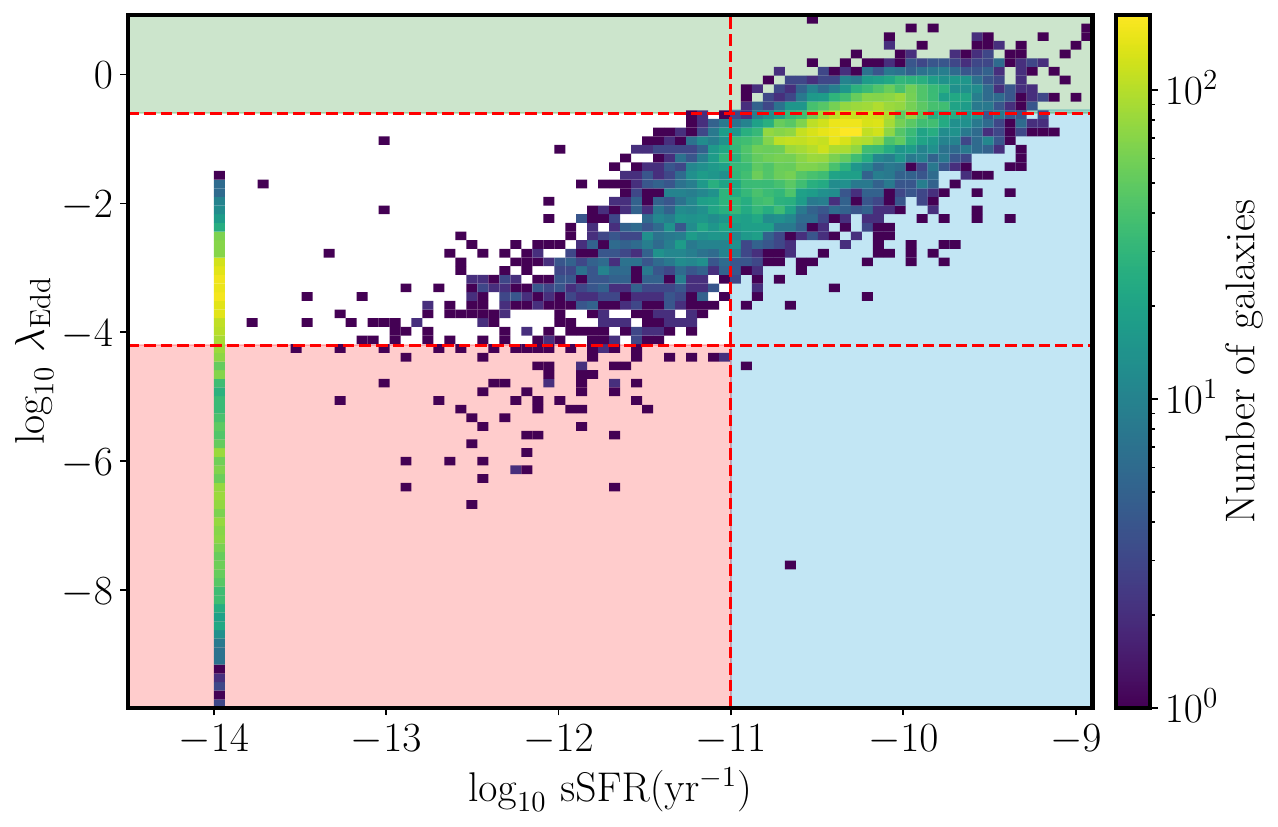}
    \caption{Galaxy classification based on $\lambda_{\rm Edd}$ and
      sSFR in simulations in EAGLE ({\sl top}) and TNG100 ({\sl
        bottom}). The light blue, green, and red regions, show our
      choice for SF, AGN, Q galaxies, respectively. Galaxies with
      zero star formation rate shown with $\log_{10} \, \mathrm{sSFR}(\mathrm{yr}^{-1}) = -14$.}
    \label{fig1}
\end{figure}

\subsection{SDSS}\label{pre-SDSS}
The continuum is removed from spectra following the high percentile method from  \citet{BMC:10}, liberally defined as the ``boosted median'' continuum (BMC). It is critical to note that continuum subtraction entails the removal of information, to eliminate systematic errors caused by dust reddening or residual flux calibrations. The PCA study focuses on the two spectral intervals of 3800 to 4200 \AA\ and 5000 to 5400 \AA\ . These intervals preserve most of the variance from galaxy to galaxy \citep[see, e.g.][]{InfoPop}. This paper will refer to these two spectral regions as the “blue” and “red” intervals, respectively. We emphasize that the variance of the input data in this work only relates to the absorption lines in the photospheres of the stellar population, so spectral regions with prominent emission lines are not included in the analysis, as presented in PCA-SDSS.

The SDSS spectra are classified with respect to nebular emission into star-forming (SF), AGN, and quiescent(Q) galaxies, following the standard BPT line ratio diagnostics \citep{BPT}, reported with parameter BPT in the {\tt galSpecExtra} table from the official SDSS database of the JHU-MPA group \citep{Brinchmann:04}. In that catalogue, a BPT flag is included for each galaxy spectrum, that characterises the type of ionisation into star formation and AGN at different levels, or quiescence defined by the absence of emission lines. Regarding the latter, we select those with a BPT flag of $-$1, along with an upper threshold on the equivalent width of $H\alpha$ emission, analogously to \cite{CF:11}. Note that a $-1$ value of this flag only means the spectrum cannot be located on the BPT diagram, therefore this classification may still include galaxies with H$\alpha$ emission (but, e.g. no measurable [N{\sc II}] due to a problem in the data). We impose that spectra with equivalent widths in emission higher than $5$\AA\ are rejected as quiescent, even if the flag is set to $-1$. For reference, \citet{CF:11} define as passive galaxies those with $\log W_{H\alpha}($\AA$)<0.5$. In our work, the BPT classification is considered as a means of selecting the strongest members of each subsample. Our motivation is to apply the variance analysis to bona fide cases of SF, AGN and Q galaxies, not to borderline cases. In this way, we ensure that the comparison between observations and simulations is as clear as possible. Therefore we select SF galaxies only with BPT = 1 (i.e. omitting composite and weak star-forming systems), and AGN as those with BPT = 4 (omitting low S/N LINERs). PCA is applied to these three subclasses of spectra independently. Separate analyses allow us to examine the individual characteristics of these three categories, as well as assess the distribution of variance between the three categories; PCA-SDSS provides a detailed discussion on this point. In addition, once PCA is applied to the data, we explore the standard deviation of the distribution as a function of wavelength to pinpoint problematic data in each spectral interval independently, resulting in a decreased size of each subsample. In the blue interval, the number of galaxies is reduced from 23168 to 17473 (Q); from 10495 to 8025 (SF), and from 3343 to 2620 (AGN). In the red interval, the samples are reduced from 23168 to 13953 (Q), from 10495 to 6319 (SF), and from 3343 to 2019 (AGN). A final homogenisation of the SDSS and synthetic spectra further changes the number of galaxies within each subsample (see section~\ref{Homog} below).

\begin{figure}
    \centering
    \includegraphics[width=0.45\textwidth]{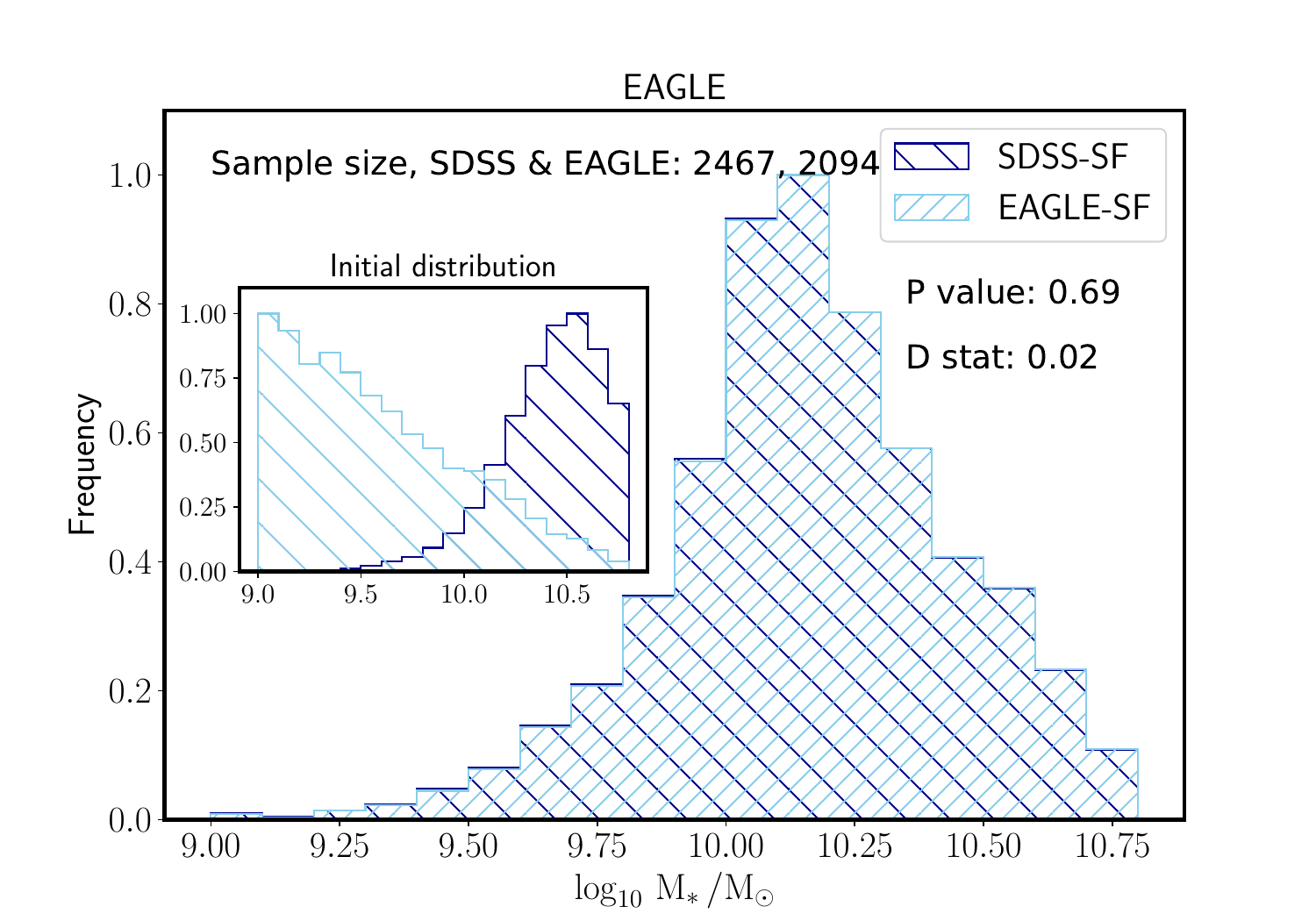}
    \includegraphics[width=0.45\textwidth]{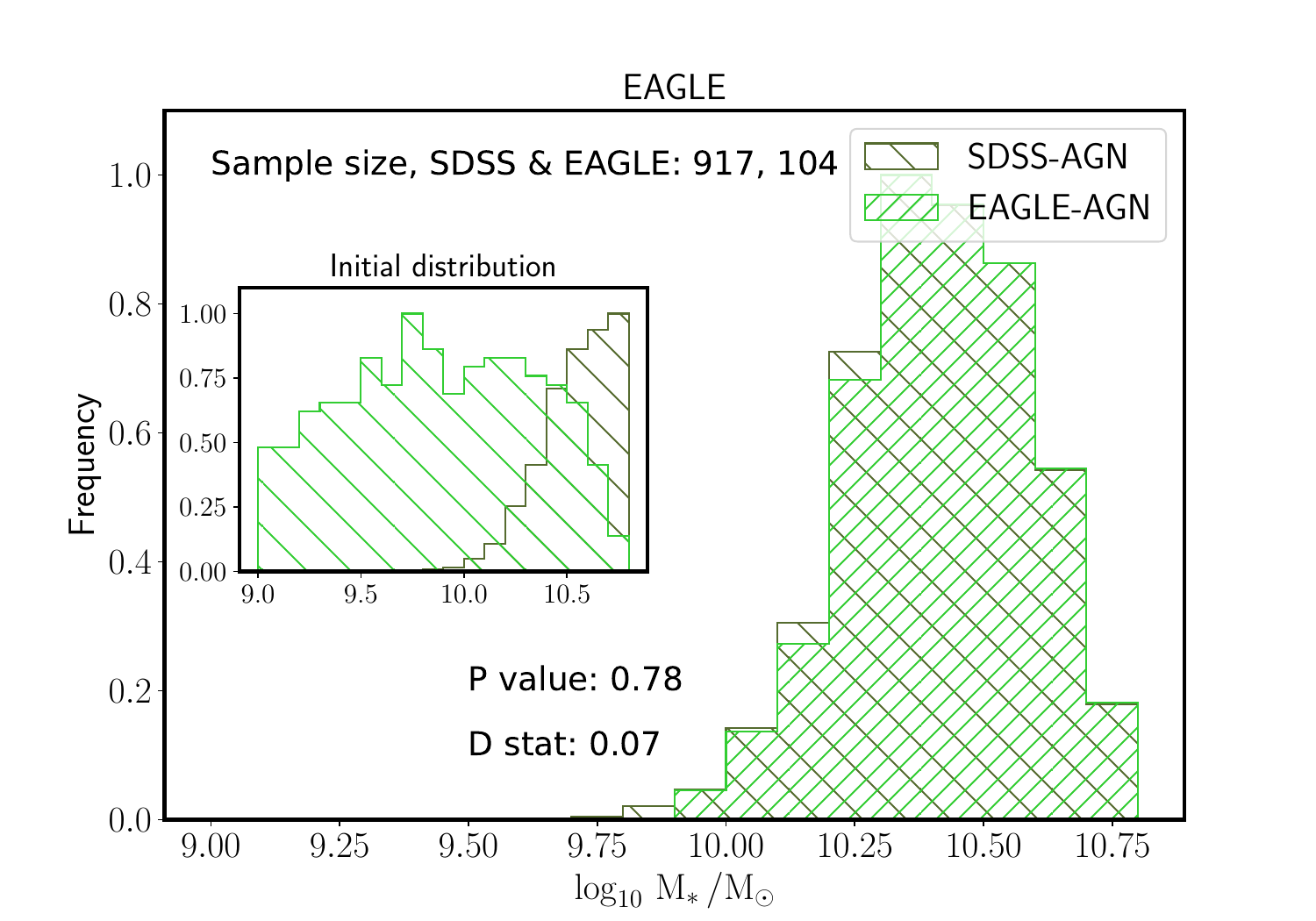}
    \includegraphics[width=0.45\textwidth]{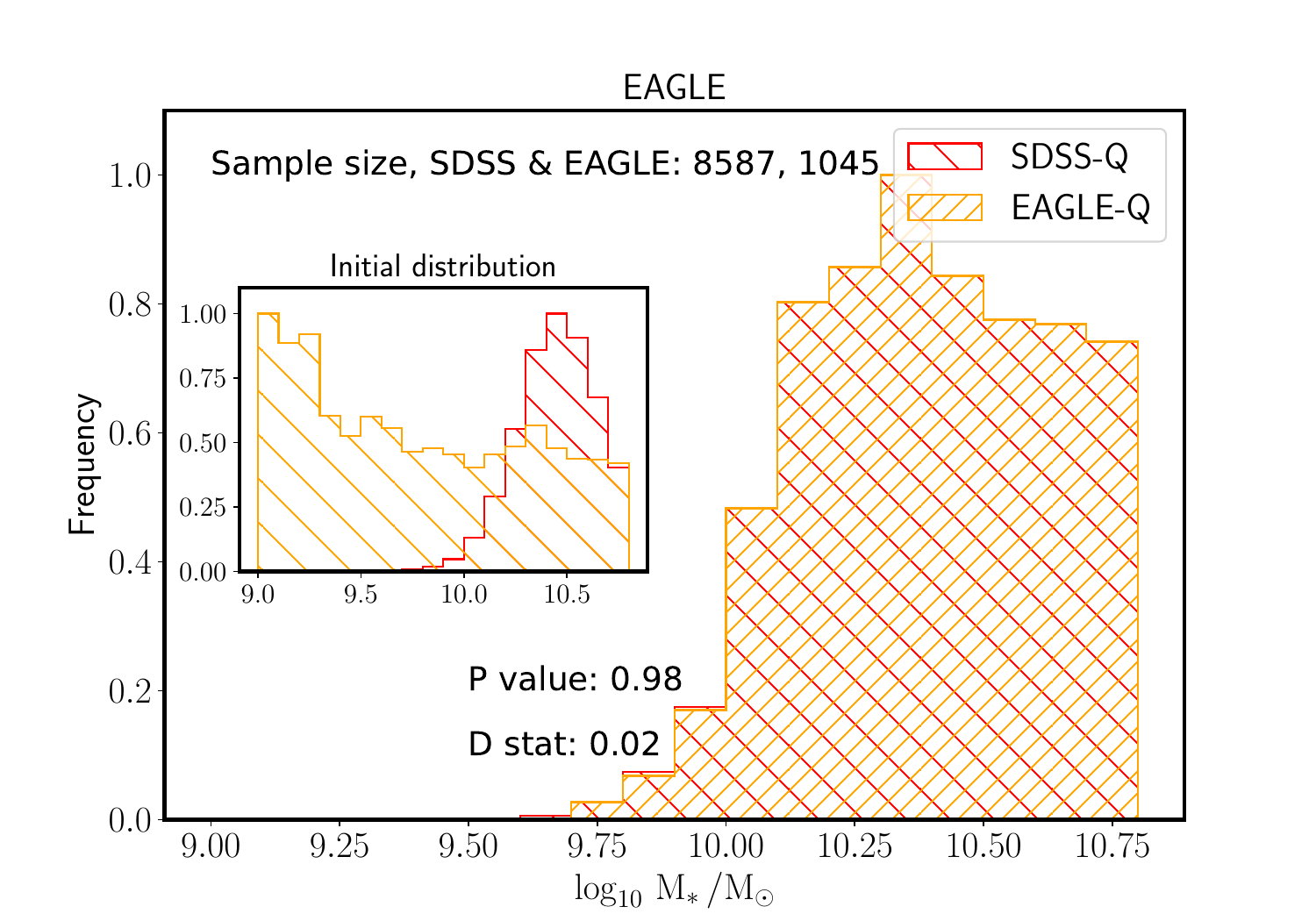}
    \caption{Distribution of stellar mass before and after homogenisation between observed (SDSS) and simulation data (EAGLE). The blue, green, and red histograms correspond to SF, AGN, and Q galaxies, respectively. The inset panels show the distribution of SDSS and EAGLE galaxies in each subsample {\sl before homogenisation}. A KS test confirms that the final distributions originate from the same parent sample. Each panel shows the corresponding $D_{\rm KS}$ and $p$ statistic. The sample sizes after homogenisation are labelled in each panel. Note that the range of stellar mass is limited to $10^{9}-10^{11}$M$_\odot$, see section~3.3 for more details.} 
    \label{fig2}
\end{figure}

\subsection{EAGLE and TNG100 classification}
From an observational perspective, a fundamental classification scheme of galactic activity concerns the presence of star formation, an active galactic nucleus, or the lack of such activity (quiescence). These stages can be readily measured in the gas phase, following standard schemes based on ratios of selected, strong emission lines  \citep{Kewley:19}. In synthetic data, an equivalent exercise involves the post-processing of the baryonic (gas and stellar particles) component with a code that incorporates the details of the sources, along with the radiative transfer that should also include dust scattering and absorption \citep[e.g.,][]{Hirschmann:2023}. While this could be a possible approach, in this work we want to minimise the underlying systematics, keeping the post-processing to a minimum. Therefore, we opted for a classification scheme of SF/AGN/Q galaxies based on the direct parameters of the simulations that track the evolution of the star formation and BH accretion.
Following \citet{Angthopo:2021}, we utilize the specific star formation rate, 
sSFR (defined as SFR/M$_s$), and the SMBH growth parameter $\lambda_{\rm Edd}$ as classification criteria. The latter is 
defined as follows:
\begin{equation}
    \lambda_{\rm Edd} = \frac{\dot{m}_{acc}}{\dot{m}_{\rm Edd}},
\end{equation}
According to observational studies, Seyfert AGN have an Eddington ratio in the range 
$-2 < \log(\lambda_{\rm Edd}) < -1$ \citep{Heckman:2004, schaye:2015, Ciotti:2017, Georgakakis:2017}, whereas lower accretion rates are associated with radiatively ineffective AGN, LINER, or even the absence of AGN. For LINER-like AGN some studies suggest a lower limit around $\log(\lambda_{\rm Edd}) \sim -6$ \citep{Heckman:2004, Li:2017}. In these systems, studies based on SDSS galaxies find the Eddington ratio from [O{\sc III}] emission, $\log(\lambda_{\rm Edd}) \sim -4$ \citep{Kewley:2006}, while others choose values as low as $\log(\lambda_{\rm Edd}) \sim -9$ \citep{Ho:2008, Ho:2009}. Regarding star formation activity, the sSFR allows us to differentiate between the star-forming and quiescent galaxies. 
The sSFR has been calculated by using the instantaneous SFR. It is also possible to use the average SFR over a period of time, however, previous studies have shown that measuring SFR in different ways has little significance at low redshift \citep{Donnari:2019, Donnari:2021}. The sSFR threshold of $\log(sSFR)\gtrsim -11$ is 
typically adopted in the literature to select star-forming galaxies \citep{Scholz:2023}. 
\begin{figure}
    \centering
    \includegraphics[width=0.45\textwidth]{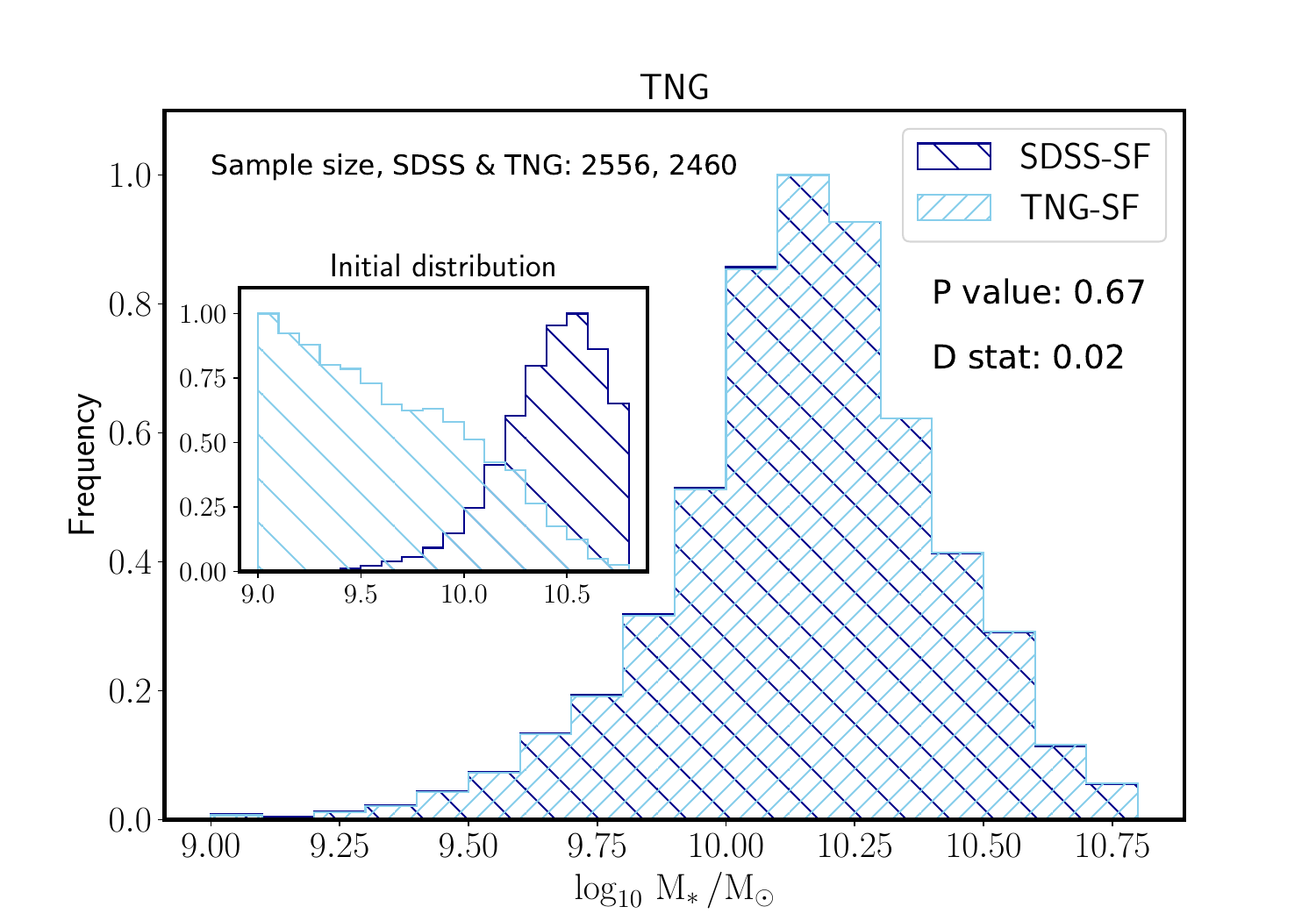}
    \includegraphics[width=0.45\textwidth]{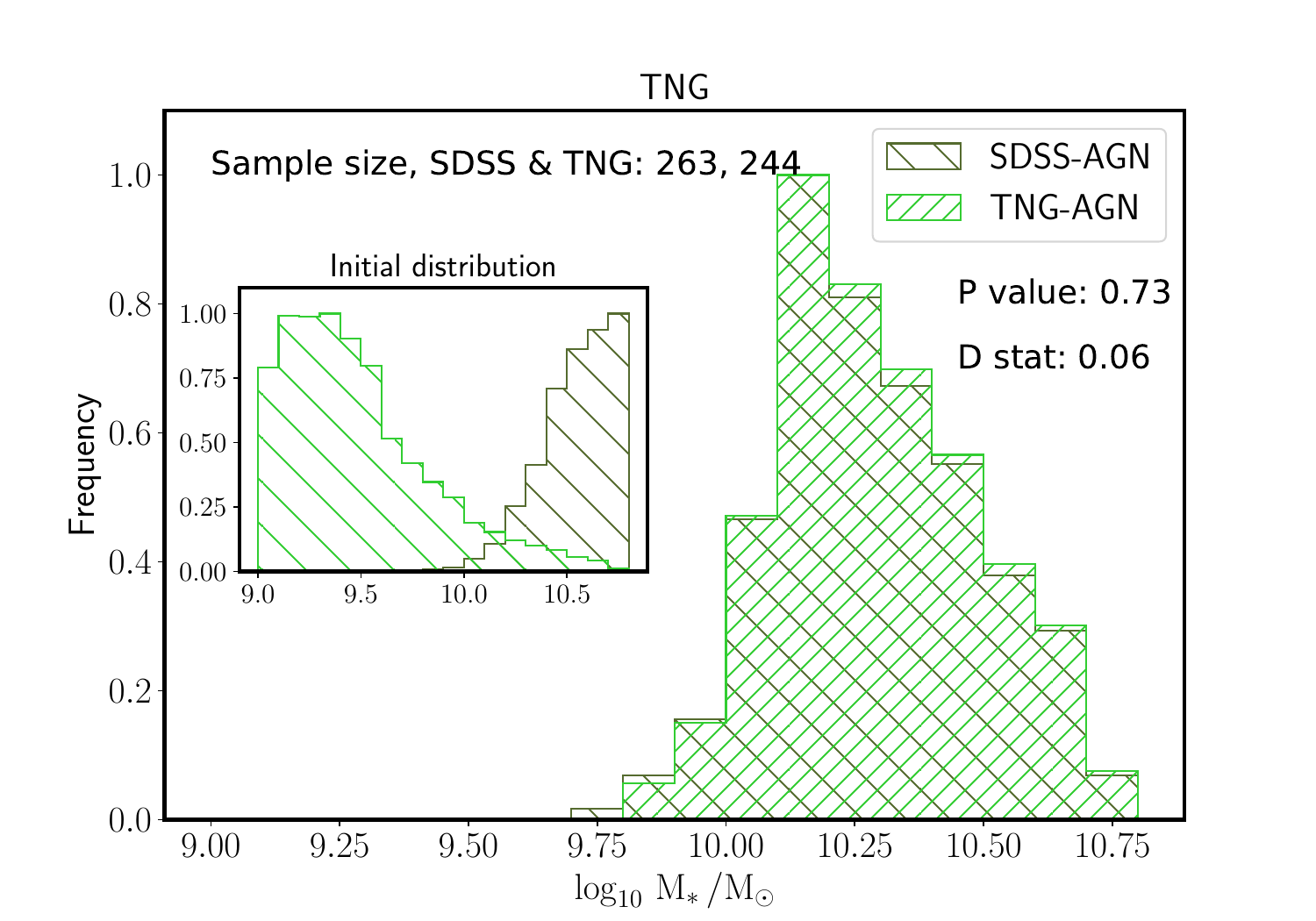}
    \includegraphics[width=0.45\textwidth]{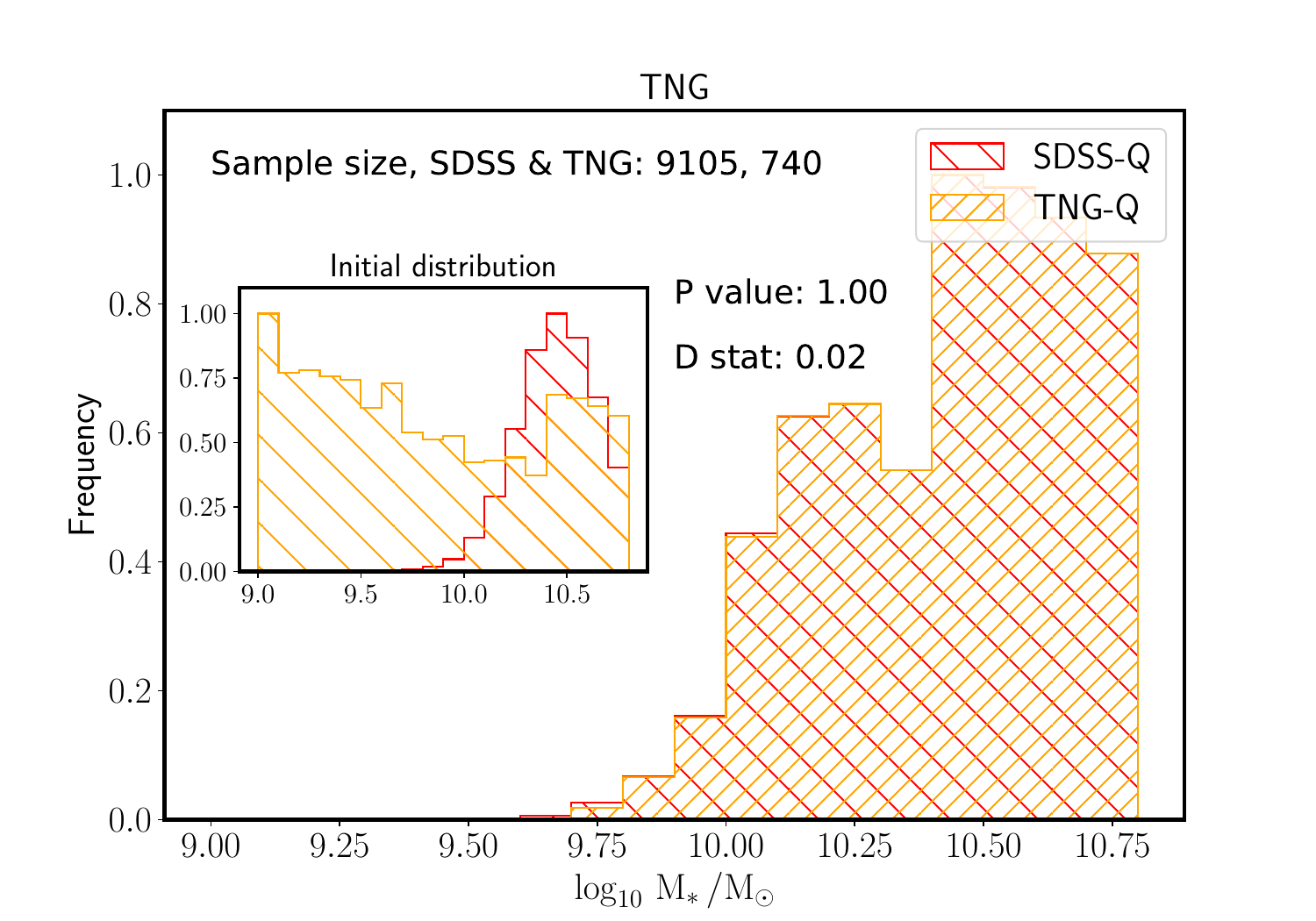}
    \caption{Equivalent of Fig.~\ref{fig2} for the homogenisation process between SDSS and TNG100.}
    \label{fig3}
\end{figure}

Figure \ref{fig1} shows the distribution of simulated galaxies on the $\lambda_{\rm Edd}$ versus sSFR plane for the EAGLE (top) and TNG100 (bottom) galaxies. The regions of high star-formation, strong AGN activity and quiescence are represented by the light blue, green, and red shaded regions, respectively. We note here that the reason for the choice of the range of parameters is to reproduce the same ratios of strong star forming (bpt flag of 1), strong AGN (bpt flag of 4) and quiescent (bpt flag of -1 and weak/non-existent H$\alpha$) galaxies as in the SDSS data \citep[see][]{Angthopo:2021}. The blue region in the top panel of figure \ref{fig1} for the EAGLE sample, is associated with the galaxies with $\log\,$sSFR$>-11$, highly star-forming galaxies, and $\lambda_{\rm Edd}<-$2, with different levels of AGN activity. In the bottom panel of Fig.~\ref{fig1}, that corresponds to TNG100, the star forming region is associated with $\log\,$sSFR$>-11$ and $\lambda_{\rm Edd}<-0.6$. Classifying the star-forming galaxies with the different levels of AGN activity is physically motivated. Note that AGN galaxies will often undergo star formation, and it is only through the comparison between sSFR and $\lambda_{\rm Edd}$ that the equivalent of a SF or AGN BPT classification is reproduced. In both panels of Fig.~\ref{fig1}, the green region indicates the range of parameters with high AGN activity: for EAGLE we choose $\lambda_{\rm Edd}>-2$ and for TNG100 $\lambda_{\rm Edd}>-0.6$. There is a difference in thresholds between the TNG100 and EAGLE samples for this parameter because the feedback associated with AGN activity turns on strongly at a specific stellar mass of $10^{10.5}$ \(M_\odot\) in TNG100, leading to large numbers of galaxies with high AGN activity and large $\lambda_{\rm Edd}$, with small scatter with respect to sSFR. The red region represents quenched galaxies, with $\log\,$sSFR$<-11$ and $\lambda_{\rm Edd}<-4.2$ for both EAGLE and TNG100. As seen in Fig.~\ref{fig1}, there is a substantially lower fraction of quiescent galaxies with a measurable sSFR (the red shaded rectangle), with respect to EAGLE. This is mainly caused by the strong correlation between sSFR and $\lambda_{\rm Edd}$ in TNG100. Therefore, the vast majority of Q galaxies in TNG100 are those fully quenched, i.e. with a zero SFR. Note the larger scatter in EAGLE between these two parameters. In the figure, galaxies with no star formation are assigned $\log\,$sSFR$=-14$. Finally, note that in the TNG100 and EAGLE samples, out of 21563 and 13173 galaxies, respectively, with stellar masses ranging from $10^{9}-10^{11}$M$_\odot$, only 18094 and 12384 galaxies, respectively, have black hole masses. We need the measurement of the black holes to determine $\lambda_{\rm Edd}$ for the assessment of the Q/AGN/SF identification. Moreover, most of those galaxies without a SMBH are at the lower mass end (i.e. whose dark matter halos did not cross the imposed threshold for black hole seeding).

\subsection{Homogenisation of simulated and observational data}
\label{Homog}

When studying the differences between observations and simulations, it is important to make sure that comparable galaxy samples are defined. Most importantly, it is necessary to ensure that the stellar mass distribution is consistent between the samples, otherwise, as a result of the well-known correlation between stellar population properties and stellar mass or velocity dispersion \citep{Bernardi:03, Gallazzi:05, Ferreras:19}, the differences will be merely caused by the systematically dissimilar distributions. As a consequence of different selection effects between observations and simulations, samples exhibit incompatible stellar mass distributions. More specifically, the Malmquist bias imposed by the SDSS-$r$$<$17.77\,AB magnitude limit for spectroscopic follow-up \citep[see, e.g.][]{Abolfathi:2018} implies that low-mass galaxies (with stellar mass $M_s\lesssim 10^{9}\,M_\odot$) are missed in the SDSS spectra, with a clear redshift-dependent trend. On the other hand, simulations are biased against high-mass galaxies ($M_s \gtrsim 10^{10} M_\odot$) due to the volume limitation \citep[see, e.g.][]{schaye:2015}. We must therefore ensure that the distributions between these two data sets are statistically compatible with respect to stellar mass, in order to make a fair comparison, and this is why homogenisation is needed.

We select sets that are 'homogeneous' from the original samples regarding stellar mass. The stellar mass is limited in the range $10^{9}-10^{11}$M$_\odot$. Although the stellar mass range of the original set of SDSS  galaxies is wider, $10^{7.2}-10^{12.3}$M$_\odot$, we limit the analysis to a more conservative range to avoid  the systematic problems of producing the galaxy population at the low mass (resolution limited) and high mass (volume-limited) ends of the galaxy distribution in simulations. After restricting  the stellar mass range, the number of galaxies in the original samples change as follows: SDSS: from 68,794 to 66,483, TNG100: from 80,323 to 21,563, and EAGLE: from 78,275 to 13,173. As expected, the change is more drastic in the simulated data.
Note that the original SDSS sample from PCA-SDSS is chosen with respect to stellar velocity dispersion, $\sigma\in$[100,150]\,km\,s$^{-1}$. This interval aims at minimising the blurring effect of the kinematic kernel that intrinsically removes information from the spectra, but also to ensure a diverse range of galaxies.
We note that in the simulations the stellar mass is a more accurate parameter to define a galaxy, instead of stellar velocity dispersion, that can only be computed for the rather massive stellar particles. Therefore, we choose stellar mass as the main selection criterion \citep[see also][]{Angthopo:2021}. As mentioned, stellar mass and velocity both correlate in a similar way with respect to the properties of the stellar populations.

\begin{figure}
    \centering
    \includegraphics[width=0.45\textwidth]{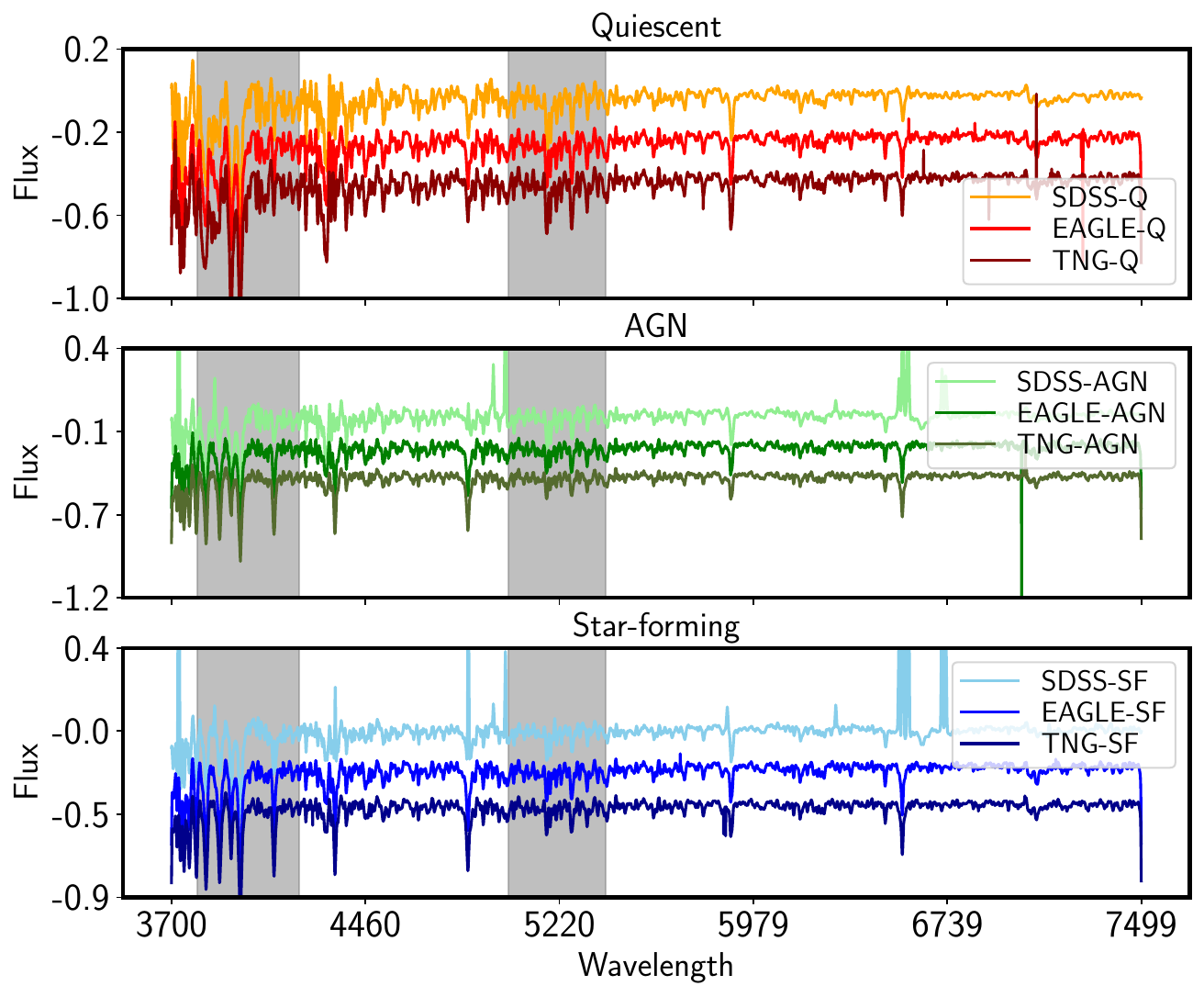}
    \caption{Comparison between stacked spectra of observed and simulated galaxies after all the processes are applied, including separation into different groups, homogenisation, and spectral synthesis. Gray areas indicate (from left) the blue and red spectral intervals explored, following PCA-SDSS.}
    \label{fig:eigenspectra}
\end{figure}

Comparisons should be made separately between SDSS and EAGLE, and between SDSS and TNG100. Our first approach for homogenisation is based on a whole galaxy sample compatible with the stellar mass range of SDSS. However, we add a new selection based on the SF/AGN/Q activity of the galaxies. For the simulations, we applied cuts on the bivariate distribution spanned by  $\lambda_{\rm Edd}$ and the specific star formation rate, to mimic the BPT classification performed in the spectra of real galaxies. After imposing this cut, we find that the stellar mass distributions of the subsamples differ, despite having homogenised the original samples. Appendix~\ref{A} provides a detailed explanation of this issue. To solve this problem, we changed the order of homogenisation and classification, so that the final, segregated samples are statistically compatible in stellar mass. 


Each subgroup of galaxies is considered in the pair of EAGLE-SDSS and TNG100-SDSS to homogenise them, following \citet{Angthopo:2020}. Figures \ref{fig2} and \ref{fig3} show stellar mass distribution before and after homogenisation between observed (SDSS) and simulation data (EAGLE, TNG100). Using a bin size of 0.1, the stellar mass is binned between $10^{9}$ and $10^{11}$\,M$_\odot$, and the galaxy fraction in simulated and observed samples is calculated in each stellar mass bin, considering the total number of galaxies in the sample. In bins of the same mass between the comparison samples, we randomly cull galaxies from the set with the higher fraction to match the other set. The homogenisation process is carried out separately for star-forming galaxies, AGN galaxies, and quiescent galaxies. A KS test confirms that these distributions originate from the same parent sample. The $D_{KS}$ and $p$ statistics for the test are shown in each panel of Figures \ref{fig2} and \ref{fig3}.  Please refer to Appendix~\ref{A} for a detailed discussion of a more general approach of homogenising the total sample first and then separating the galaxies into different categories. A comparison of the number of galaxies in each subsample before and after homogenisation is given in table \ref{tab:homogenisation}, which compares the EAGLE-SDSS pair with TNG100-SDSS pair.

\begin{table}
\centering
\resizebox{\columnwidth}{!}{%
\begin{tabular}{lccc}
\toprule
\textbf{Sample} & \textbf{Subgroup} & \textbf{Before Homogenisation} & \textbf{After Homogenisation} \\
\midrule
\multirow{6}{*}{EAGLE-SDSS} & SF  & 8743  & 2094 \\
                              & AGN & 357   & 104  \\
                              & Q   & 2794  & 1045 \\
                              & SF (SDSS) & 10289 & 2467 \\
                              & AGN (SDSS) & 3087 & 917  \\
                              & Q (SDSS) & 22895 & 8587 \\
\midrule
\multirow{6}{*}{TNG100-SDSS} & SF  & 9916  & 2460 \\
                               & AGN & 2860  & 244  \\
                               & Q   & 1867  & 740  \\
                               & SF (SDSS) & 10289 & 2556 \\
                               & AGN (SDSS) & 3087 & 263  \\
                               & Q (SDSS) & 22895 & 9105 \\
\bottomrule
\end{tabular}%
}
\caption{Number of galaxies in each subsample before and after homogenisation for EAGLE-SDSS and TNG100-SDSS pairs.}
\label{tab:homogenisation}
\end{table}


\section{Defining SDSS-like synthetic spectra}
\label{resol-noise}

Once the simulated samples are chosen and homogenised, we need to produce the synthetic spectra to project on to the eigenvectors derived from the SDSS data. A stellar particle in the simulations represent a population with a well defined age and metallicity. Therefore a simple stellar population (SSP) gives an ideal representation for each particle. For every galaxy in the EAGLE and TNG100 datasets, the SDSS-equivalent spectra are produced by combining the SSPs for all the stellar particles within a R=3\,kpc galactocentric radius (2D projected radius from simulation), weighed by the corresponding stellar mass of each one, corrected by the effect of the returned fraction, i.e. the loss of mass from stellar ejecta as the population ages. Note our motivation in Sec.~\ref{synthetic-data} for choosing this aperture estimate. The mixture of the different stellar particles produce the composite population of the galaxy (for a detailed explanation see \citealt{Negri:2022}). This approach is well justified since the stellar particles have a mass over $10^6 M_\odot$, avoiding IMF sampling issues. For the SSPs, we use the E-MILES models \citep{EMILES} -- based on a fully empirical stellar library \citep{sanchez:2006}, Padova isochrones \citep{girardi:2000}, and the \citet{chabrier:2003} IMF. These models extend from the far-UV to the mid-IR (1680\AA$<\lambda<$5$\mu$m),  with constant sampling $\Delta\lambda$=0.9\AA, spanning stellar ages from 6.3\,Myr to 17.8\,Gyr (we restrict the oldest ages to the cosmological age of the Universe at the fiducial redshift, $\sim$12.4\,Gyr), and metallicity ranging from [Z/H] = $-$1.71 to $+$0.22. We perform a bilinear interpolation in the age and metallicity grid of E-MILES spectra. We note that the spectral resolution of the E-MILES models is comparable to that of the SDSS spectrograph, i.e. ${\cal R}\equiv \lambda/\Delta\lambda\sim$2000.

The spectra corresponding to the composite population mimicking the 3\,arsec fibre observations from SDSS are then convolved with a Gaussian kernel to bring them to the effective lower resolution caused by the stellar velocity dispersion. Our SDSS sample is restricted to the range $\sigma\in[100,150]$\,km\,s$^{-1}$, so we apply a kernel that produces a resolution in this range of velocities, roughly of order $\Delta\lambda\sim\lambda\sigma/c\sim 2$\AA. The spectra are also rebinned to the 1\AA\  bin size adopted in the PCA of the SDSS data. To produce a distribution of synthetic data as close as possible to the SDSS sample, for each EAGLE or TNG100 galaxy, we randomly select one from the homogenised SDSS dataset and pick its velocity dispersion for the Gaussian convolution. This process ensures that the distribution of velocity dispersion is compatible with the original data. Note the interval in $\sigma$ is rather small, and the dispersion between stellar mass and velocity dispersion is large enough to make this method accurate enough.

Finally, synthetic spectra are modified by adding noise to produce data that are equivalent to the observed SDSS data. The noise component includes Poisson, instrumental and background (sky) noise, along with the procedures involved in the data reduction pipeline. While simpler noise models can be adopted -- for instance adding a Gaussian component normalized by the expected S/N in a reference spectra window, or using the variance of the flux data as a proxy of noise plus flux differences from the absorption lines -- we find that any PCA-based comparison of simulations and real spectra should model as realistically as possible the noise from the observations.

Our adopted noise component was directly determined from the inverse variance (ivar) provided for each SDSS spectrum in the FITS file. This inverse variance encapsulates all the different contributors to the noise. Since the data are normalized, we opt instead to define the noise by the signal to noise ratio S/N$\equiv \Phi/\Delta\Phi = \Phi({\rm ivar})^{1/2}$. For each synthetic spectrum, we randomly choose one from the equivalent, homogenised SDSS sample, and add Gaussian noise, in quadrature, that produce the same S/N in each wavelength bin. Therefore, by construction, we produce a dataset with idential distribution of noise properties, including its wavelength dependence. The final synthetic spectra appear very similar to the observed data: Fig.~\ref{fig:eigenspectra} shows the stacked and continuum-subtracted spectra of each subgroup for the two simulations, and include the SDSS stacks for reference. It is important to emphasize that in \citet{variance} as well as in this paper, we exclude all flux values located at the position of prominent emission lines: 3869\AA\ ([Ne{\sc III}]); 3889\AA\ (H$\zeta$); 3969\AA\ (H$\epsilon$); 4100\AA\ (H$\delta$); and 5007\AA\ ([O{\sc III}]). To distinguish between the two spectra, we add a constant offset of $-$0.2 ($-$0.4) in the flux values of the EAGLE (TNG100) spectra. Note that for the AGN sample, the synthetic spectra look more jagged, i.e. noisier. This might arise from the fact that the sample size is smaller in this group. Stacking large samples has the benefit of removing galaxy-to-galaxy variations within the same subset, and increasing the total S/N. We emphasize that this is the most realistic way to produce samples that are comparable to the SDSS data.

\begin{figure*}
    \centering
    \includegraphics[width=65mm]{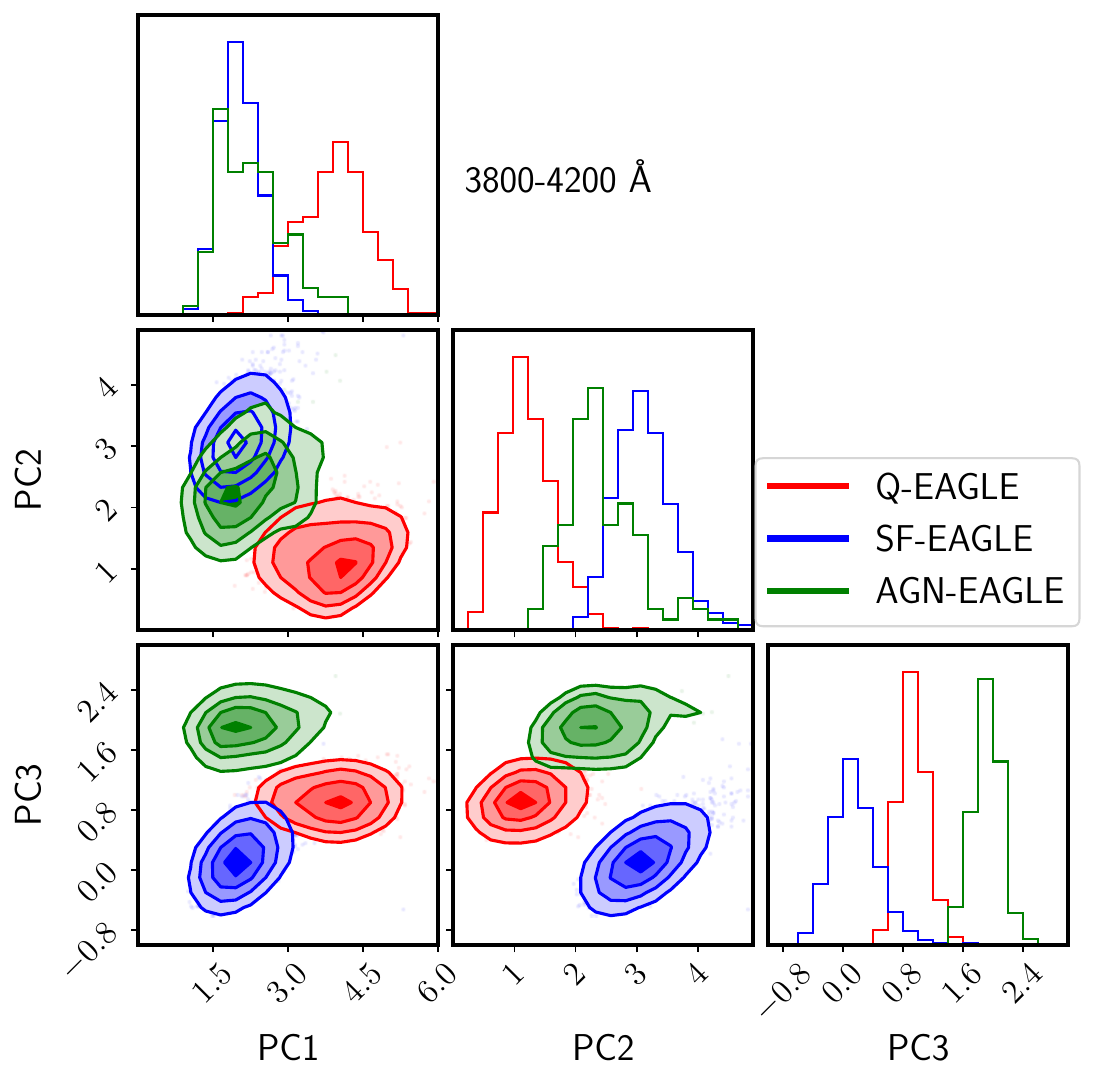}
    \includegraphics[width=65mm]{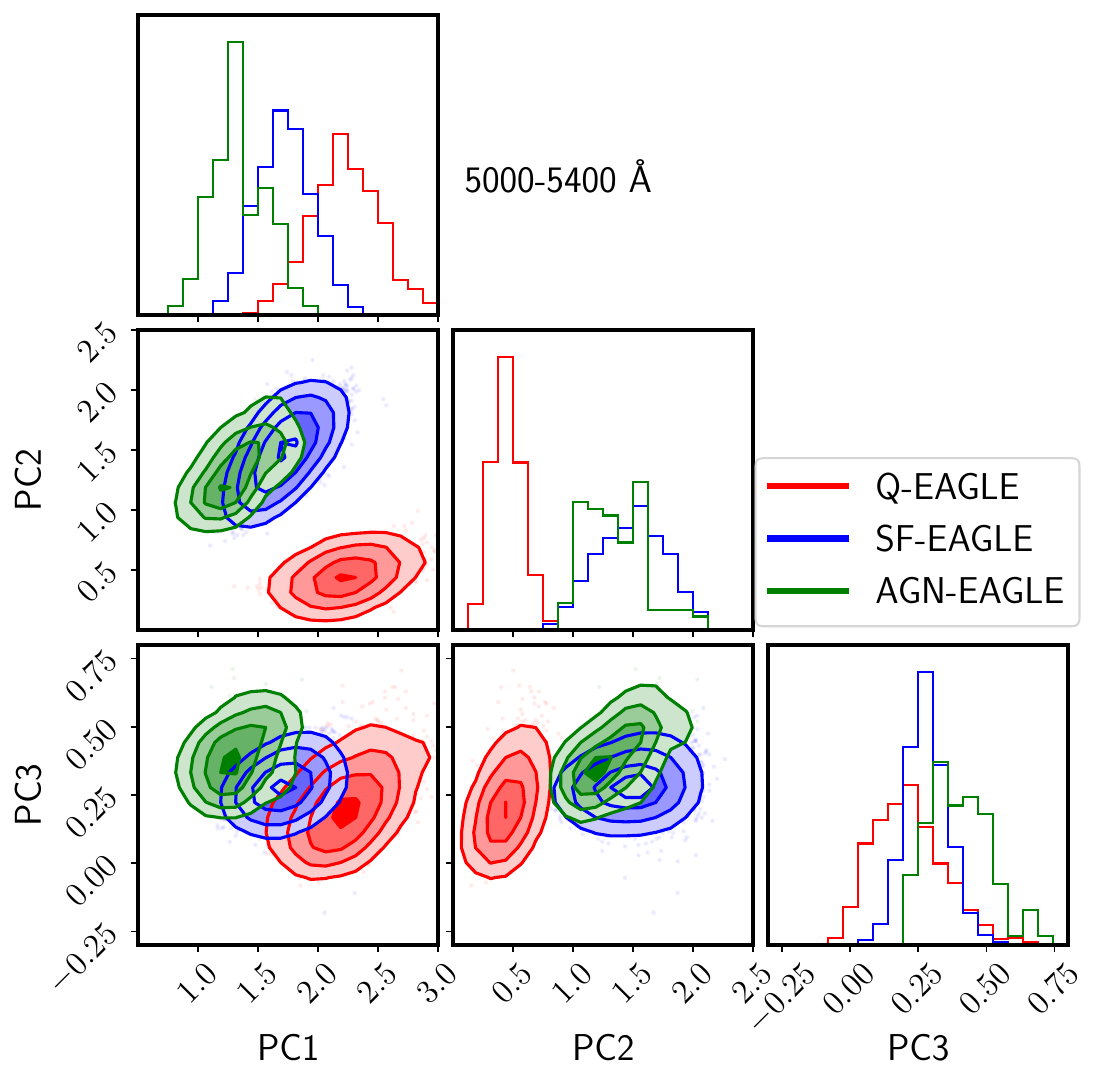}
    \includegraphics[width=65mm]{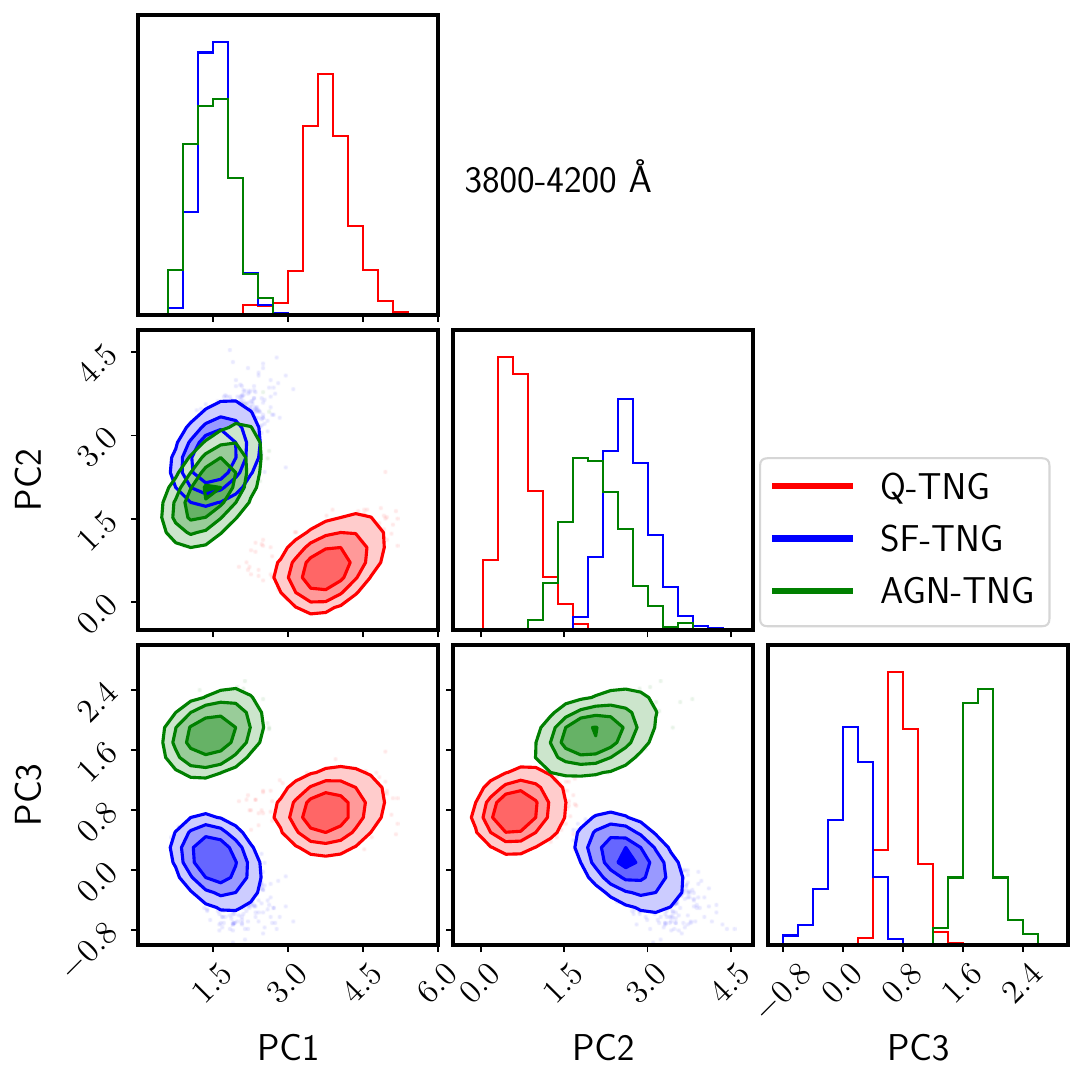}
    \includegraphics[width=65mm]{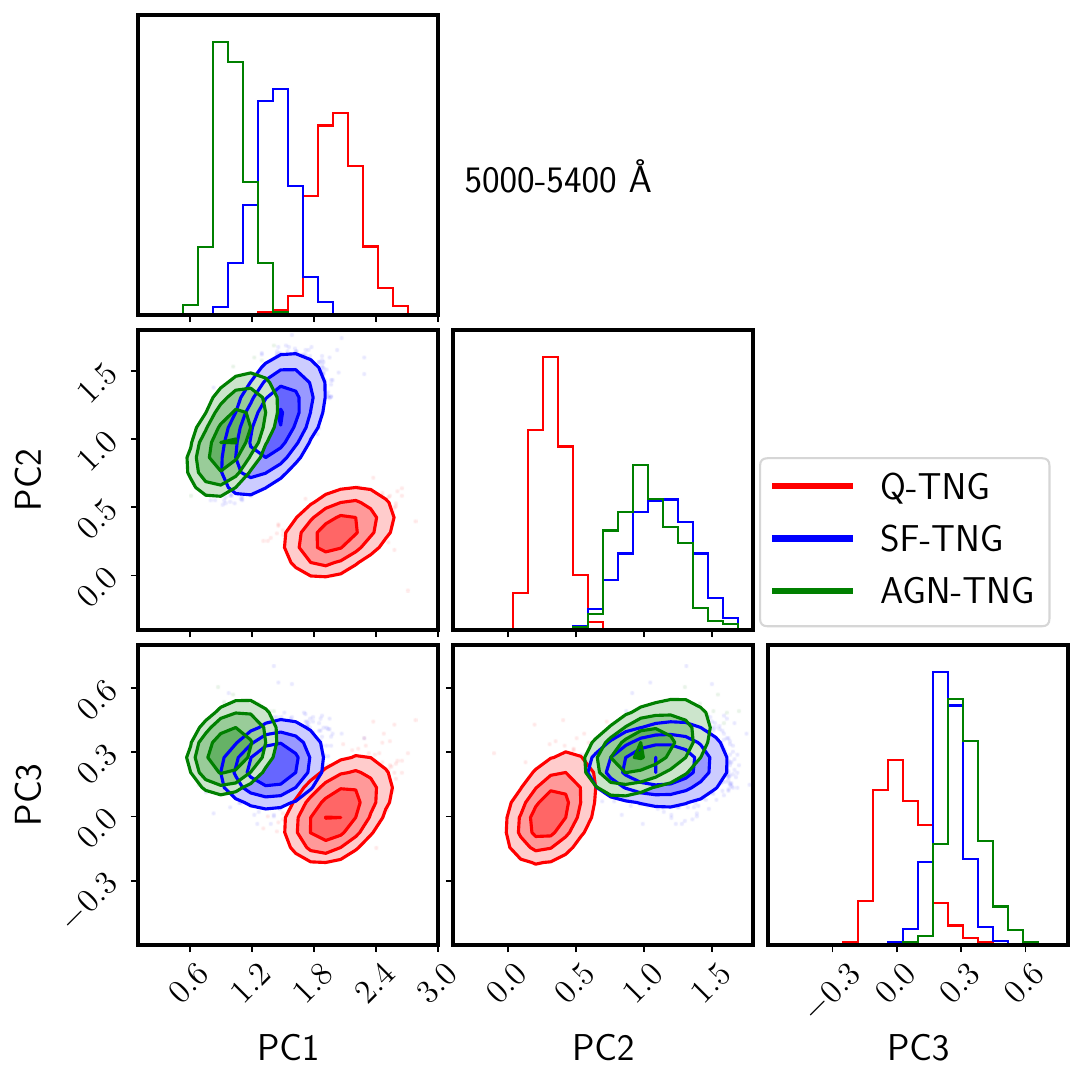}
    \includegraphics[width=65mm]{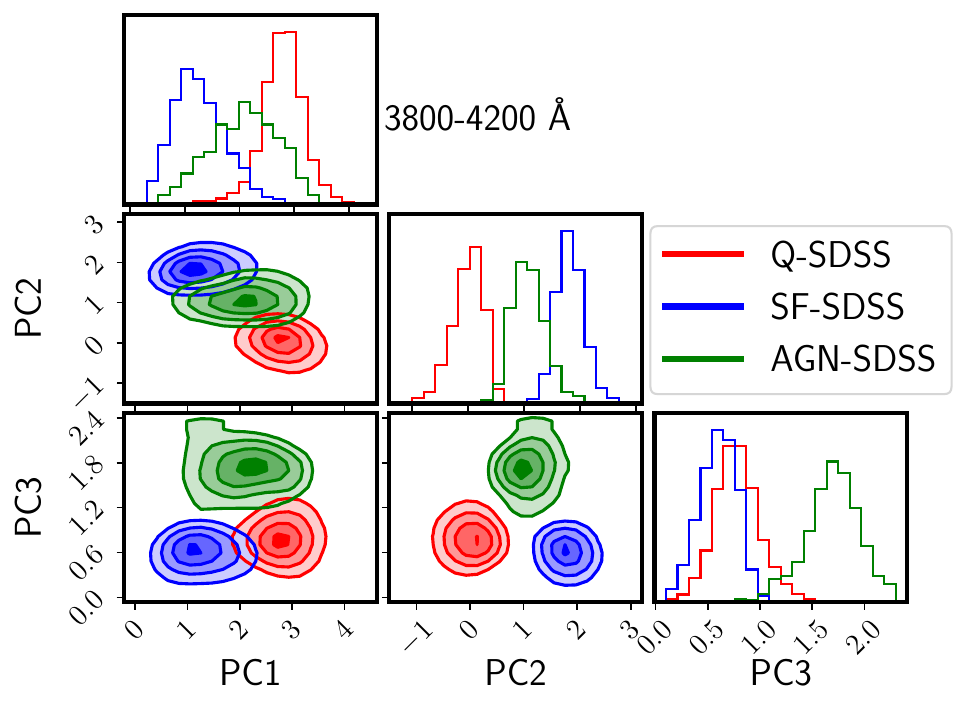}
    \includegraphics[width=65mm]{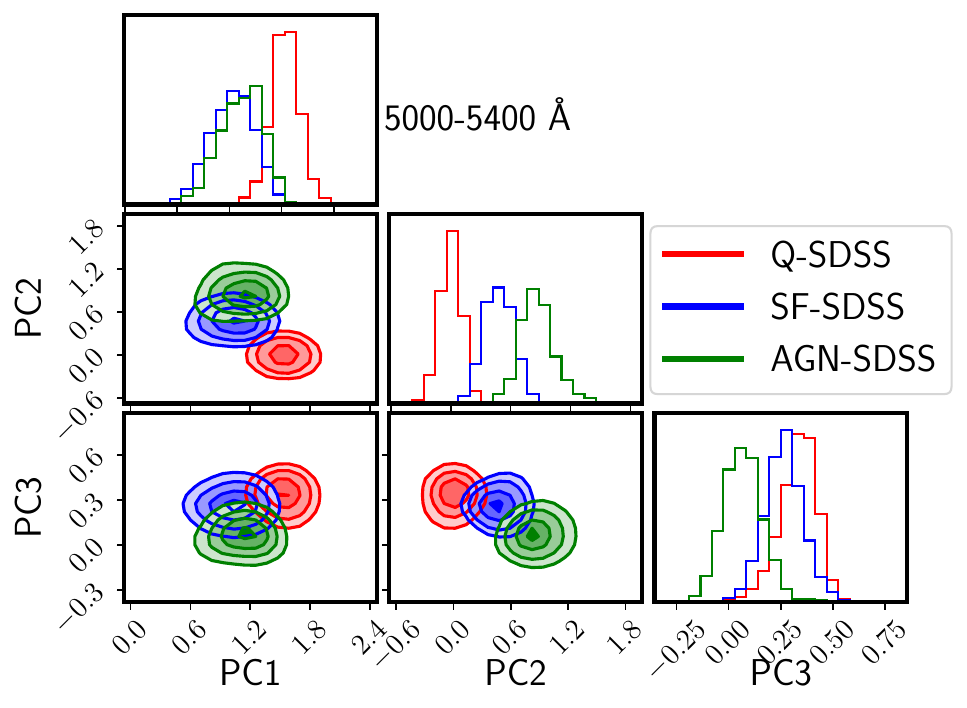}
    \caption{Distribution of the projections of the EAGLE ({\sl top}), TNG100 ({\sl middle}) and SDSS ({\sl bottom}) spectra onto the first three principal components of PCA derived by the SDSS sample. The galaxies are separated into star-forming (blue), AGN (green), and quiescent (red). The left (right) panels correspond to the results of the blue (red) spectral interval. The contours engulf 25, 50, 75, and 90\% of each subsample. See Fig.~\ref{fig:3D_sims} as the 3D equivalent latent space.}
    \label{fig:corner_sims}
\end{figure*}

\section{Principal component analysis (PCA)}\label{method}
We conduct a variance analysis of subsets of continuum-subtracted Sloan Digital Sky Survey (SDSS) spectra and evaluate equivalent synthetic EAGLE and TNG100 spectra. Principal component analysis (PCA) is a covariance analysis method \citep{Pearson:1901}. Using PCA, the input data is rearranged into a ranked set of variables, the principal components, by performing rotations in the N-dimensional parameter space. Each spectrum -- input data -- is defined by N numbers, in this case, the fluxes within a range of N wavelength intervals, so that each spectrum represents a vector in this N-dimensional space. The whole set of spectra define an N$\times$N covariance matrix, and rotations simply correspond to matrix transformations ${\cal M}\in SO(N)$. Out of the possible rotations, PCA focuses on the one that diagonalises the covariance matrix. We determine the eigenvectors (information vectors) of the data covariance matrix and the eigenvalues that give the weight of each one. Eigenvalues represent the individual contribution to the variance of the eigenvectors (also known as principal components). Typically, these are ranked in decreasing order of variance, expressed as a fractional contribution, so that higher-order components contribute progressively less to the total variance. When the spectra are projected onto the eigenvectors, the ``coordinates'' of latent space are produced, and the first few components retain most of the information, in the sense of variance. Thus, the dimensionality of the input data is reduced. 

Although PCA is mostly used for classification purposes \citep[e.g.][]{folkes1996, Madgwick:03, Nersesian:21}, in this study we investigate how the latent space encodes differences in the underlying stellar populations of different groups of galaxies following PCA-SDSS. A significant specificity of this is that PCA is applied independently to three subsets of spectra, classified by their nebular emission into three groups, namely star-forming, AGN, and quiescent galaxies (see section \ref{pre-SDSS} and PCA-SDSS). While the input spectra originate from the same dataset, we have conducted a separate analysis to explore the differences between the three groups and to assess how variance is distributed between them. Additionally, the analysis presented in PCA-SDSS is unique in that the continuum has been removed to avoid reddening and flux calibration systematics, although this is at the cost of losing information. PCA-SDSS has shown that almost 30-70 percent of the variance, depending on the galaxy group is encoded on the continuum. Moreover, strong nebular emission lines are removed from spectral windows. As a result, the study focuses on the spectral absorption features of different galaxies or stellar populations and their properties. Also, conservative culling is applied to remove discordant data before calculating the covariance matrix in order to concentrate on stellar population-driven variations. We refer the interested reader to \citet{variance} for details about our sample culling. This cleaning process does not bias the velocity dispersion distributions in the three subgroups. The projections onto the principal components yield a larger separation between the three subgroups and hence a more clearly defined latent space when culling the sample, but our main conclusions remain unaffected. We emphasise that PCA (or any other method based on variance) is sensitive to the presence of outliers (comparable to, e.g. linear least squares fitting). In general, these methods can be used either: 1) to detect such anomalous objects as interesting, or 2) to explore the properties of the majority of the galaxies. Our goal focuses on the latter. We want to assess the general trend of galaxies when classified as SF/AGN/Q, so culling is an important step to avoid spurious signals from ``anomalous systems''. Our spectral study focuses on two wavelength intervals, namely [3800,4200]\AA\ and [5000,5400]\AA. These intervals preserve most of the variance from galaxy to galaxy \citep{InfoPop}.

In this study, we aim to use the information vectors or eigenvectors produced in PCA-SDSS. We can use the analysis of the SDSS sample to evaluate the simulation and synthetic spectra. After all, realistic simulations should be able to produce data with the same covariance as the SDSS spectra. Having synthetic spectra comparable to observed spectra (see sections \ref{pre-process} and \ref{resol-noise}), we project the synthetic spectra onto the PCA-SDSS eigenvector of the corresponding type to create the PC ``coordinates'' in latent space, for instance:
\begin{equation} \label{eq:PC1}
{\rm PC1}_{j,{\rm syn}} = \Phi_{j,{\rm syn}}\cdot\hat{{e}}_{1,{\rm SDSS}} = \sum\limits_{i=1}^N\Phi_{j,{\rm syn}}(\lambda_{i})\hat{{e}}_{1,{\rm SDSS}}(\lambda_{i}),
\end{equation}
where $\Phi_{j,{\rm syn}}$is the flux of the spectrum for the $j$th
simulated galaxy and $\hat{{e}}_{1,{\rm SDSS}}$ is the first
eigenvector from the SDSS covariance analysis of the corresponding
sub-group. We focus our analysis on the first three principal
components. In PCA-SDSS we show that the first three PCs capture most
of the variance. We note that the covariance matrix is sign invariant,
i.e. a change in the sign of a principal component (and thus its
projections) does not affect the matrix. Therefore, the numerical code
can arbitrarily produce $\hat{e}_1$ and $-\hat{e}_1$ as
solutions to the same eigenvector.  To mitigate this
issue, we enforce a positive sign for the median of the projections of
all simulated galaxies of the corresponding type. When restricting to
the first three principal components, this means the data are
statistically constrained to the first octant of the latent space
spanned by {PC1, PC2, PC3}.

\section{Exploring the spectral latent space of simulated galaxies}
\label{sec-proj}

The projections of the simulated data in latent space are shown in Figure \ref{fig:corner_sims}, as a density plot, with the usual 2D cuts and 1D histograms, adopting the {\sc Python} corner module of \citet{emcee}. A three dimensional, easier to visualize version, is presented in Appendix~\ref{B}. We emphasize that the results are obtained from continuum subtracted spectra, rejecting regions dominated by nebular emission lines, so that the analysis does not depend on the dust or gas components. Our goal is to assess whether the absorption line information from the stellar populations in the synthetic data are located in the same regions of parameter space as the real (SDSS) data. Such a result would imply that at the most fundamental level, the majority of the spectral variance of real galaxies is reproduced by the simulations. The data are shown in blue (SF), green (AGN), and red (Q), in the blue (left) and red (right) spectral windows, in both Figs.~\ref{fig:corner_sims} and \ref{fig:3D_sims}. In these figures, the SDSS sample is a homogenised version (see \S\ref{Homog}): in Fig.~\ref{fig:corner_sims} it corresponds to the EAGLE homogenisation, and for Fig.~\ref{fig:3D_sims} each SDSS set is shown for the corresponding simulation, as labelled. The SDSS sample shown here corresponds to the one homogenised with EAGLE, but the distribution in latent space of both homogenised SDSS subsamples is similar to the original SDSS data, as can be compared with Figure~4 in PCA-SDSS. We emphasize that although these figures show projections onto different eigenvectors depending on the SF/AGN/Q classification\footnote{PCA is independently applied to three different covariance matrices, resulting in different sets of eigenvectors}, the underlying data always relate to the stellar populations of the galaxies. Therefore Figs.~\ref{fig:corner_sims} and \ref{fig:3D_sims} are not a disjoint comparison of PCA projections, and the eigenvectors of the three groups are not completely independent. See App.~\ref{C} for a check on this point. The three sets of eigenvectors reflect, instead, the typical stellar populations found in each group. The trends in the three latent spaces support the hypothesis of an evolutionary sequence from star formation to AGN to quiescence. This is reminiscent of the bimodal distribution into blue cloud (i.e. star-forming) and red sequence (i.e. quiescent) galaxies \citep[e.g.,][]{Strateva:01, Baldry:04}, with AGN preferentially populating the green valley \citep[see, e.g.,][]{Schawinski:07,Salim:14, Angthopo:19}. 

The separation is more pronounced in the blue interval, where PC1 is the component carrying most of the variance, showing a clear separation between the three groups in the SDSS data, with AGN located between the other two, while for both simulated samples, there is an overlap in the distribution of PC1 projections between the star-forming and AGN samples. When comparing TNG100 and EAGLE data, PC1 puts the Q sample at a greater distance from the AGN and SF samples, and TNG100 shows the least overlap between Q and AGN galaxies. PC2 for both observed and simulated samples shows a separation between three groups with AGN galaxies located between the other two. According to PC3, for the SDSS sample, SF and Q galaxies occupy the same region, while AGN stands out with higher values; and for the simulated samples, there is a clear separation between three groups with Q galaxies lying between the other two. In the red interval, the distributions appear separately when projected onto PC1 for the synthetic spectra, with the SF galaxies in between the other two, whereas, for the SDSS spectra, PC1 appears to “isolate” Q galaxies. PC2 shows the separation of the three groups with the SF galaxies in between the other two for the SDSS spectra and appears to “isolate” Q galaxies for the synthetic spectra. Concerning PC3, it sets aside AGN for the optical spectra and Q galaxies for the TNG100 synthetic spectra while separating the three groups with SF galaxies between and overlaps between the three groups for the EAGLE synthetic spectra. We note that the red interval includes more of the metallicity-sensitive indicators in the Mgb-Fe complex around 5100-5300\AA, whereas the blue interval includes more prominent age-sensitive features. We believe that comparisons in the blue interval give a more fundamental, lower order interpretation, whereas the variane in the red interval encodes more detailed information related to chemical enrichment.

Fig.~\ref{fig:3D_sims} helps visualize the distribution of principal component projections in the 3D latent spaces. Note that while PCA is independently applied to three different covariance matrices, which result in different eigenvectors, their clustering is significant even considering the different orientations. Further, it is shown in the appendix of PCA-SDSS that the results of using three covariance matrices are comparable to those with a single covariance matrix, whereas the use of three subsets reduces the overlap and allows us to focus on the stellar population of each subsample. Further proof can be found in Appendix~\ref{C} of this paper, in which we performed a test projecting the spectra from one group onto the eigenvectors of another group, and still find a separation in latent space. By construction, the input data lack emission lines or the continuum, thus the clustering can only reflect variations in the properties of the underlying stellar populations. It is difficult to ascribe such differences to specific physical/observational factors, and the PCA-SDSS study examined the correlation between these components and galaxy properties (see Figures~6 and 7 of PCA-SDSS). The eigenvectors in Fig.~\ref{fig:corner_sims} have familiar absorption features but cannot be associated, say, with specific populations, or phases of evolution. There is a notable difference between these subclasses in terms of average age and chemical composition. However, more subtle differences might be due to details of the underlying stellar populations. In this work, we are mostly interested in how different the stellar populations are in well-defined areas of the latent space spanned by the covariance analysis of the real data. Following the approach presented in PCA-SDSS, we will compare via spectral fitting the EAGLE and TNG100 synthetic spectra.

\subsection{Spectral fitting of the projected data}
\begin{figure*}
  \includegraphics[width=.05\linewidth]{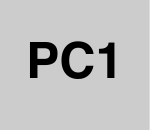}
  \includegraphics[width=.3\linewidth]{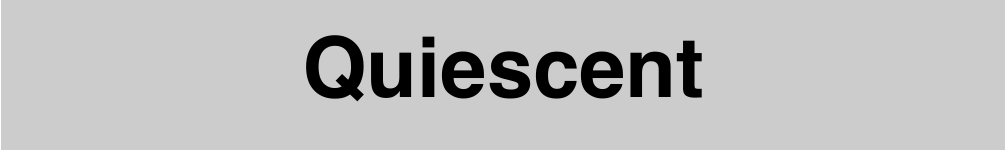}
  \includegraphics[width=.3\linewidth]{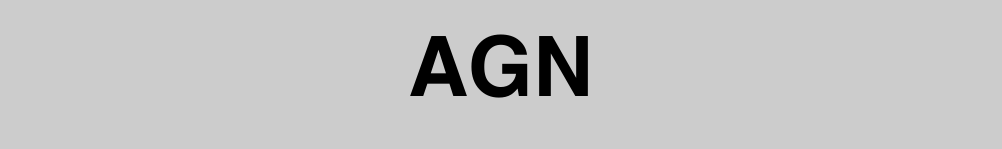}
  \includegraphics[width=.3\linewidth]{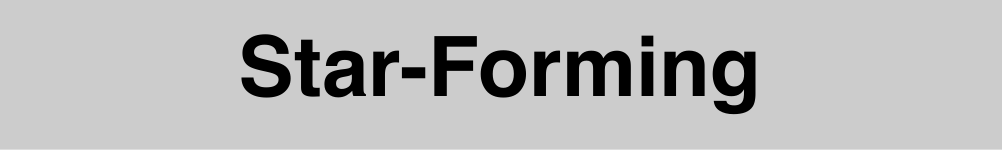}
  \includegraphics[width=.05\linewidth]{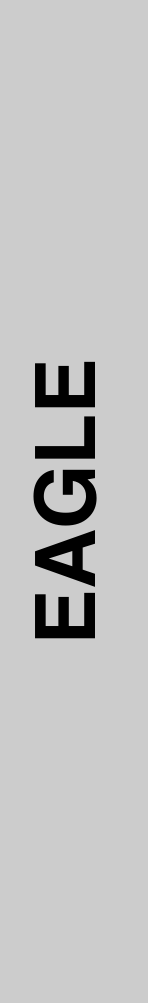}
  \includegraphics[width=.3\linewidth]{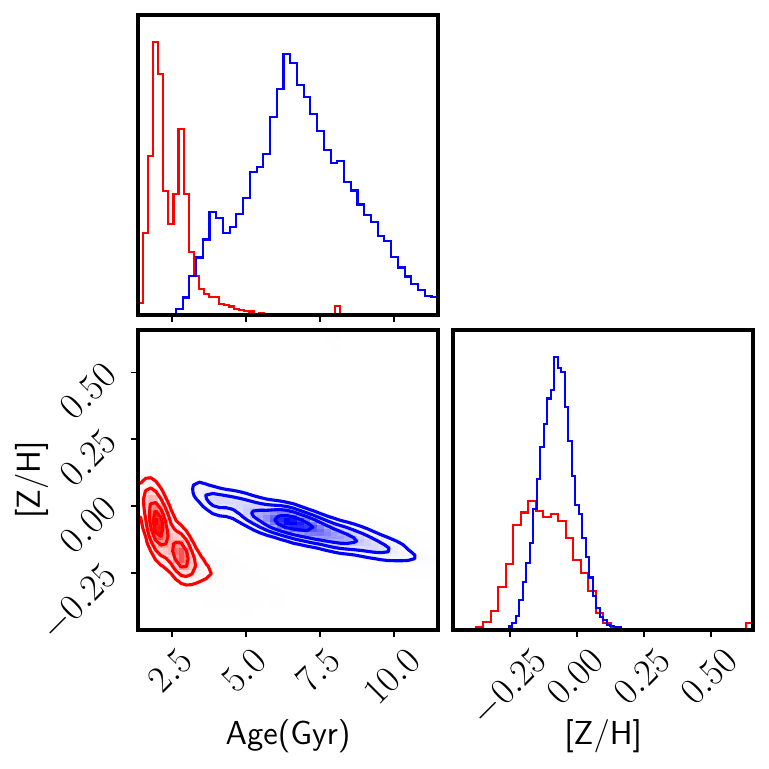}
  \includegraphics[width=.3\linewidth]{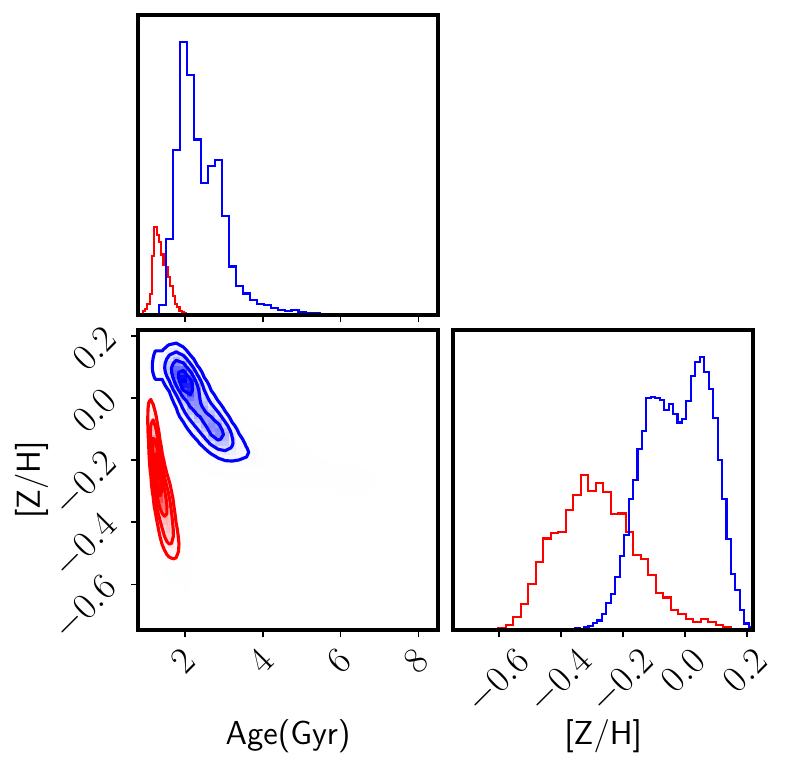}
  \includegraphics[width=.3\linewidth]{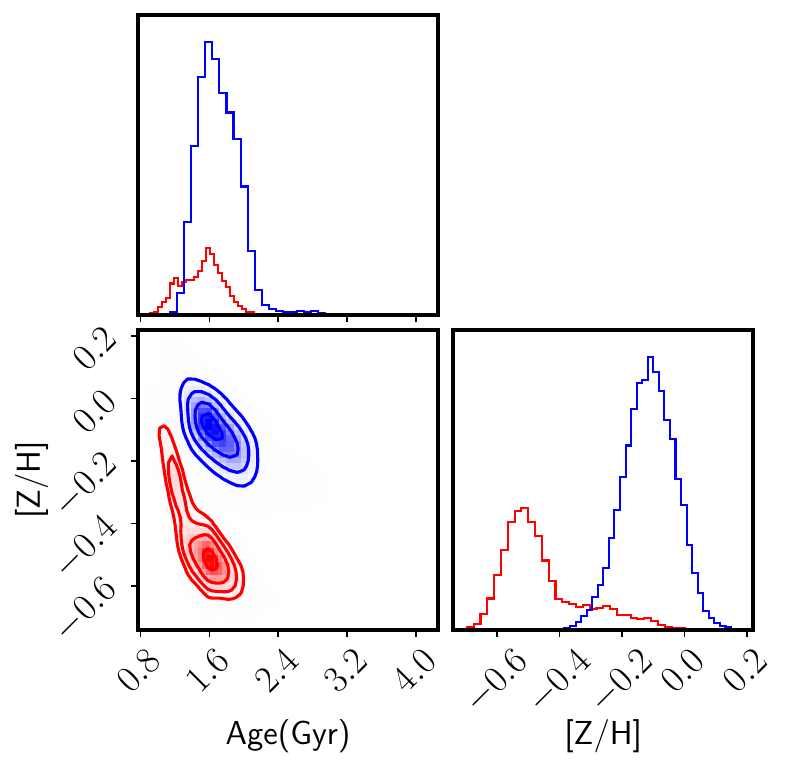}
  \includegraphics[width=.05\linewidth]{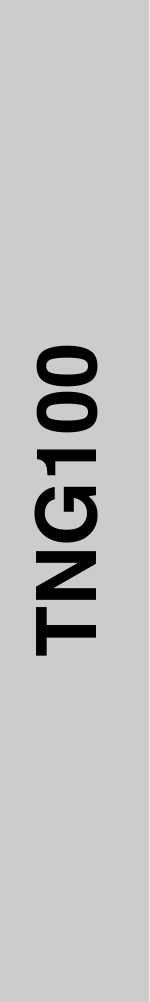}
  \includegraphics[width=.3\linewidth]{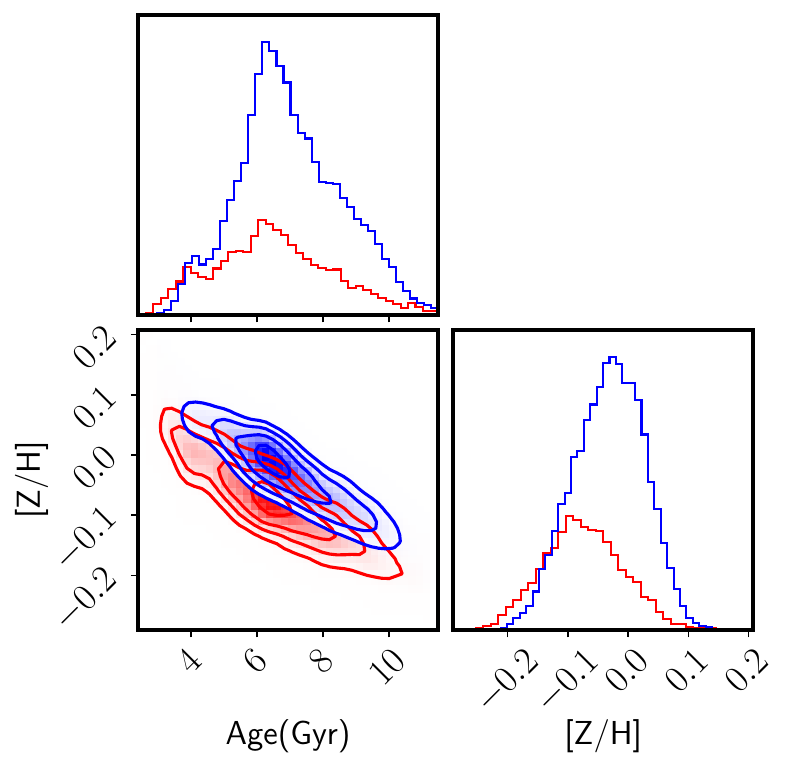}
  \includegraphics[width=.3\linewidth]{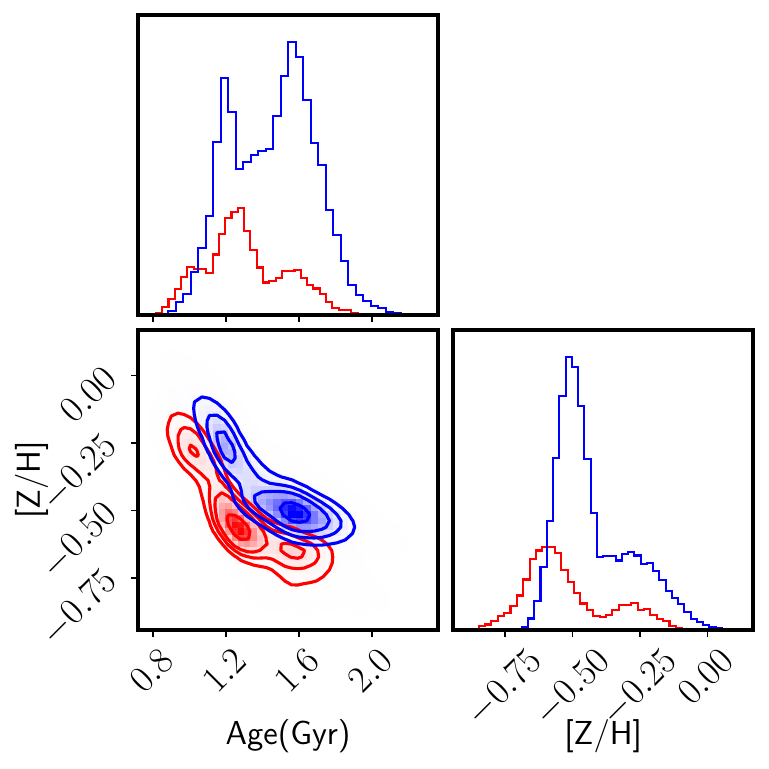}
  \includegraphics[width=.3\linewidth]{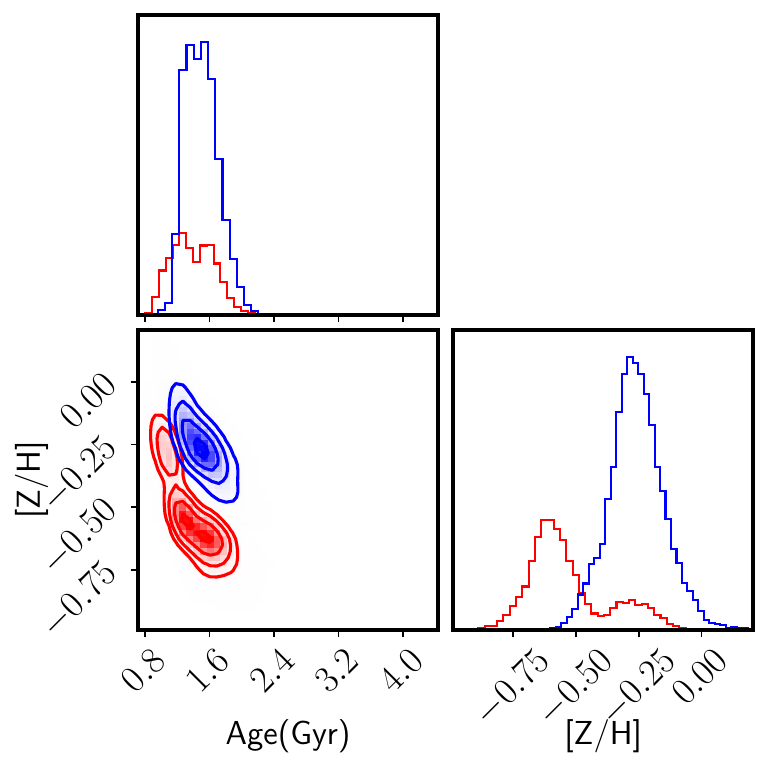}
  \includegraphics[width=.05\linewidth]{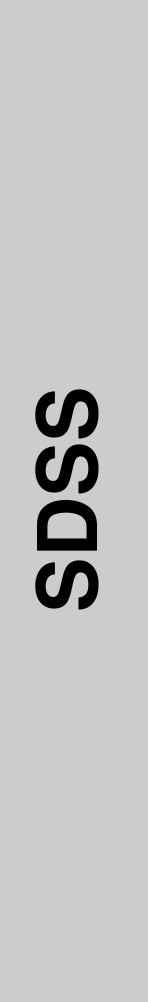}
  \includegraphics[width=.3\linewidth]{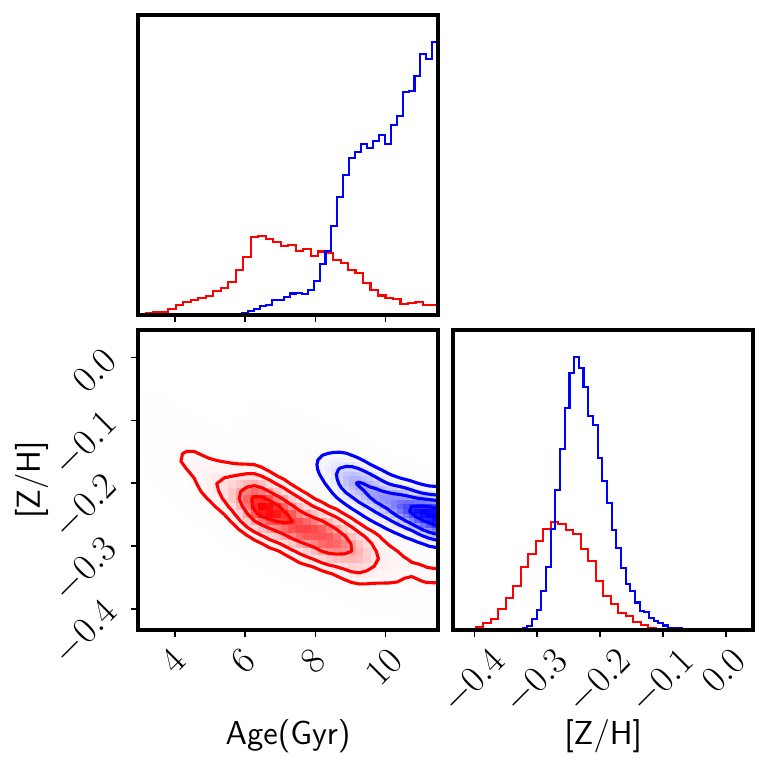}
  \includegraphics[width=.3\linewidth]{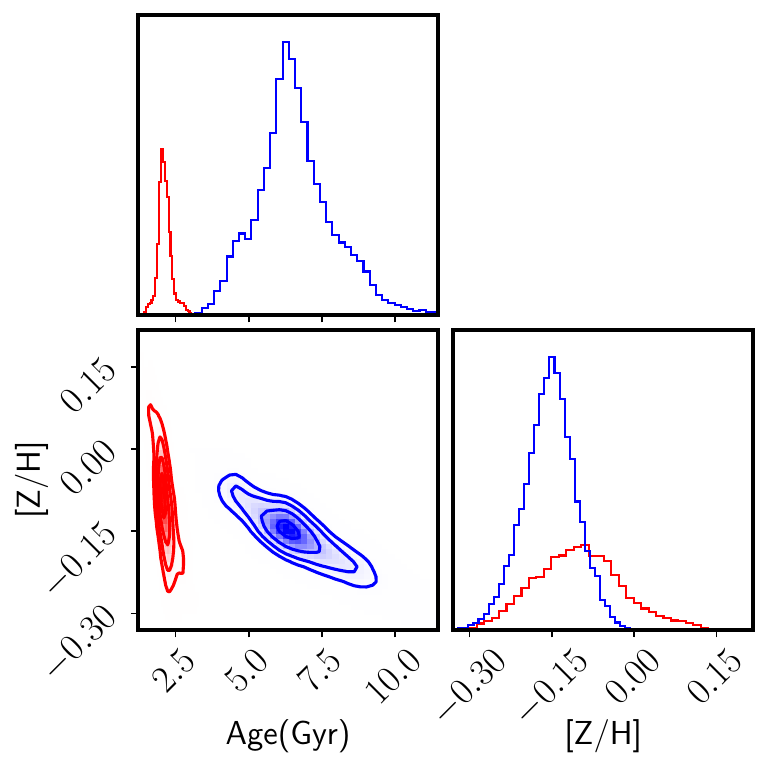}
  \includegraphics[width=.3\linewidth]{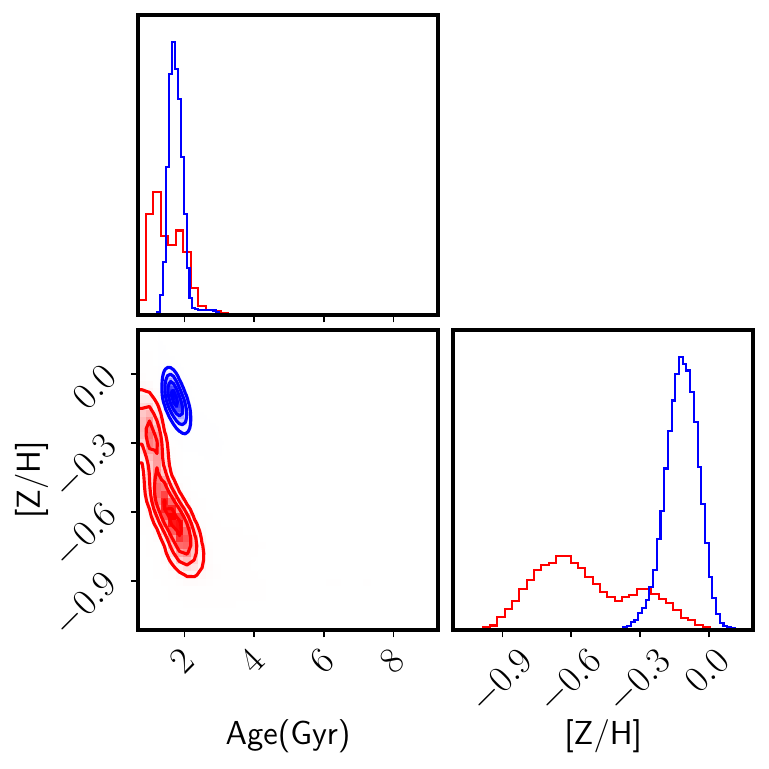}
  \caption{Confidence levels of the SSP-equivalent age (in Gyr) and metallicity ([Z/H]) obtained from fitting the continuum subtracted spectra with the E-MILES population models \citep{EMILES}. We fit the stacked spectra produced by combining data from the lowest (33 percentile, red) and highest (67 percentile, blue) PC1 projections of the blue interval. The results are shown, from top to bottom, for EAGLE, TNG100, and SDSS data, and from left to right for Q, AGN, and SF galaxies.}
\label{fig:MCMC-pc1}
\end{figure*}

\begin{figure*}
  \includegraphics[width=.05\linewidth]{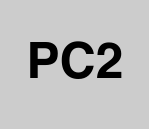}
  \includegraphics[width=.3\linewidth]{Figs/SSPFits/Col1.pdf}
  \includegraphics[width=.3\linewidth]{Figs/SSPFits/Col2.pdf}
  \includegraphics[width=.3\linewidth]{Figs/SSPFits/Col3.pdf}
  \includegraphics[width=.05\linewidth]{Figs/SSPFits/Row1.pdf}
  \includegraphics[width=.3\linewidth]{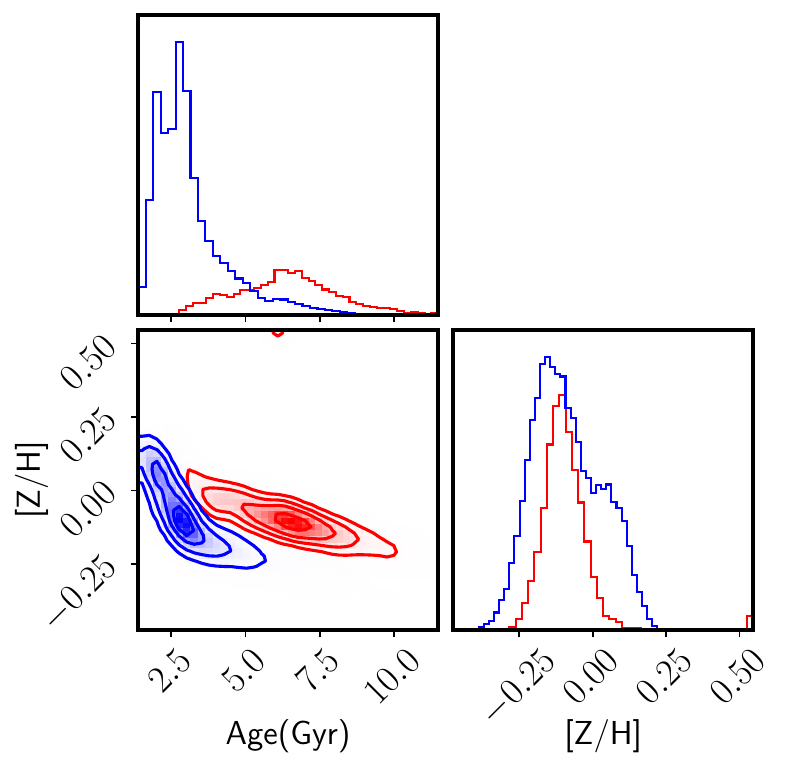}
  \includegraphics[width=.3\linewidth]{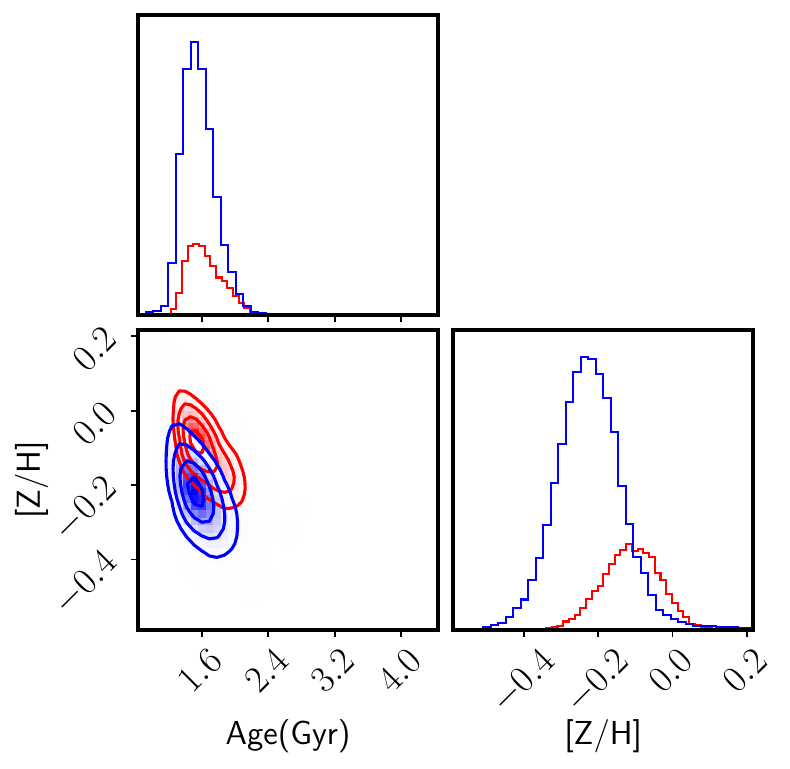}
  \includegraphics[width=.3\linewidth]{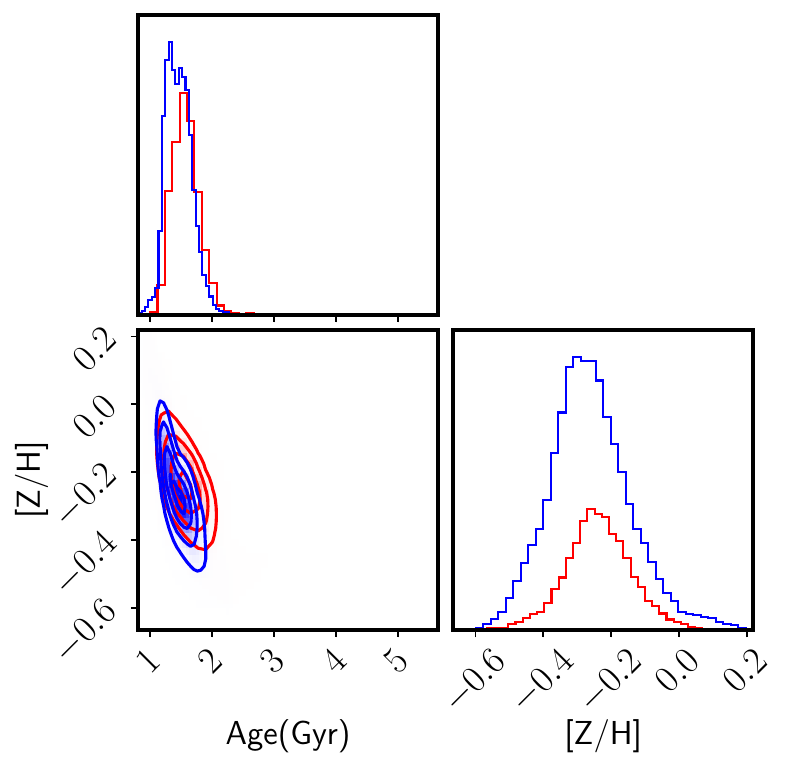}
  \includegraphics[width=.05\linewidth]{Figs/SSPFits/Row2.pdf}
  \includegraphics[width=.3\linewidth]{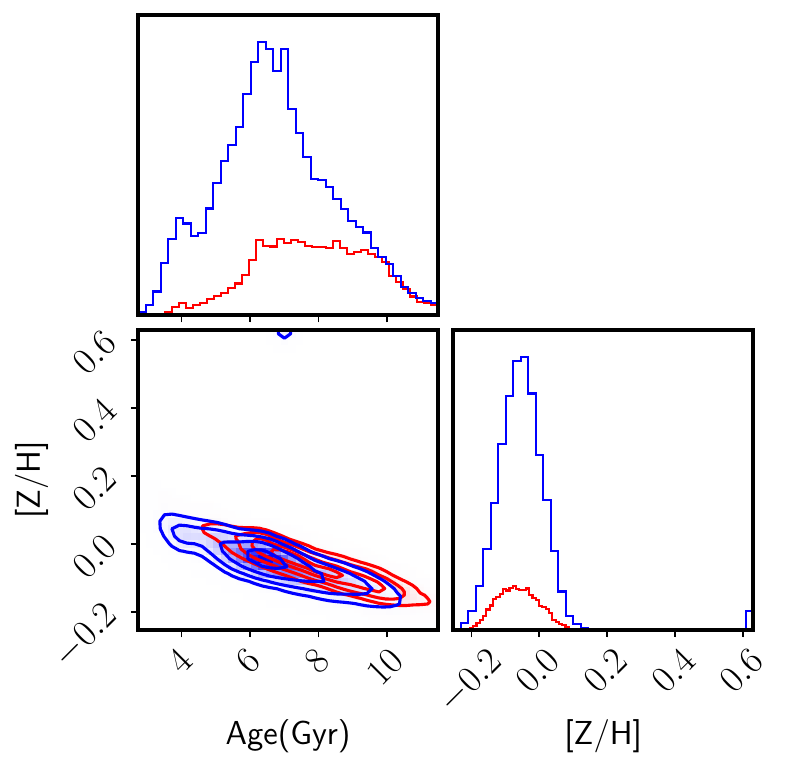}
  \includegraphics[width=.3\linewidth]{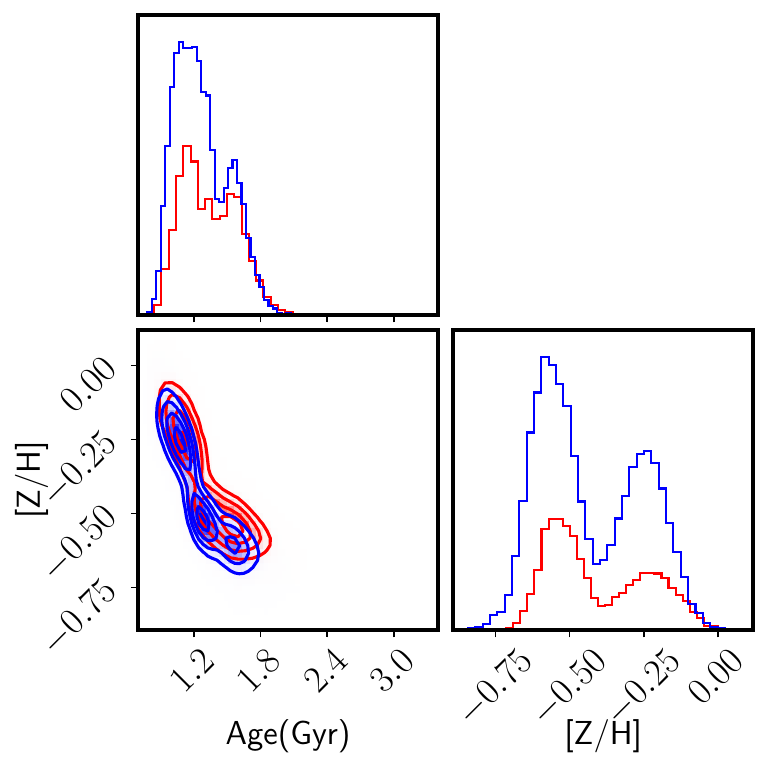}
  \includegraphics[width=.3\linewidth]{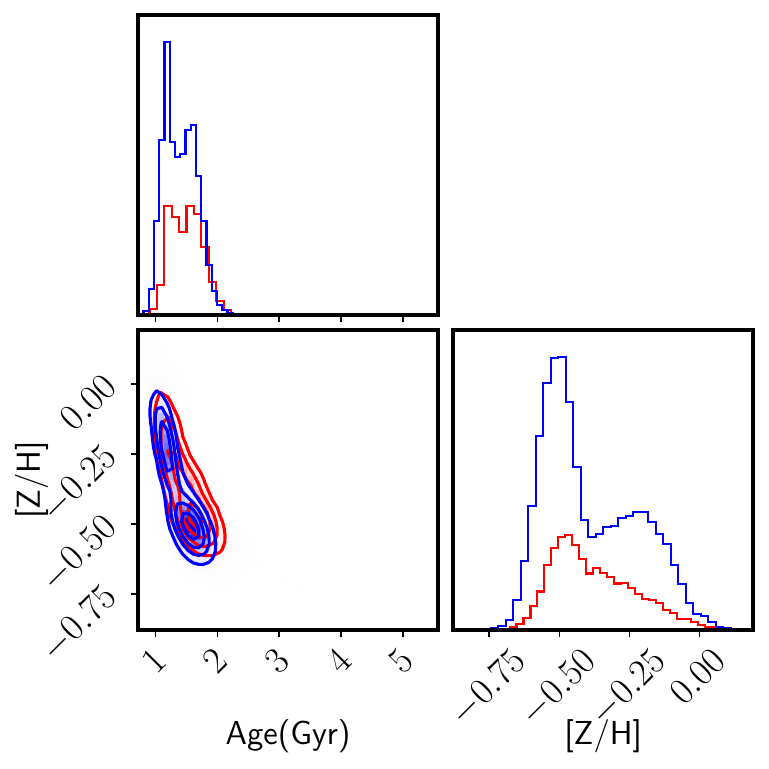}
  \includegraphics[width=.05\linewidth]{Figs/SSPFits/Row3.pdf}
  \includegraphics[width=.3\linewidth]{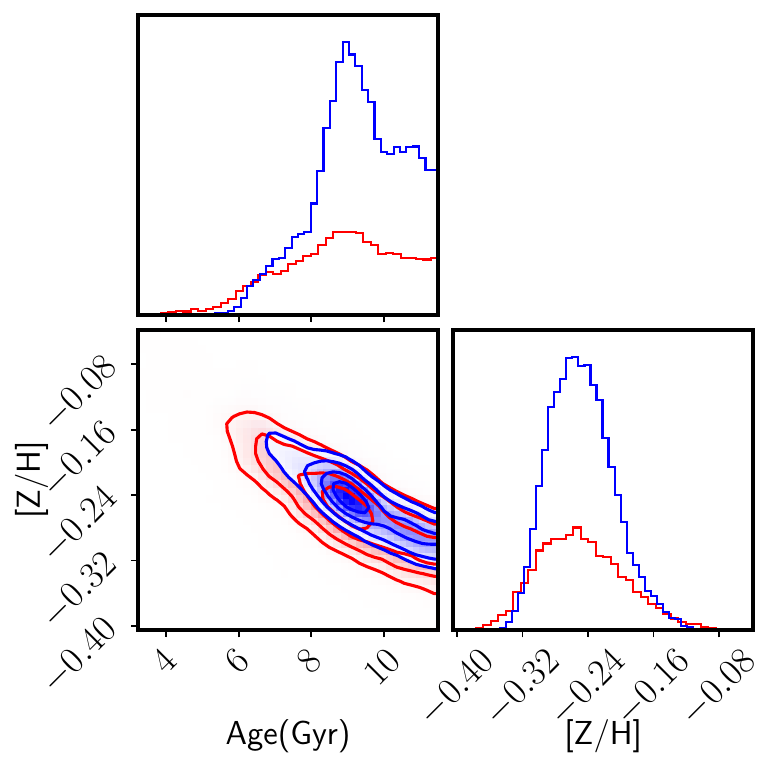}
  \includegraphics[width=.3\linewidth]{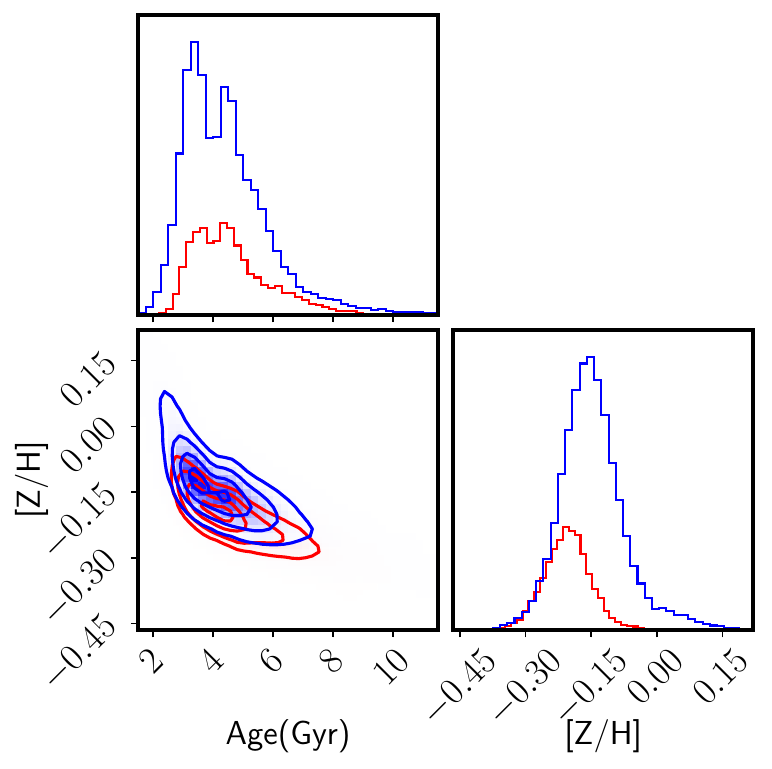}
  \includegraphics[width=.3\linewidth]{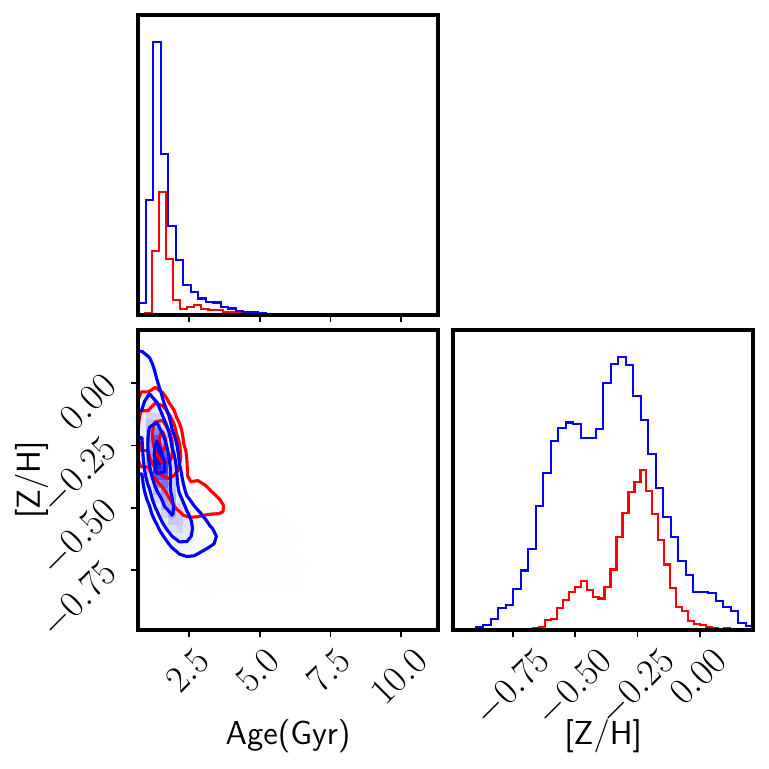}
  \caption{Equivalent of Fig.~\ref{fig:MCMC-pc1} for galaxy spectra stacked according to their projections on PC2 in the blue interval.}
\label{fig:MCMC-pc2}
\end{figure*}

\begin{figure*}
  \includegraphics[width=.05\linewidth]{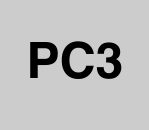}
  \includegraphics[width=.3\linewidth]{Figs/SSPFits/Col1.pdf}
  \includegraphics[width=.3\linewidth]{Figs/SSPFits/Col2.pdf}
  \includegraphics[width=.3\linewidth]{Figs/SSPFits/Col3.pdf}
  \includegraphics[width=.05\linewidth]{Figs/SSPFits/Row1.pdf}
  \includegraphics[width=.3\linewidth]{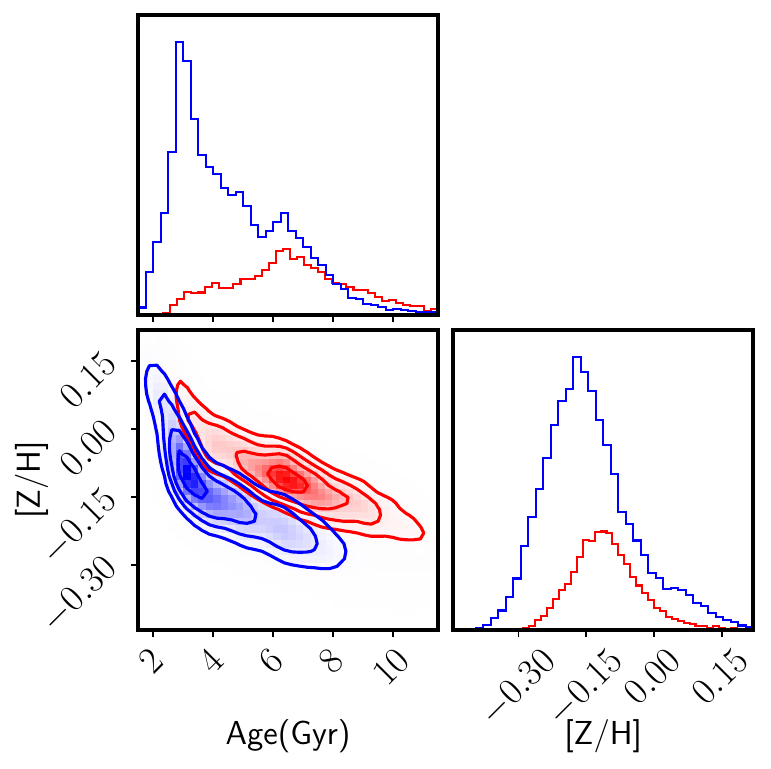}
  \includegraphics[width=.3\linewidth]{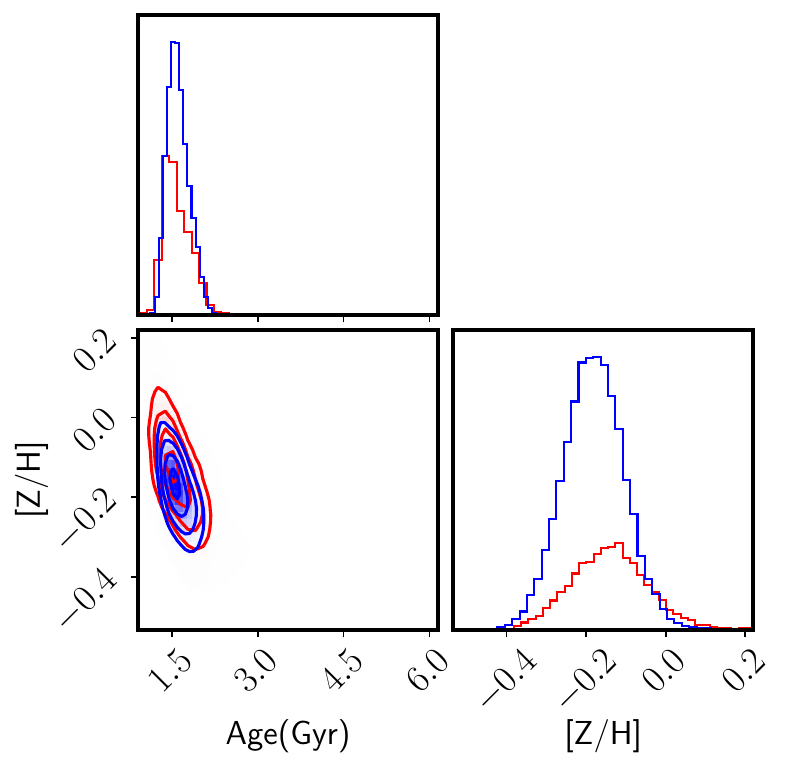}
  \includegraphics[width=.3\linewidth]{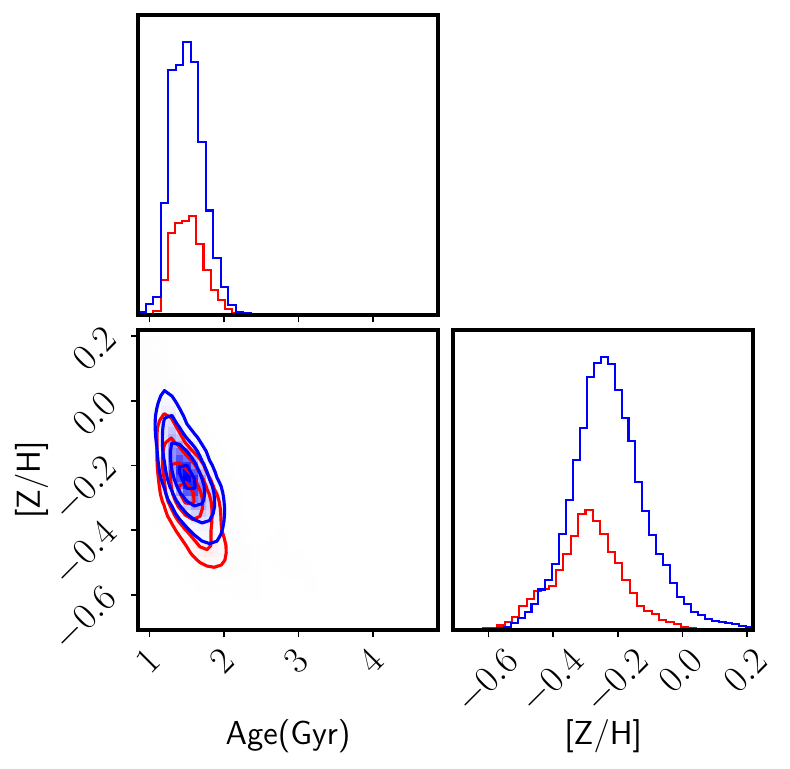}
  \includegraphics[width=.05\linewidth]{Figs/SSPFits/Row2.pdf}
  \includegraphics[width=.3\linewidth]{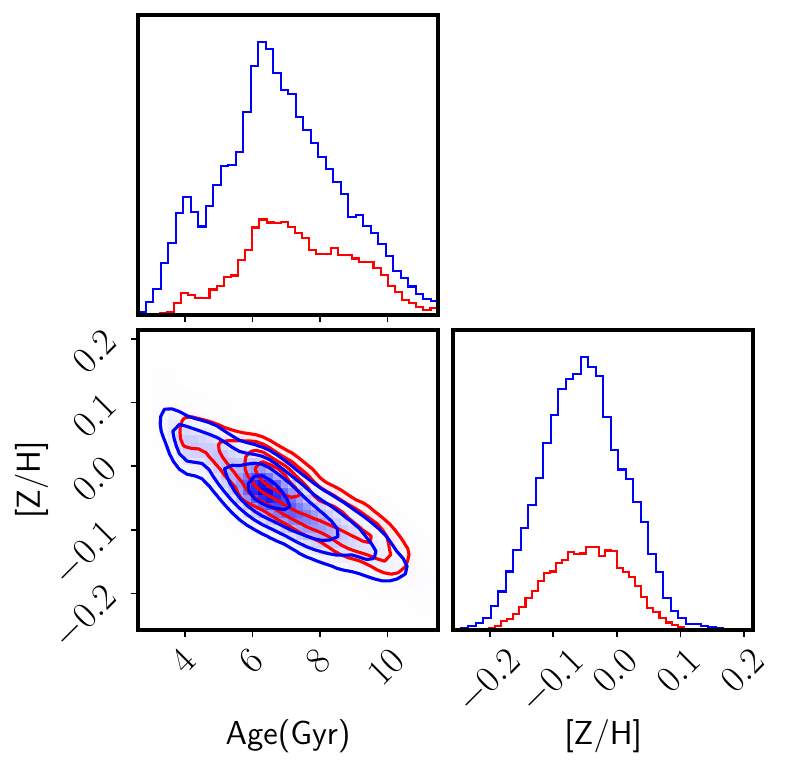}
  \includegraphics[width=.3\linewidth]{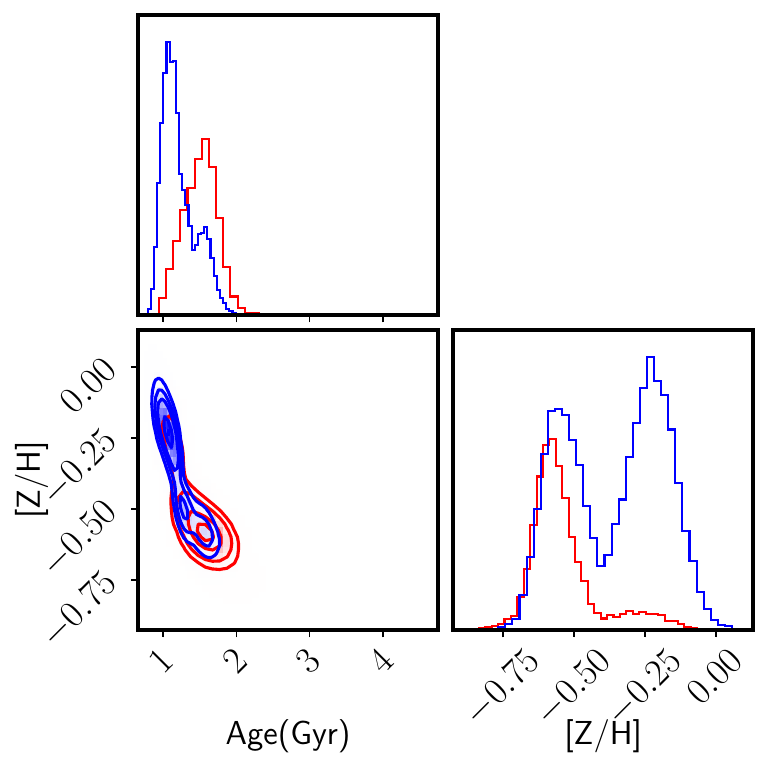}
  \includegraphics[width=.3\linewidth]{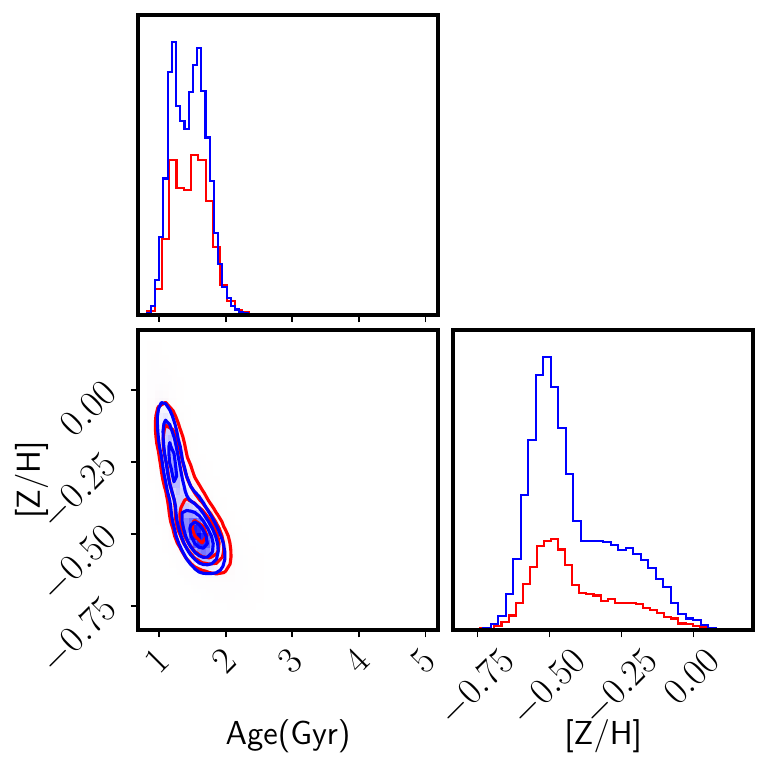}
  \includegraphics[width=.05\linewidth]{Figs/SSPFits/Row3.pdf}
  \includegraphics[width=.3\linewidth]{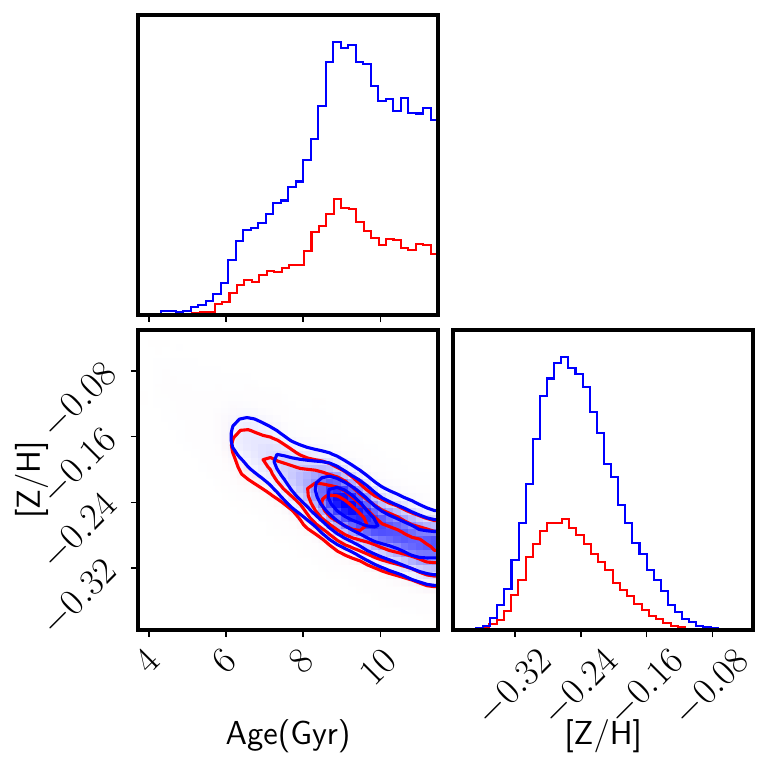}
  \includegraphics[width=.3\linewidth]{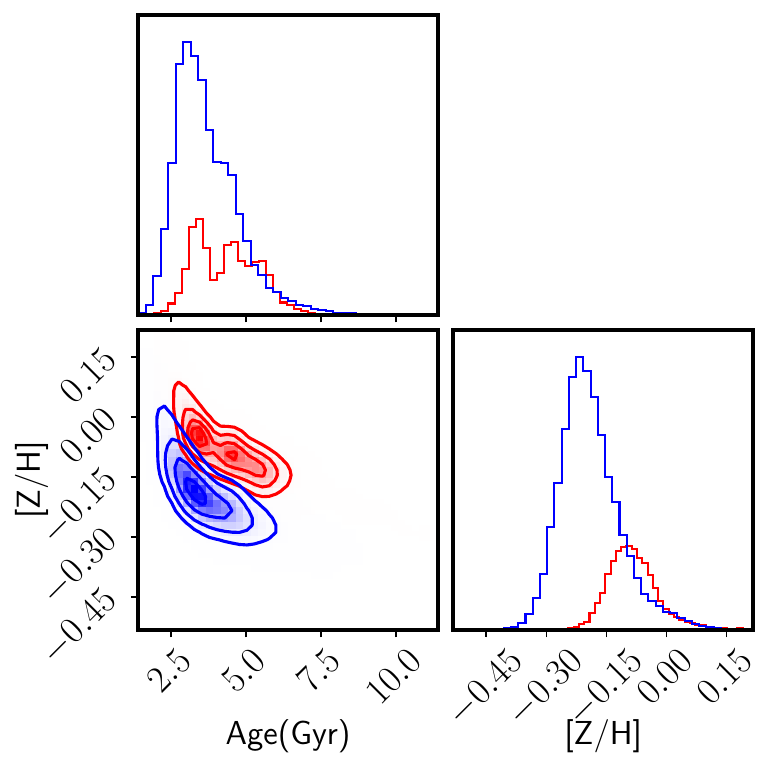}
  \includegraphics[width=.3\linewidth]{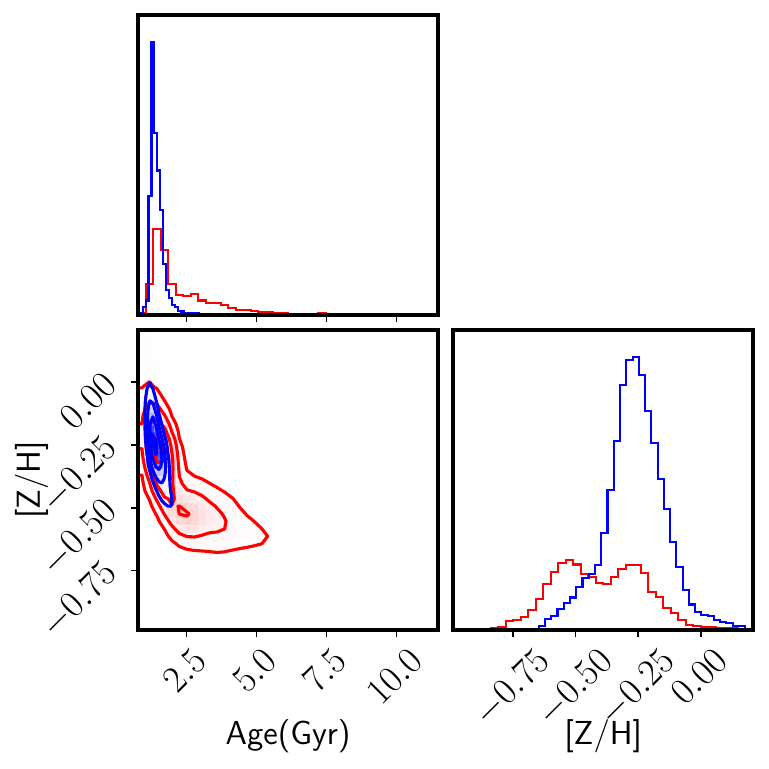}
  \caption{Equivalent of Fig.~\ref{fig:MCMC-pc1} for galaxy spectra stacked according to their projections on PC3 in the blue interval.}
\label{fig:MCMC-pc3}
\end{figure*}

As a follow-up to our previous analysis, we now study the simulated spectra within a given (SF/AGN/Q) subclass regarding their projections on (SDSS) latent space. For each choice of spectral inteval and nebular activity, we produce two stacked spectra by combining the data from galaxies whose projections on a given PC lies either in the 33rd (lowest) or the 67rd (highest) percentiles of the distribution. The resulting stacks are  compared with model SSPs using a standard method based on the $\chi^2$ statistic, fitting the continuum-subtracted data -- for consistency with our analysis, the continuum does not play any role in the fitting. The spectral window for fitting is 4000$<\lambda<$6000\AA. We define the standard likelihood  ${\cal L}(t,[Z/H])\propto\exp(-\chi^2(t,[Z/H])/2)$. We take the SSPs from the E-MILES models \citep{EMILES}, and the only two free parameters are the stellar age (exploring the 0.01-11.5\,Gyr range) and total metallicity (between [Z/H]=$-$2.3 and $+$0.22), with the stellar initial mass function of
\citet{KroupaU:01}\footnote{When performing spectral fitting, there is no significant difference between this choice of IMF and \citet{chabrier:2003}.}. The fitting process is implemented with the {\sc Python}-based MCMC solver {\tt emcee} \citep{emcee} to produce the confidence levels of the fits, shown in Figs.~\ref{fig:MCMC-pc1}, \ref{fig:MCMC-pc2}, and \ref{fig:MCMC-pc3}, respectively for PC1, PC2 and PC3 (these figures show results for the blue interval; check the supplementary material for the red interval plots). The contours are shown at the 1, 2, 3, and 4\,$\sigma$ levels, with the lowest PC projections in red and the highest in blue.  The columns, from left to right, represent Q, AGN, and SF spectra, and the rows, from top to bottom, correspond to the EAGLE, TNG100, and SDSS samples, respectively. The reader may note that the selection of galaxies for stacking differs from those in PCA-SDSS, where the 10th and 90th percentiles were used to identify extreme values. Our choice in this work is made because of the smaller number of galaxies.

In the three sub-groups of the EAGLE-SDSS comparison, the highest (blue) and lowest (red) values of the projection onto the first eigenspectrum (PC1, Fig.~\ref{fig:MCMC-pc1}), correspond to the oldest and youngest galaxies, respectively, with the average populations being, unsurprisingly, older in the sequence SF$\rightarrow$AGN$\rightarrow$Q. However, for the TNG100 sample, the highest and lowest values of PC1 correspond to a similar age; Q galaxies have an older average age, and although SF and AGN galaxies have younger average ages, their age distribution is similar. Concerning metallicity, no significant difference is found between Q galaxies in the three samples. Note that for the AGN galaxies in the TNG100 and SDSS samples there is no significant difference regarding metallicity, however in the EAGLE sample, the AGN stacks show different metallicities in the opposite direction to the age-metallicity degeneracy (the latter traced by the elongated confidence levels, from top left to bottom right in these plots). SF stacks in the three samples show different metallicities in the opposite direction to the age-metallicity degeneracy. Consequently, it is in the AGN sample where we find discrepancies between the observed and synthetic spectra concerning the variance distribution of PC1. It is worth emphasising that for the SF subset, all three cases (SDSS and both simulations) produce compatible results in these SSP fits, whereas the discrepancies arise with the AGN sample, and -- down the evolutionary path -- with the Q galaxies. This would suggest that the subgrid presciptions regarding star formation produce consistent results with real galaxies, whereas the AGN feedback requires more work.

Regarding PC2 (Fig.~\ref{fig:MCMC-pc2}), we find similar results in TNG100 and SDSS: Q spectra do not show any appreciable difference, while in the EAGLE sample, there is a significant correlation with age, in the sense that Q galaxies with low PC2 projections have older ages. AGN galaxies in the SDSS sample show a mild correlation with metallicity, towards low PC2 projections having lower [Z/H], and an opposite trend in the EAGLE sample, however, the differences are small. In contrast, the TNG100 sample does not show any substantial difference with respect to PC2 in either of the three subgroups. SDSS behaves similarly to the TNG100 stacks, but note that the metallicity in the AGN and especially SF groups shows different trends with respect to [Z/H].

The lowest variance component of our latent space, PC3 (Fig.~\ref{fig:MCMC-pc3}) behaves similarly to PC2 in the Q subgroup for the three samples, except that the correlation with age in the EAGLE sample is milder than in PC2. For the AGN subgroup and in the SDSS sample there is an anti-correlation with metallicity, in the TNG100 sample there is a correlation with metallicity (note these opposing trends were subtle but also present in SF galaxies from SDSS and TNG100).  In the SDSS sample, SF galaxies with the lowest PC3 values tend to be older; in TNG100 and EAGLE data no difference is apparent. At this level, we can only confirm that the simulated data give reliable results just at the highest level of variance. The Supplementary Material available in the online version includes the equivalent of Figs.~\ref{fig:MCMC-pc1}, \ref{fig:MCMC-pc2}, and \ref{fig:MCMC-pc3} for the red spectral interval. The trends agree well with the projections of the first principal component, demonstrating the substantial covariance across a wide range of wavelengths \citep[as illustrated in][]{InfoPop}. PC2 and PC3 have more subtle differences.

\subsection{Star formation history of the projected data}
\begin{figure*}
    \centering
    \includegraphics[width=0.35\textwidth]{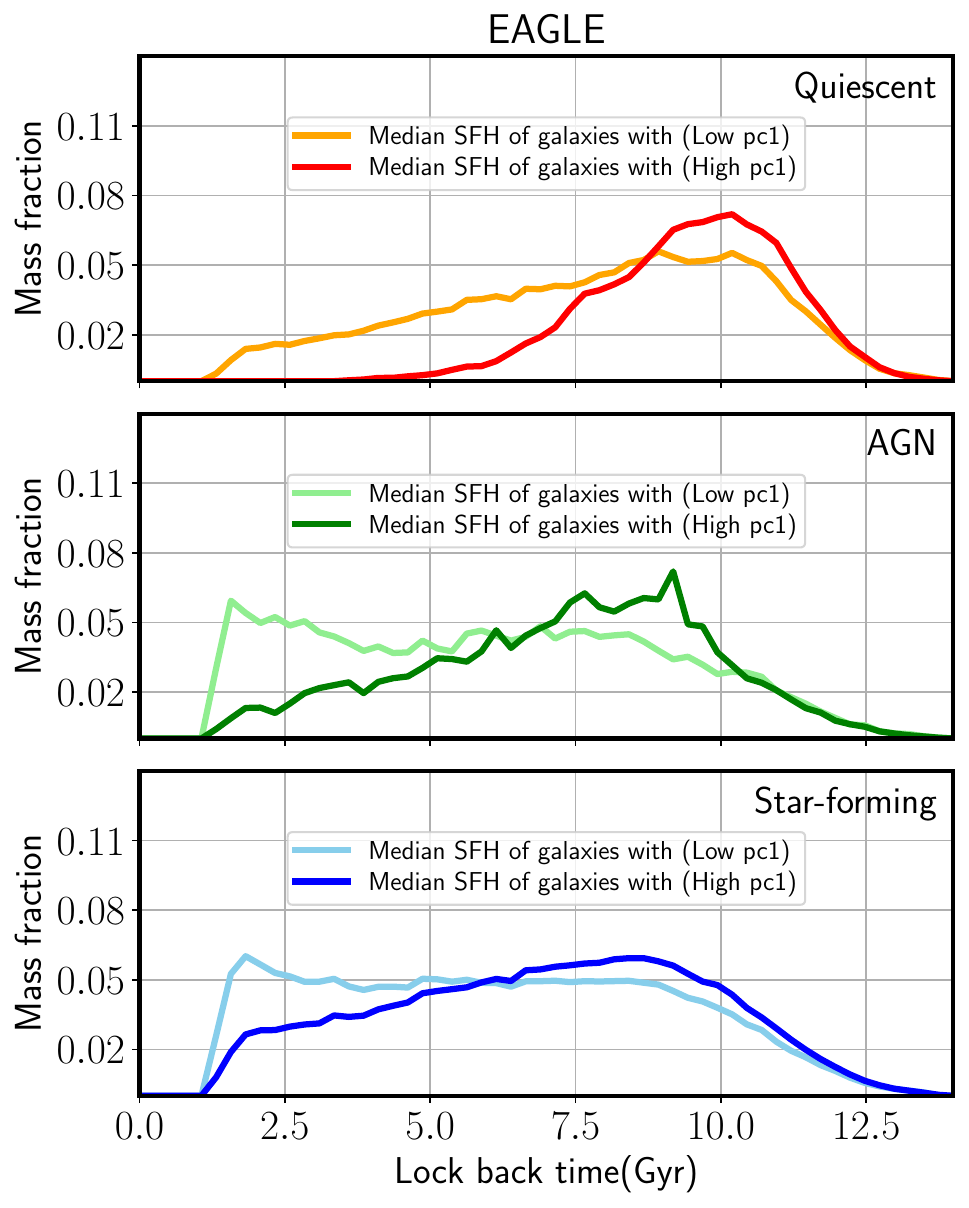}
    \includegraphics[width=0.35\textwidth]{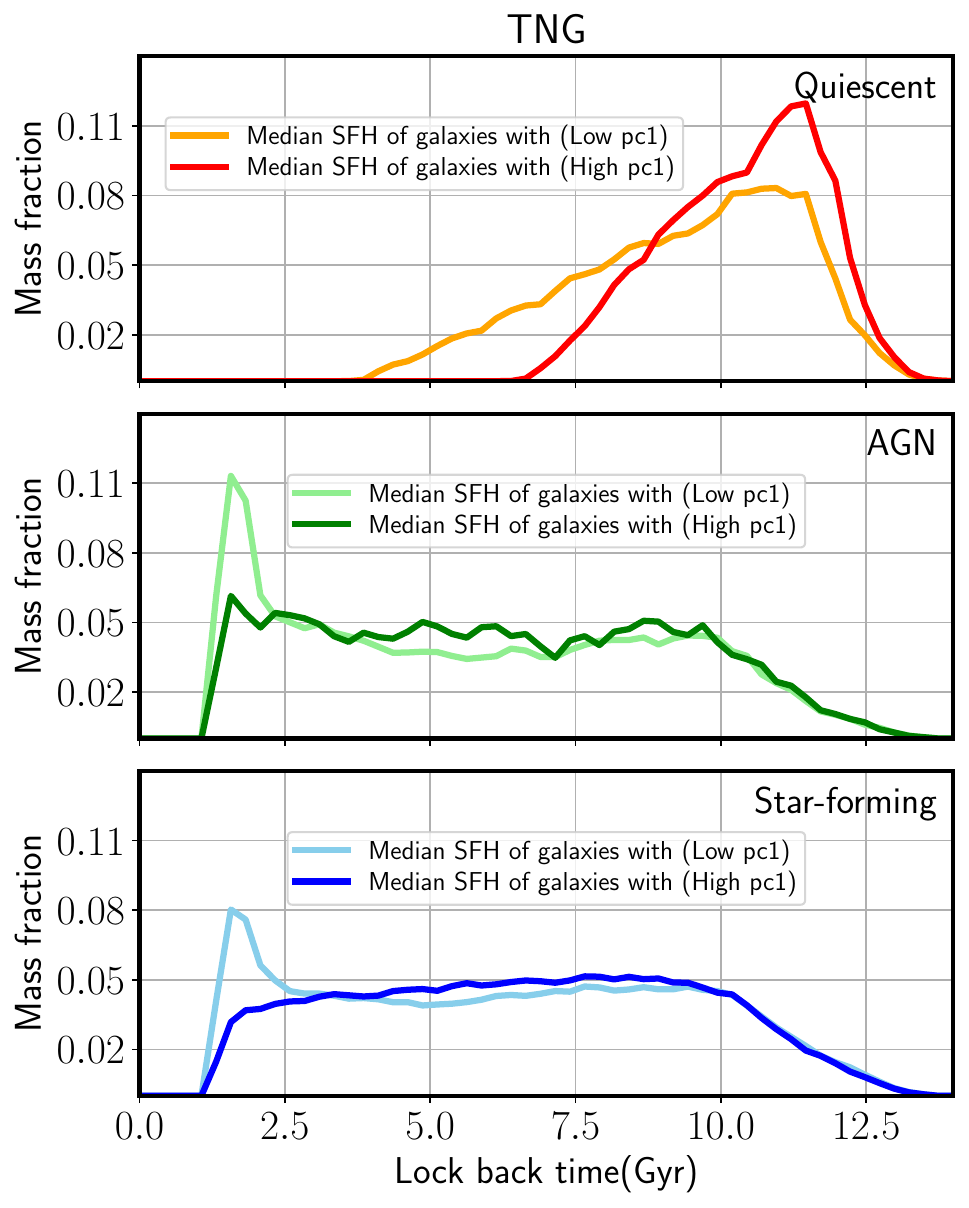}
    \caption{Median star formation history of galaxies with the lowest (33 percentile) and highest (67 percentile) value of PC1 in the blue interval, for Q, AGN, and SF sub-classes (from top to bottom). The EAGLE (TNG100) sample is shown in the left (right) column. } 
    \label{fig:SFH_sims_PC1}
\end{figure*}

\begin{figure*}
    \centering
    \includegraphics[width=0.35\textwidth]{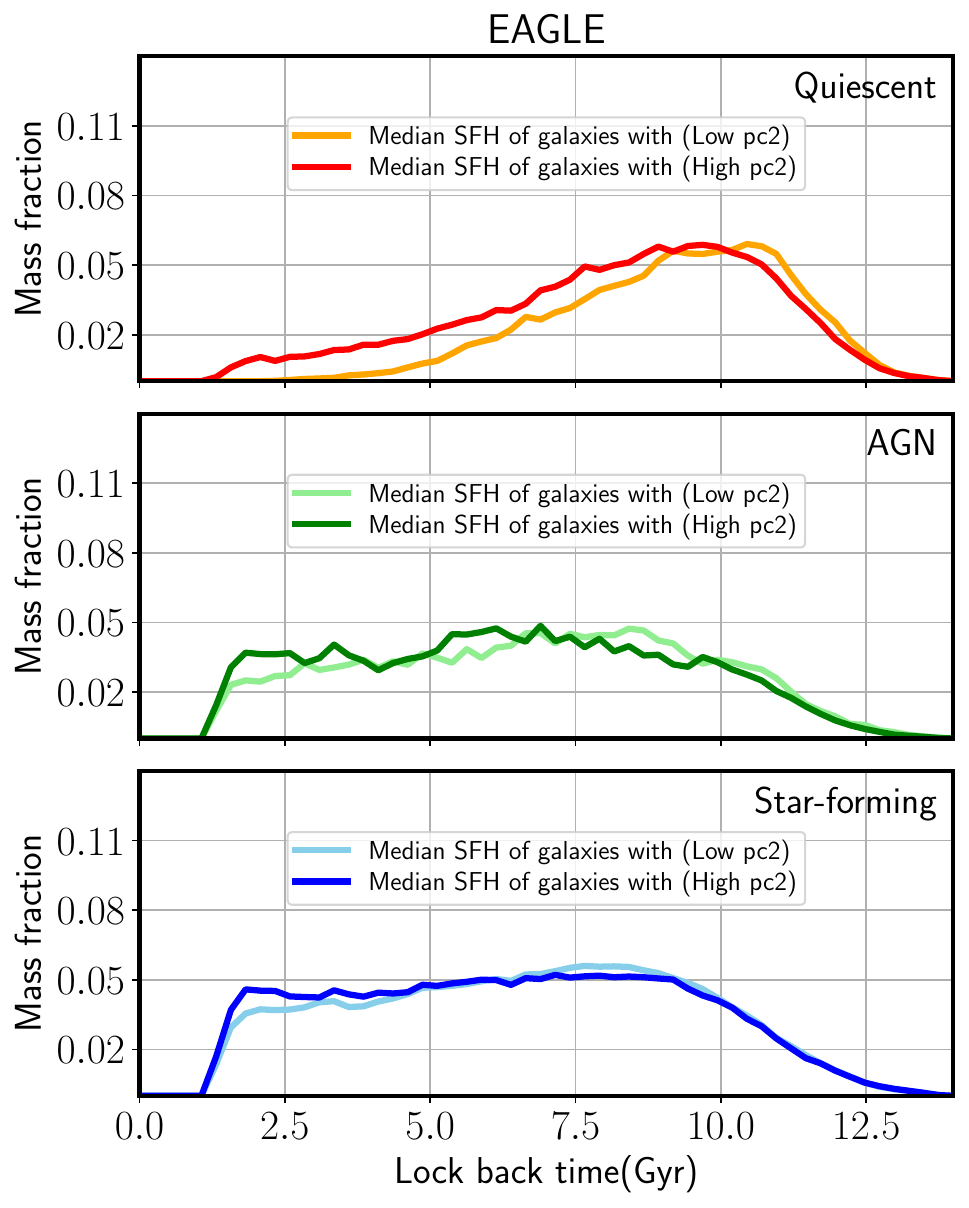}
    \includegraphics[width=0.35\textwidth]{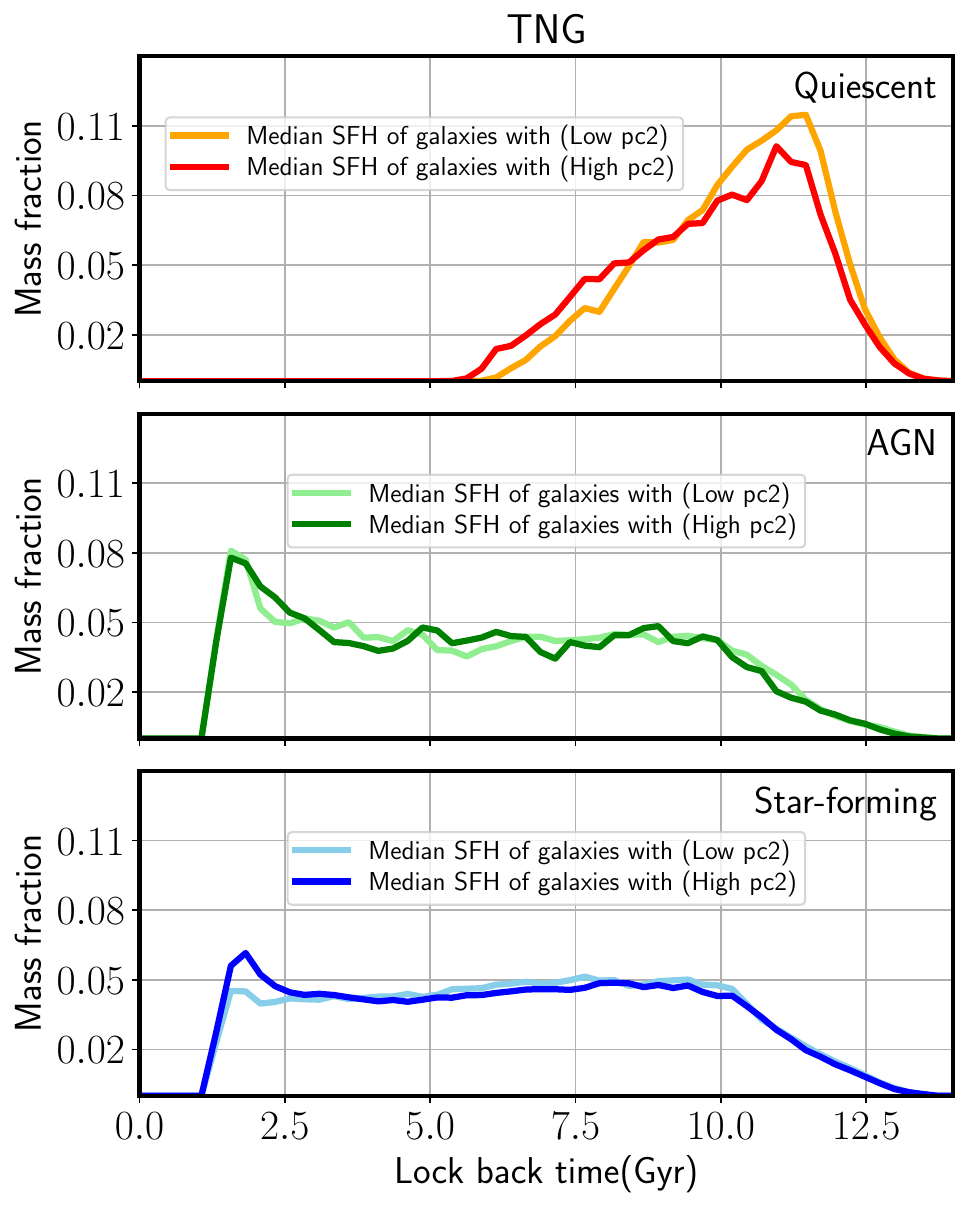}
    \caption{Equivalent of Fig.~\ref{fig:SFH_sims_PC1} for the projections onto PC2 in the blue interval.}
    \label{fig:SFH_sims_PC2}
\end{figure*}

\begin{figure*}
    \centering
    \includegraphics[width=0.35\textwidth]{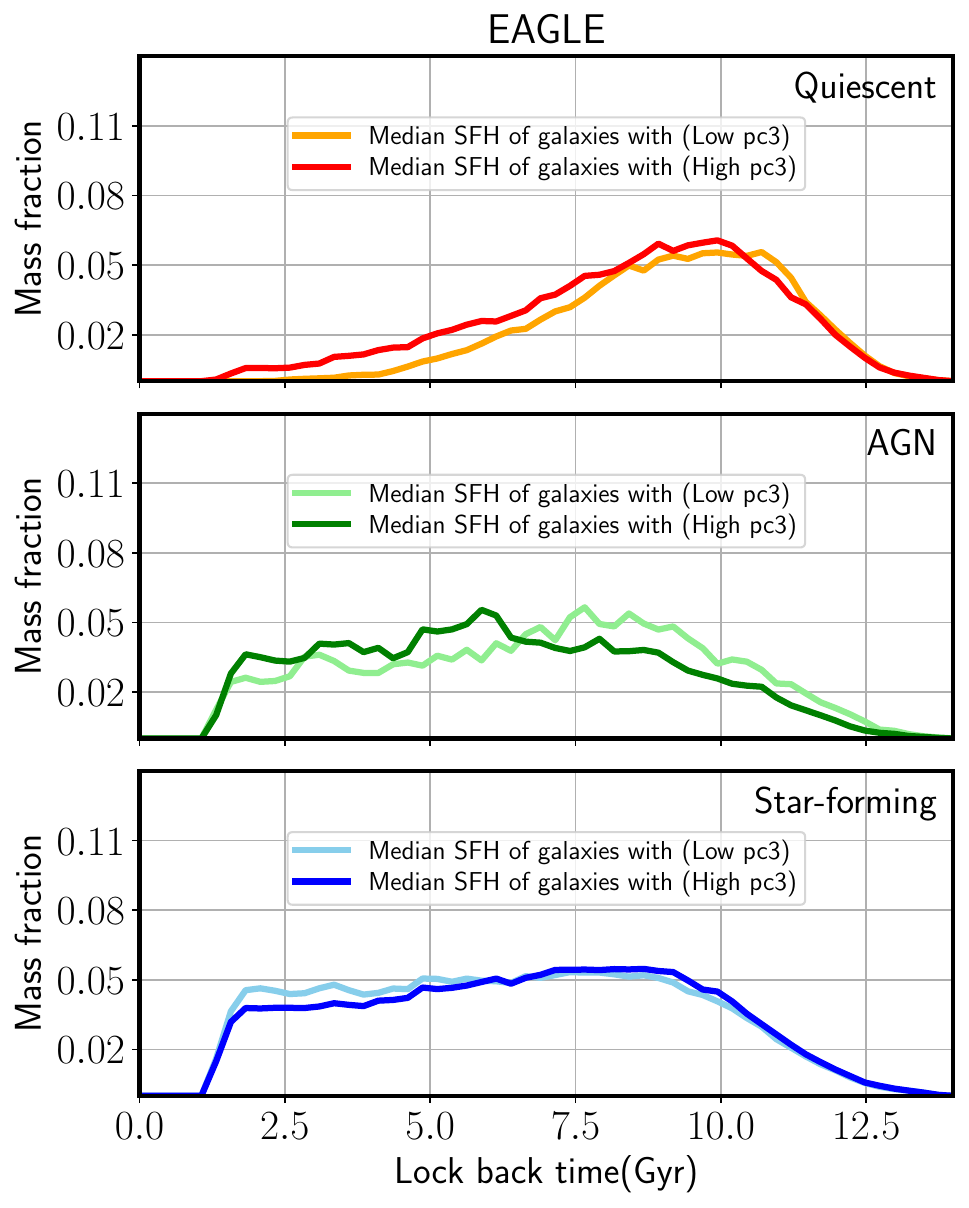}
    \includegraphics[width=0.35\textwidth]{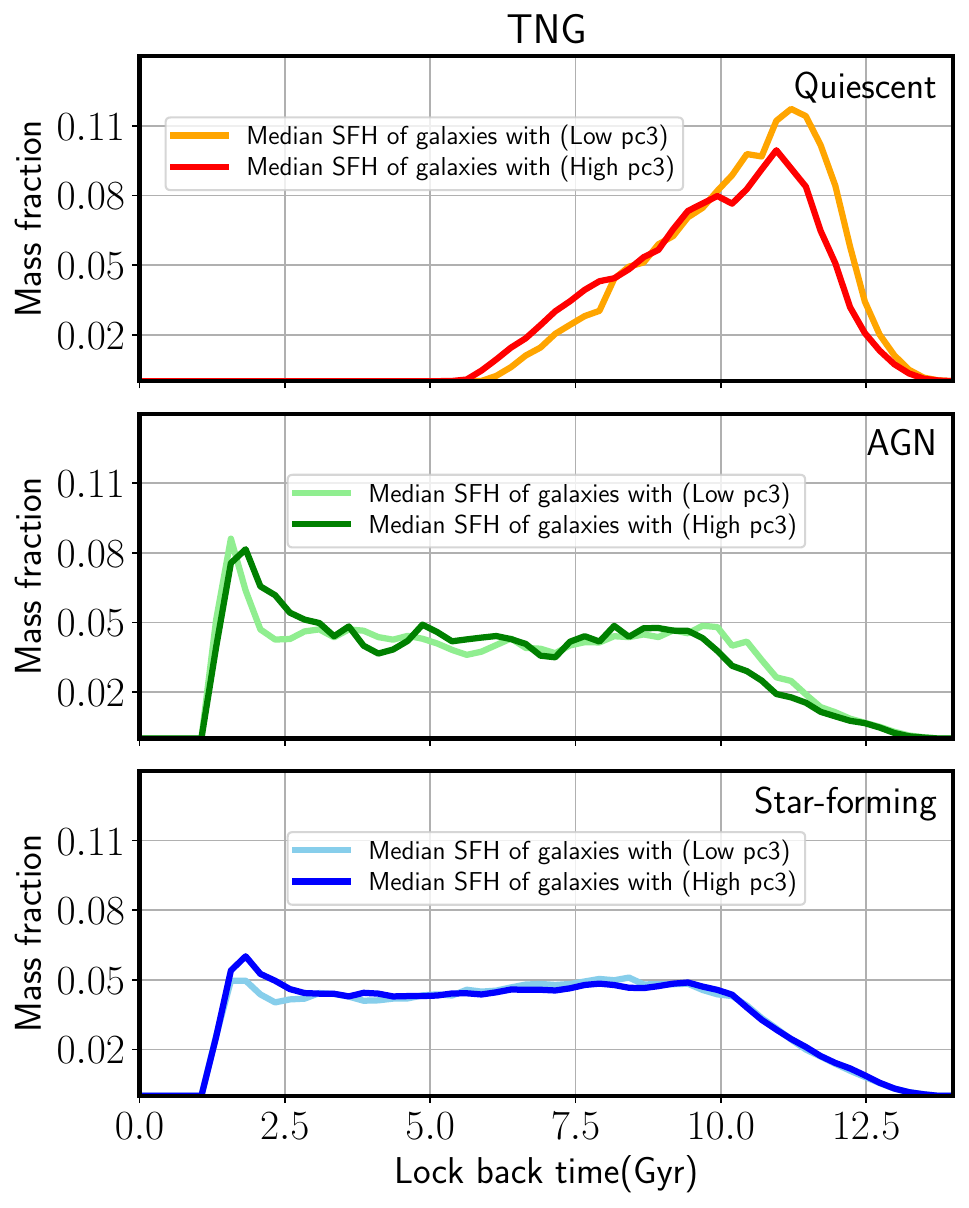}
    \caption{Equivalent of Fig.~\ref{fig:SFH_sims_PC1} for the projections onto PC3 in the blue interval.}
    \label{fig:SFH_sims_PC3}
\end{figure*}

The advantage of using simulation data in this project is that we can now direct the analysis towards the actual star formation histories (SFHs) produced by the models. The segregation is performed as above, i.e. regarding the nebular activity (SF/AGN/Q) and depending on the projection on the first three principal components obtained from the SDSS spectra. We split again the sample for each component into the lowest (33rd) and highest (67rd) percentiles of the distribution. Figs.~\ref{fig:SFH_sims_PC1}, \ref{fig:SFH_sims_PC2}, and \ref{fig:SFH_sims_PC3} show the median star formation history for each subclass for galaxies with the highest and lowest values of PC1, PC2, and PC3, respectively. The panels on the left (right) of the figures column represent the EAGLE (TNG100) sample. The formation histories are produced by binning the formation time weighed with the initial mass of the stellar particles formed in steps of  50\,Myr (the size of a time bin). These figures take into account the effect of the returned fraction, i.e. the fraction of mass ejected from stars back into the gas phase. This fraction allows us to transform the observed stellar mass distribution into a star formation history. 

In Fig.~\ref{fig:SFH_sims_PC1}, the median SFH is shown for galaxies in the 33rd and 67rd percentiles of the distribution of PC1. Among Q galaxies, the overall shape of star formation is different for galaxies with the highest and lowest values of PC1, while the peak of star formation occurs at approximately the same (early) time. Galaxies with the highest PC1 values formed rapidly, and their mass accumulated faster than galaxies with low PC1 values. The EAGLE and TNG100 samples both exhibit similar behaviour, although the difference between high and low PC1 in the TNG100 sample is more subtle, having fewer Q galaxies with later star formation. AGN galaxies in EAGLE with the lowest PC1 values formed in a more extended period with respect to those with higher PC1 projection, and interestingly, show an increased star formation at later cosmic time. In the TNG100-AGN sample both subgroups with the lowest and highest PC1 values have approximately constant star formation for over 8\,Gyr, with a late spike in star formation for galaxies with low PC1. In both simulations, the SF galaxies show very similar SFHs with respect to the AGN population for both subgroups. The fact that SF and AGN galaxies share the same SFH may explain the overlap between these two groups in latent space. Although there may be subtle differences between galaxies with the lowest and highest PC distributions, PC2 and PC3 (Figs.~\ref{fig:SFH_sims_PC2} and \ref{fig:SFH_sims_PC3}), both groups feature similar star formation histories. Most of the differences can be attributed to the projection onto the first principal component. Notably, this component is purely driven by the variance of the data, and it can identify differences the stellar populations and in the SFHs. In the Supplementary Material available in the online version of this paper, similar results are shown for the red interval. They are consistent with those shown for the blue interval.

\section{Discussion and Conclusions}\label{sec-discussion}

This paper examines cosmological hydrodynamic simulations of galaxy formation following an innovative methodology based on the covariance of observed spectra, as presented in \citet[][referred throughout this paper as PCA-SDSS]{variance}. This approach produces the fundamental units for comparison (the principal components) in a fully data-driven manner. The eigenvectors are independently derived from the continuum-subtracted data, i.e. focusing on the absorption features of the stellar populations in three different sub-groups of galaxies: star-forming (SF), AGN, and quiescent (Q). In Sec.~\ref{synthetic-data}, we show how the synthetic spectra are generated, with the same instrumental and observational effects as the SDSS sample. Additionally, different methods have been tested to add sky noise to the synthetic spectra, as described in Sec.~\ref{resol-noise} and Appendix~\ref{D}. A realistic simulation is expected to produce the same distribution of spectral variance as in the observed sample. This is a powerful way to test the validity of simulations that go beyond the standard comparison of scaling relations, spectral fitting, etc. In Sec.~\ref{method}, we project the synthetic spectra of each different subgroup to the equivalent principal components derived from the SDSS subgroups and investigate the distributions in latent space. While the overall appearance of the synthetic data is comparable to the observations, there are subtle differences that can help researchers improve the modelling of galaxy formation at the subgrid level. In more detail, Sec.~\ref{sec-proj} compares the PCA-based projections with population synthesis models. There is an (unsurprising) overall agreement in the trends with a clear age sequence in all cases in the direction SF$\rightarrow$AGN$\rightarrow$Q. While an SSP analysis is inherently limited, it allows us to find that it is the AGN subset where the differences between simulations (and with observations) arise. Most of the agreement resides in the SF subset, suggesting that the star formation subgrid physics is less subject to discrepancies between simulations. As we will see below, differences in the AGN feedback are more pronounced, especially as it relies on the more uncertain processes involving the formation and growth of the central black hole.

\begin{figure}
    \centering
    \includegraphics[width=0.4\textwidth]{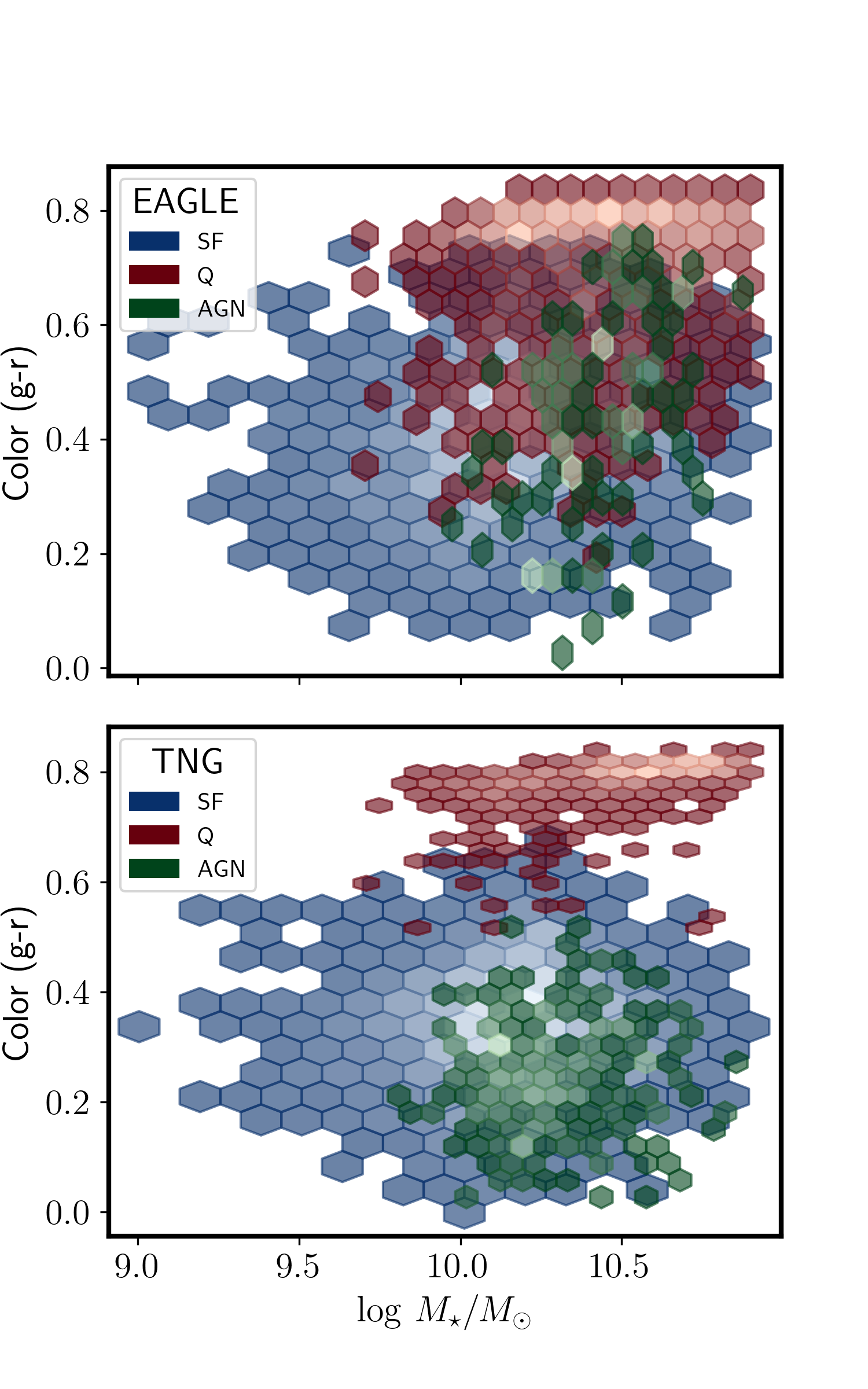}
    \caption{SDSS ($g-r$) colour vs stellar mass relation of the homogenised sub-samples of the EAGLE (top) and TNG100 (bottom) simulations. They are colour-coded with respect to their activity, into star-forming (blue), AGN (green) and quiescent (red). The SDSS $g$ and $r$ (dust-free) magnitudes are taken directly from the simulations.}
    \label{fig:color}
\end{figure}

As this analysis is a comparison between observed and synthetic spectra, it relies on the homogenisation of the samples. For a comparison with the analysis of SDSS spectra presented in PCA-SDSS, we need to classify the synthetic galaxies into three subgroups that were defined in the observational data with respect to nebular activity, and then homogenise them based on each pair of subgroups between simulation and observation. Initially, it may seem more logical to homogenise each pair of SDSS-EAGLE or TNG100-SDSS spectra and then group them according to their characteristics \citep{Angthopo:2021}. However, this procedure produces subsamples with different stellar mass distributions (Fig.~\ref{fig5}, see App.~\ref{A}) between simulation and observation and, consequently, different stellar populations, due to the well-known correlation between stellar mass and stellar population properties \citep{Bernardi:03, Gallazzi:05, Gallazzi:14, Ferreras:19}. The SDSS spectra have been classified with respect to the standard BPT \citep{BPT} emission line ratio diagram. In the simulations, we decide not to rely on the modelling of the emission lines \citep[as in, e.g.][]{Hirschmann:2023} and opt instead for a more fundamental classification of nebular activity based on the parameters that control star formation and AGN activity. To do so, the bivariate distribution of $\lambda_{\rm Edd}$ (that parameterise the growth of the SMBH) and specific star formation rate are used to classify the synthetic spectra into different sub-groups. In Fig.~\ref{fig1}, a notable difference is found between TNG100 and EAGLE simulations, with the former featuring a substantially lower scatter between SF and AGN activity. Several studies have shown that there is small scatter in the TNG100 simulations, reflected in a variety of relationships, for example between stellar mass and black hole mass \citep{Habouzit:2021}. TNG100 and EAGLE samples differ in this relation between $\lambda_{\rm Edd}$ and sSFR because in EAGLE, quiescent galaxies exhibit a wide range of sSFR ($10^{-13} \leqslant\log\,$sSFR(yr$^{-1}$)$\leqslant 10^{-11}$), in contrast with the lower values of sSFR for quiescent galaxies in TNG100 \citep{Habouzit:2021}. Note that galaxies with $\log\,$sSFR (yr$^{-1}$)$= 10^{-14}$ in Figure \ref{fig1}, have SFR $=0$ and this value is defined so that they can be plotted. Different feedback models used in these two simulations produce different quenching levels. Star formation and AGN are arguably the main two feedback mechanisms that quench star formation in galaxies. Depending on the stellar mass range of the sample, either can be responsible for quenching. A clustering of highly quenched galaxies can be found in TNG100 with practically zero SFR for galaxies whose stellar mass exceeds $\sim 10^{10.5}$M$_\odot$. In contrast, in the EAGLE simulation, they can exist throughout the entire stellar mass range \citep{Habouzit:2021}. Fig.~\ref{fig:color} illustrates the relationship between the SDSS $(g-r)$ colour, used as a proxy of the quenching level, vs. stellar mass for the homogenised sub-samples.  In EAGLE, quiescent galaxies (red dots) have a large scatter, showing a wide range of quenching levels despite the clustering of highly quenched galaxies. The TNG100 sample exhibits a much smaller colour scatter for Q galaxies especially at higher stellar mass. Our definition of AGN galaxies includes those with different star-formation levels and high BH accretion rates. AGN  galaxies are known to contribute substantially to the Green Valley region \citep[e.g.,][]{Angthopo:19}. AGN activity may trigger star formation in the host galaxies bringing them towards rejuvenation, or alternatively quench star formation \citep{2Mulcahey:2022, Lammers:2023}. Our results (see Figs.~\ref{fig1} and \ref{fig:color}) indicate, for the EAGLE galaxies, that the AGN regime can be defined as described above. In contrast, in TNG100 galaxies, due to the lack of scatter in the parameter space as well as the sudden onset of AGN activity, we only find AGN galaxies with a large amount of star formation, and fewer AGN on the Green Valley.

\begin{figure}
    \centering
    \includegraphics[width=0.42\textwidth]{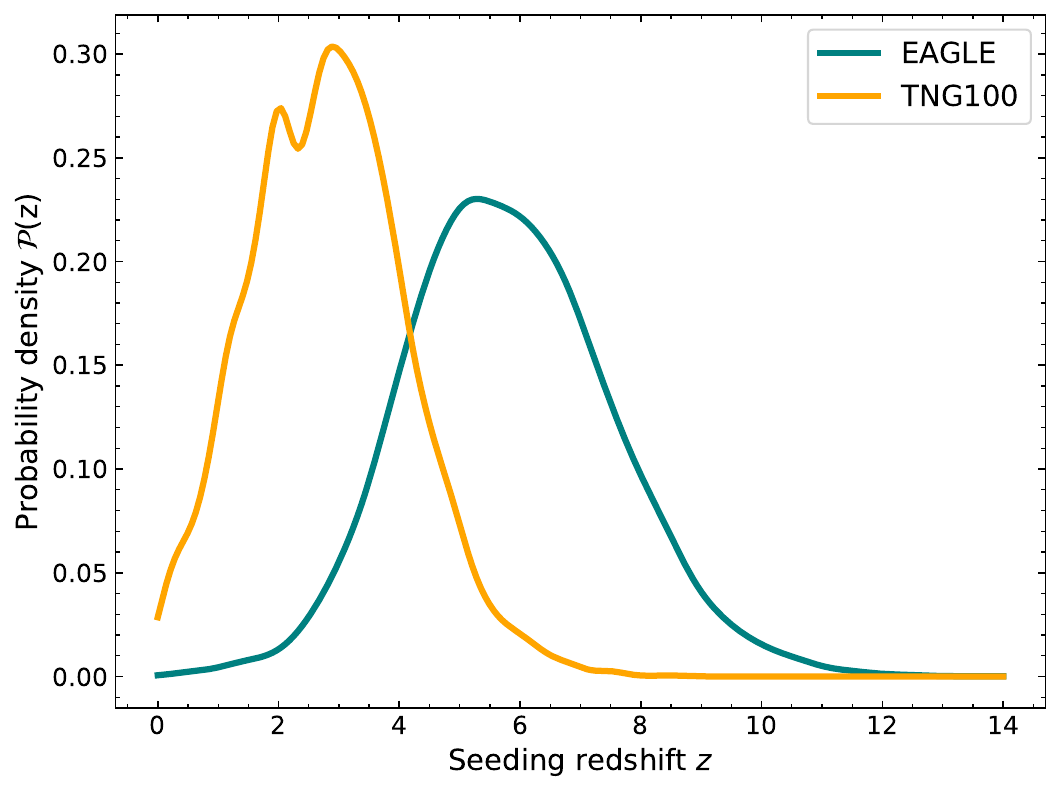}
    \includegraphics[width=0.42\textwidth]{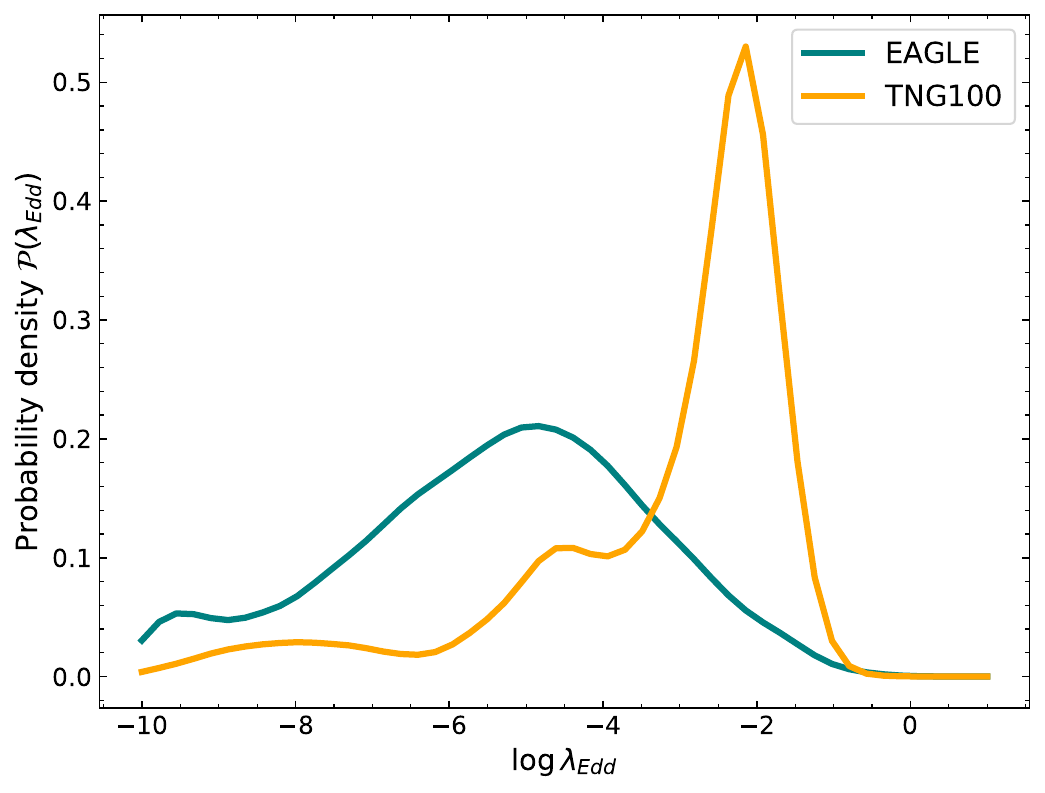}
    \caption{Top: Comparison of the SMBH seeding times -- expressed as redshift -- between the EAGLE and TNG100 simulations. Bottom: Distribution of $\lambda_{\rm Edd}$ in EAGLE and TNG100, measured in the $z=0.1$ snapshot.
    }
    \label{fig:BHs}
\end{figure}

By projecting the synthetic spectra onto the first three principal components obtained by the SDSS data, we reduce the dimensionality of the problem, describing each galaxy by just three numbers that keep the highest amount of variance. We aim to understand how these components relate to the physical properties and how they differ between observed and simulated samples. The latent spaces for observed and simulated data are shown in Figs.~\ref{fig:corner_sims} and \ref{fig:3D_sims}. The trends found support the hypothesis of an evolutionary sequence from star formation to AGN to quiescence, reflecting the bimodality between galaxies \citep{Strateva:01, Baldry:04}. We emphasize that our results provide a complementary approach to comparisons between simulations and observations, with respect to the more standard techniques based on model fitting. Regarding the location of the different subgroups in latent space, TNG100, EAGLE, and SDSS samples appear to be mostly in good agreement. However, there is an overlap between the AGN and star-forming subgroups of the simulated samples, whereas quiescent galaxies appear separated from these in the TNG100 data. This can be related to the quenching mechanisms implemented in the simulations.  EAGLE finds shallow AGN activity in galaxies with stellar mass below $10^{9.7}$M$_\odot$, with most of the quenching attributed to stellar feedback or the environment \citep{Crain:15}. Above $10^{9.7}$M$_\odot$, EAGLE galaxies quench star formation via AGN feedback \citep{Bower:2017}. In the stellar mass range of $10^{9.7}<$M$_s$/M$_\odot < 10^{10.3}$, EAGLE mimics radio-mode AGN feedback, while more massive galaxies undergo a rapid increase in the SMBH accretion rate \citep{Wright:2019}. In the TNG100 simulation, above $10^{10.5}$M$_\odot$\  the kinetic BH-driven winds suppress star formation. At the stellar mass of $10^{10.5}$M$_\odot$, AGN feedback switches from predominantly thermal feedback to kinetic/radio mode feedback \citep{weinberger:2017, nelson:2018, Terrazas:2020, Donnari:2021, Davies:2020}. AGN feedback in TNG100 is more drastic, with  a stronger quenching effect. This is due to the higher accretion rate in many BHs and the shorter time scale in galaxy life. To illustrate a possible cause, we contrast in the top panel of Fig.~\ref{fig:BHs} the difference between the BH seeding times (expressed as the redshift at which the BH is inserted in the galaxy) in both simulations. We consider in this figure all SMBHs at the centres of galaxies in the z=0.1 snapshot. Note that the ``formation'' of black holes in EAGLE occurs earlier in cosmic time and appears more evenly distributed in redshift, whereas the black holes in TNG100 are formed much later in time with a sharp peak approximately at z$\sim$2. Such a difference can lead to an implementation of stronger AGN feedback if the data are to match the observational constraints, as suggested by the distribution of the $\lambda_{\rm Edd}$ parameter, shown in the bottom panel of Fig.~\ref{fig:BHs}. It is worth noting that both the black hole seeding mass and the halo mass threshold for seeding are about one order of magnitude higher in TNG100 with respect to EAGLE. This would mean that statistically, later BH seeding is expected in the former, as shown in the figure.

The analysis of the distribution in PC1 projection shown in PCA-SDSS demonstrated that this component has a strong correlation with stellar age in the observed data. In this paper, we consistently show, using SSP fitting (Fig.~\ref{fig:MCMC-pc1}), that EAGLE galaxies also differ primarily due to their ages in PC1 projection, suggesting a good match with the observed spectra. In contrast, TNG100 data appear more indistinguishable. We ascribe this difference to the quenching mechanisms adopted by the simulations. Regarding metallicity, PC1 does not appear to be correlated with metallicity in the SDSS and TNG100 data, but AGN galaxies in EAGLE show difference in metallicity. Although SDSS and EAGLE galaxies have some agreement on the SSP-equivalent age distributions, the way EAGLE produces AGN galaxies does not properly match the spectral variance of the observational data, with discrepancies possibly at the level of chemical enrichment.

Our results (Fig.\ref{fig:MCMC-pc2}) indicate that the second-order component of variance in galaxy spectra differs from the synthetic spectra of EAGLE among the Q and AGN groups. In contrast, the three subgroups of the TNG100 sample behave similarly to the observed spectra. In terms of the third-order component (PC3), we are unable to attribute the difference between the observed and synthetic spectra to any particular group (Fig.~\ref{fig:MCMC-pc3}). Each sample appears to have distinct PC3 variance properties, as well as significant discrepancies from the observed spectra. Overall, the analysis reveals that as far as the general difference between different groups of galaxies, EAGLE produces the most realistic set, but even in this case we do find a mismatch in the covariance with respect to real (SDSS) galaxies.

Finally, we extract directly details about the formation histories from the simulations. We explore the star formation history of galaxies segregated with respect to their projections in latent space. The SFH of galaxies selected with respect to PC1 projection (Fig.~\ref{fig:SFH_sims_PC1}) shows a clear separation in PC1, in the sense that high PC1 projections in all cases correspond to earlier formation, whereas lower values of PC1 include later star formation. Interestingly, Q galaxies appear more homogeneous (and old) in TNG100, whereas EAGLE galaxies show a strong trend towards more extended SFHs at lower projections of PC1. AGN galaxies also feature a clear difference, with EAGLE galaxies with high PC1 showing a decreasing star formation rate with cosmic time, whereas TNG100 show overall constant SFH with an intringuing spike of strong star formation at later times. Note that the SFHs shown in Fig.~\ref{fig:SFH_sims_PC1} correspond to a median of all galaxies located within the prescribed two percentiles regarding PC1 projection, thus cannot be interpreted as a single galaxy. SF galaxies also show differing SFHs in EAGLE and TNG100, with the former once more showing a stronger trend between PC1 and the presence of extended star formation. The results for the higher order components PC2 (Fig.~\ref{fig:SFH_sims_PC2}) and PC3 (Fig.~\ref{fig:SFH_sims_PC3}) do not show substantial differences.

This paper illustrates the power of spectral variance as a way to constrain simulations of galaxy formation. While these simulations are typically tested/calibrated with well-established scaling relations (e.g. colour-magnitude, Tully-Fisher) or distribution functions (stellar mass function, effective radius), our approach adds a complementary way that focuses on the way the star formation and chemical enrichment histories realistically map the behaviour of real galaxies (i.e. SDSS) through their spectra. In the spirit of \citet{Disney:08}, we find that a single parameter catches the essence of the difference and we identify this as the overall dependence on stellar age, with high (low) values of PC1 projection mostly represented by older (younger) populations. However, in more detail, the latent three-dimensional space spanned by the three components with the highest spectral variance provide a powerful benchmark where simulations need to be tested. 

\section*{Acknowledgements}
IF and ZS acknowledge support from the Spanish Research Agency of the Ministry of Science and Innovation (AEI-MICINN) under grant PID2019-104788GB-I00. AN and CDV acknowledge grant PID2021-122603NB-C22 from the same funding agency. OL acknowledges STFC Consolidated Grant ST/R000476/1 and a Visiting Fellowship at All Souls College and at the Physics Department, Oxford. ZS and IF thank Prof. Marina Trevisan and Dr. Elham Eftekhari for their valuable comments and suggestions.  Funding for SDSS-III has been provided by the Alfred P. Sloan Foundation, the Participating Institutions, the National Science Foundation, and the U.S. Department of Energy Office of Science. The SDSS-III web site is http://www.sdss3.org/. We acknowledge the Virgo Consortium for making their simulation data available. The EAGLE simulations were performed using the DiRAC-2 facility at Durham, managed by the ICC, and the PRACE facility Curie based in France at TGCC, CEA, Bruy\`eres-le-Ch\^atel. We also thank all members of the IllustrisTNG collaboration for making their data publicly available.

\section*{Data Availability}
The spectra used in the analysis presented here can be downloaded from \href{https://zenodo.org/records/14044867}{Zenodo}. The repository includes {\sc Python} code that illustrates the PCA decomposition. This work has been fully based on publicly available data: the original galaxy spectra were retrieved from the \href{https://www.sdss.org/dr16/}{SDSS DR16 archive} and stellar population synthesis models can be obtained from the respective authors. The synthetic spectra are retrieved from the \href{https://icc.dur.ac.uk/Eagle/index.php}{EAGLE} and \href{https://www.tng-project.org/}{IllustrisTNG} simulations.

\bibliographystyle{mnras}
\bibliography{Sims-SDSS}

\appendix

\section{Homogenisation and classification of synthetic spectra}\label{A}
\begin{figure}
    \centering
    \includegraphics[width=0.52\textwidth]{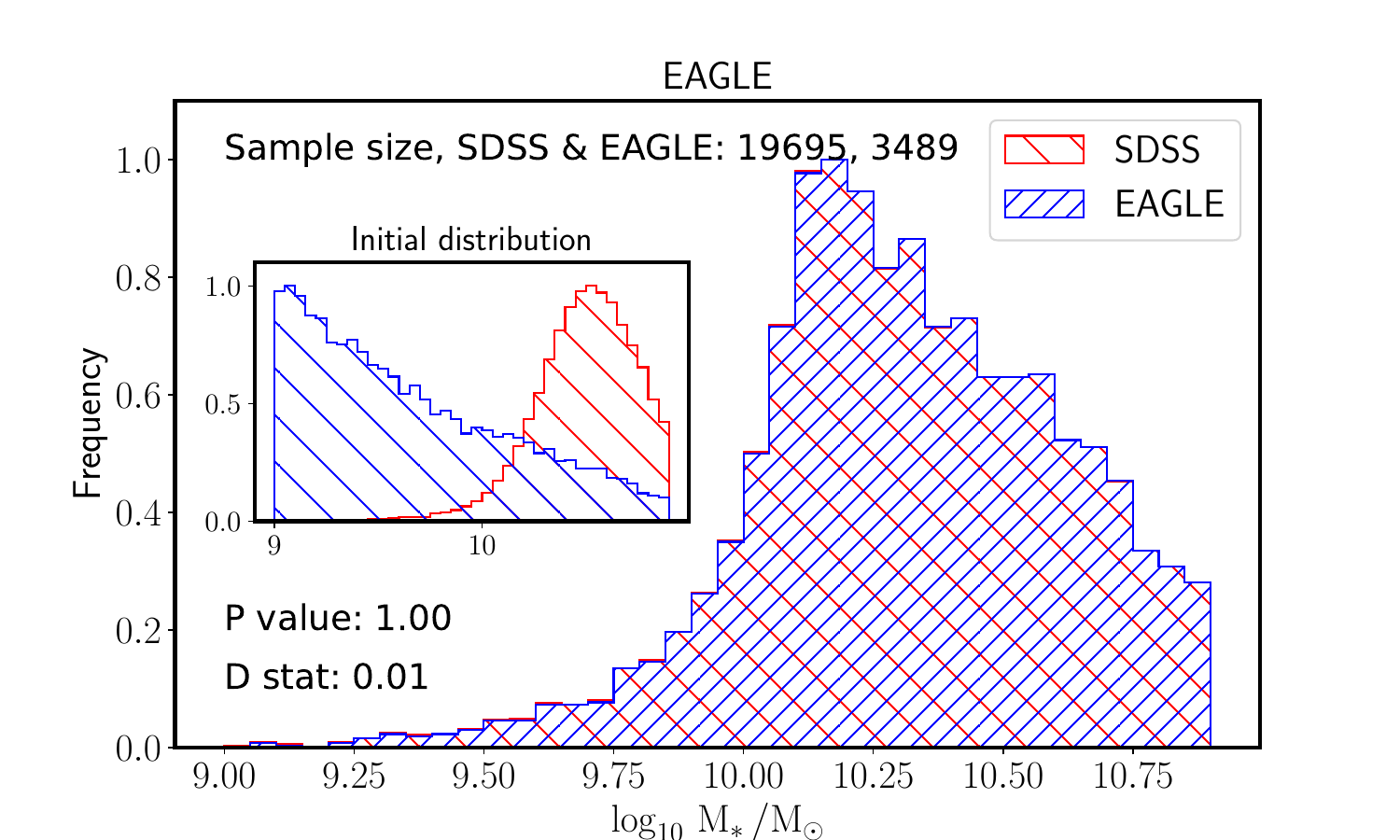}
    \caption{Distribution of stellar mass before and after homogenisation between the whole observed sample, SDSS, and the whole simulation sample, EAGLE. The size of the samples is the size of the sample after homogenisation.  A KS test confirms that these distributions originate from the same parent sample.}
    \label{fig:A1}
\end{figure}

\begin{figure}
    \centering
    \includegraphics[width=0.52\textwidth]{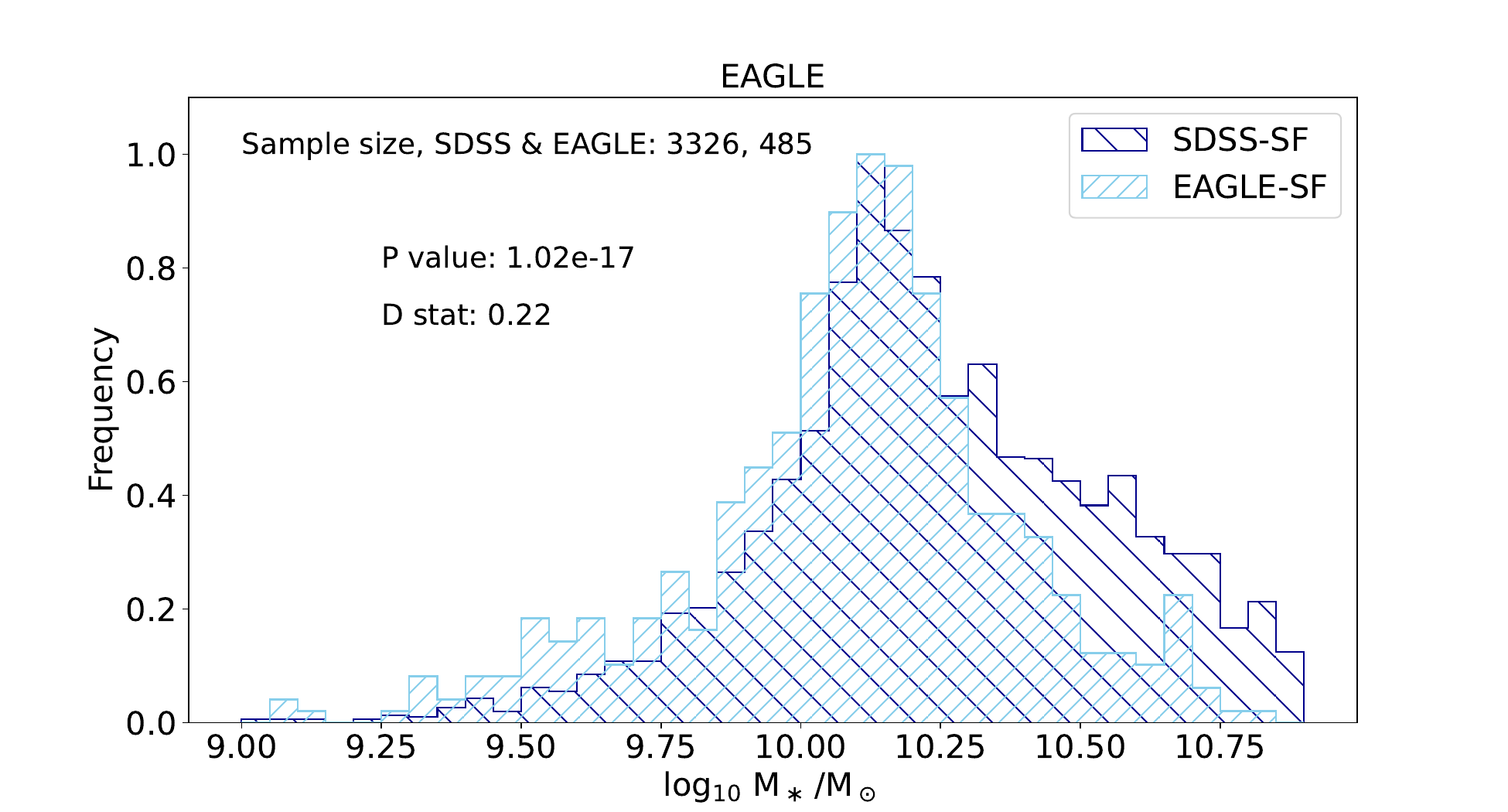}
    \includegraphics[width=0.52\textwidth]{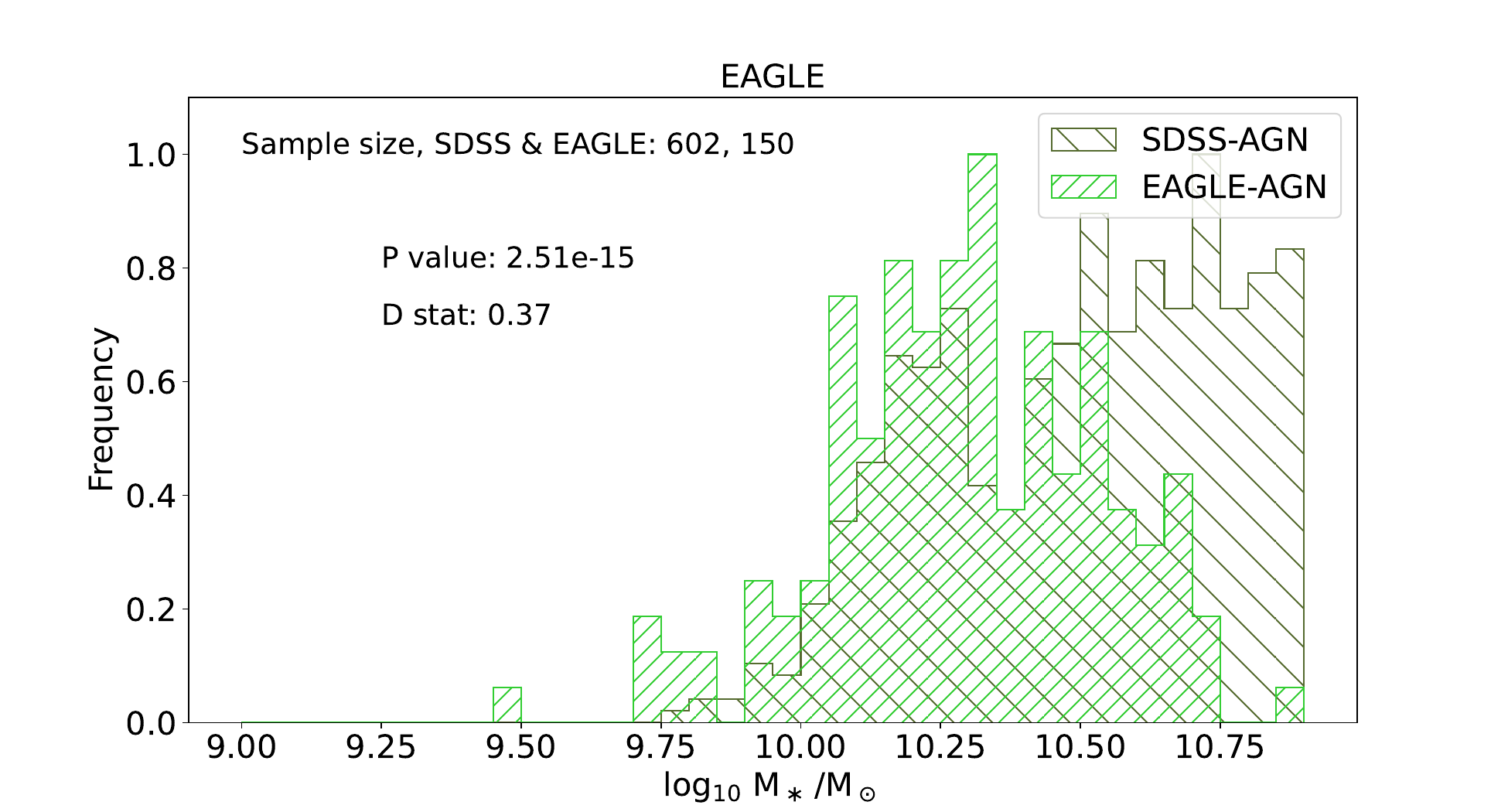}
    \includegraphics[width=0.52\textwidth]{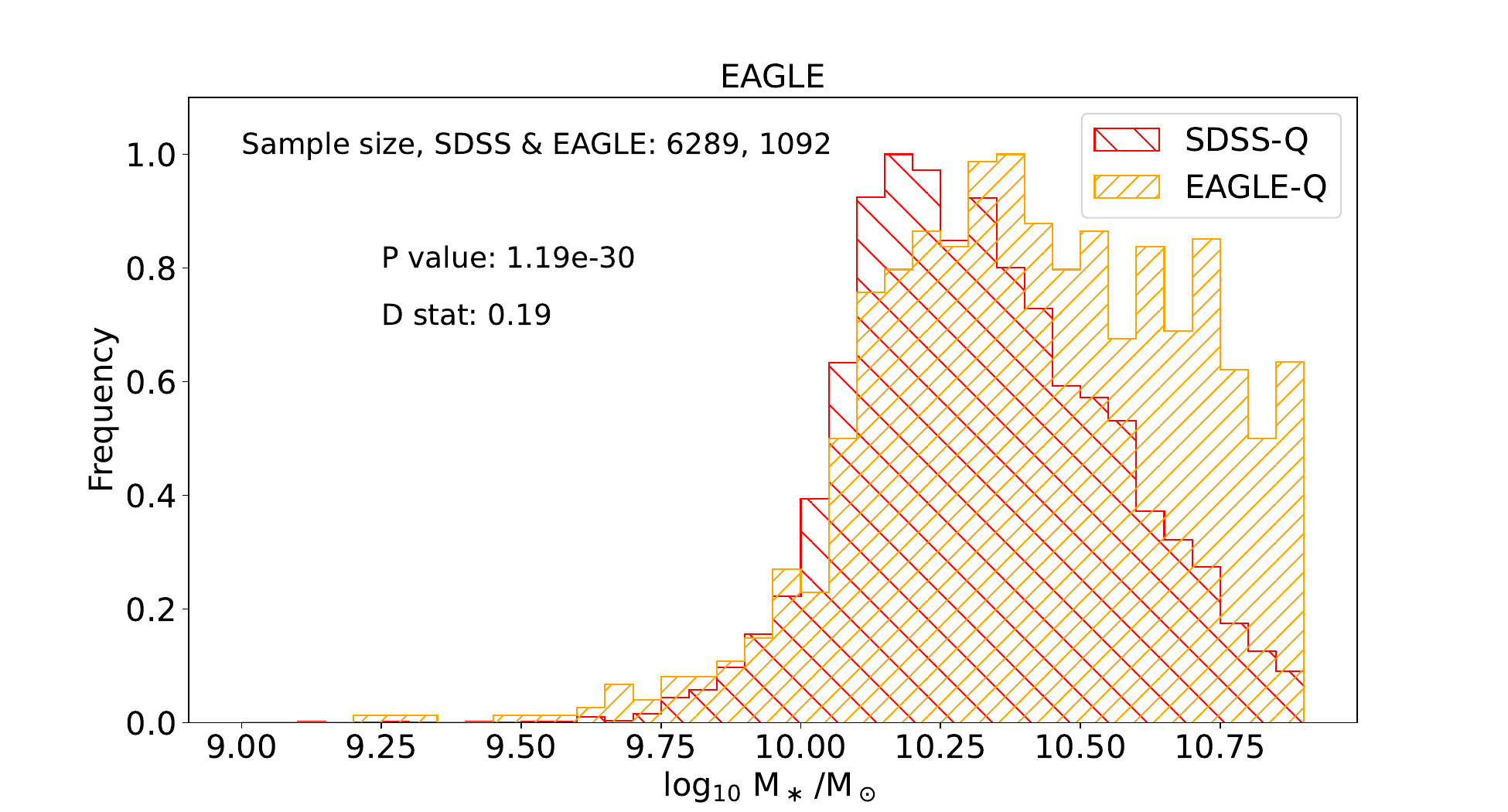}
    \caption{Stellar mass distribution of the star-forming (blue), AGN(green), and quiescent (red) galaxies of EAGLE and SDSS homogenised sample. The classification of the SDSS sample into different subgroups is based on the BPT, and for the EAGLE sample is based on the bivariates of the $\lambda_{\rm Edd}$ and sSFR. A KS test doesn't confirm that these distributions originate from the same parent sample.}
    \label{fig5}
\end{figure}

Following the homogenisation method of \citet{Angthopo:2021}, we tried to first homogenise the total sample of each simulation with the observed galaxy spectra from SDSS. Figure~\ref{fig:A1} shows the initial stellar mass distribution, as well as the distribution after homogenisation, for the EAGLE-SDSS pair. A KS test applied to the distributions after homogenisation confirms that they originate from the same parent sample. After homogenising the samples, we separated the synthetic spectra into different subgroups using the bivariate spanned by $\lambda_{\rm Edd}$ and sSFR. Threshold values were imposed on $\lambda_{\rm Edd}$ and sSFR to produce equivalent global ratios of Seyfert, (strong) SF, and Q to those found in the fully homogenised SDSS samples. Table~\ref{tab1} shows the selection criteria and the fraction of galaxies in each subsample. Note that as the SF/AGN/Q subsets have been chosen as clear cases regarding their activity, the sum of all do not add up to 100\%, but rather to approximately half of the total sample. Despite our interest in the relative mass-dependent variation between the different groups, we also need to have comparable stellar mass distributions within each subsample between SDSS and its respective simulation. The method just described unfortunately generates different stellar mass distributions for each subsample\footnote{Despite the great success of simulations such as EAGLE and TNG100, we attribute this issue to a combination of sample selection bias along with the failure of numerical simulations to produce a fully consistent picture of galaxy formation.}. Such differences will produce systematic differences in the stellar populations. See Fig.~\ref{fig5} as an example between EAGLE and SDSS sets. There can be several reasons why stellar mass distributions are different when we match the global fractions of galaxies between simulation and observation. One is the non-trivial subgrid physics in simulations at the low mass end. It is important to note that we have a low-velocity dispersion or a low-mass sample because it is more meaningful for the covariance analysis -- i.e. higher spectral resolution. So due to this incompatibility, we choose to separate the galaxies first, without taking into account the global fractions of galaxies among the simulation and observation, and then homogenise the galaxies of each subsample based on the stellar mass. While not fully satisfactory, this pragmatic approach allows us to produce the most consistent sets of galaxies between SDSS and the simulations.

\begin{table}
  \centering  
  \begin{tabular}{|c|c|c|c|c|}
    \hline
    Type & $x$=$\log(\lambda_{\rm Edd})$ & $y$=$\log(\text{sSFR} \, \text{yr}^{-1})$ & EAGLE (\%) & SDSS (\%) \\
    \hline
    SF & $x < -2$ & $y > -10.5$ & 14.03 & 16.88\\
    AGN & $x > -2$ & $y > -11.5$ & 4.34 & 3.05 \\
    Q & $x < -4.2$ & $y < -11$ & 31.60 & 31.93 \\
    \hline
  \end{tabular}
  \caption{The criteria used to classify galaxies, as Seyfert AGNs, quiescence, or star formations. Considering the total population of galaxies in the homogenised SDSS and EAGLE samples, the last two columns are indicative of the fraction of galaxies in each subsample. Here x represents the parameter at the top of each column. The $\lambda_{\rm Edd}$ and the sSFR are chosen in order to approximate the fraction of galaxies in each subsample with the fraction of galaxies (classified based on the BPT) in the SDSS sample (in our study as an example of the real universe).}
  \label{tab1}
\end{table}

Concerning the effect of the definition of stellar mass on the homogenisation process,
we note that there is a linear relation between the stellar mass measured inside the fibre and the total stellar mass. Regardless of this choice, the overall behaviour of the stellar mass remains mostly unchanged in the homogenisation process. Given the well-established relations between total stellar mass and stellar population properties measured within the SDSS fibres \cite[e.g.,][]{Gallazzi:05}, we decided to keep total stellar mass as the main parameter for homogenisation.

\section{3D rendering of latent space}\label{B}

\begin{figure*}
    \centering
    \includegraphics[width=60mm]{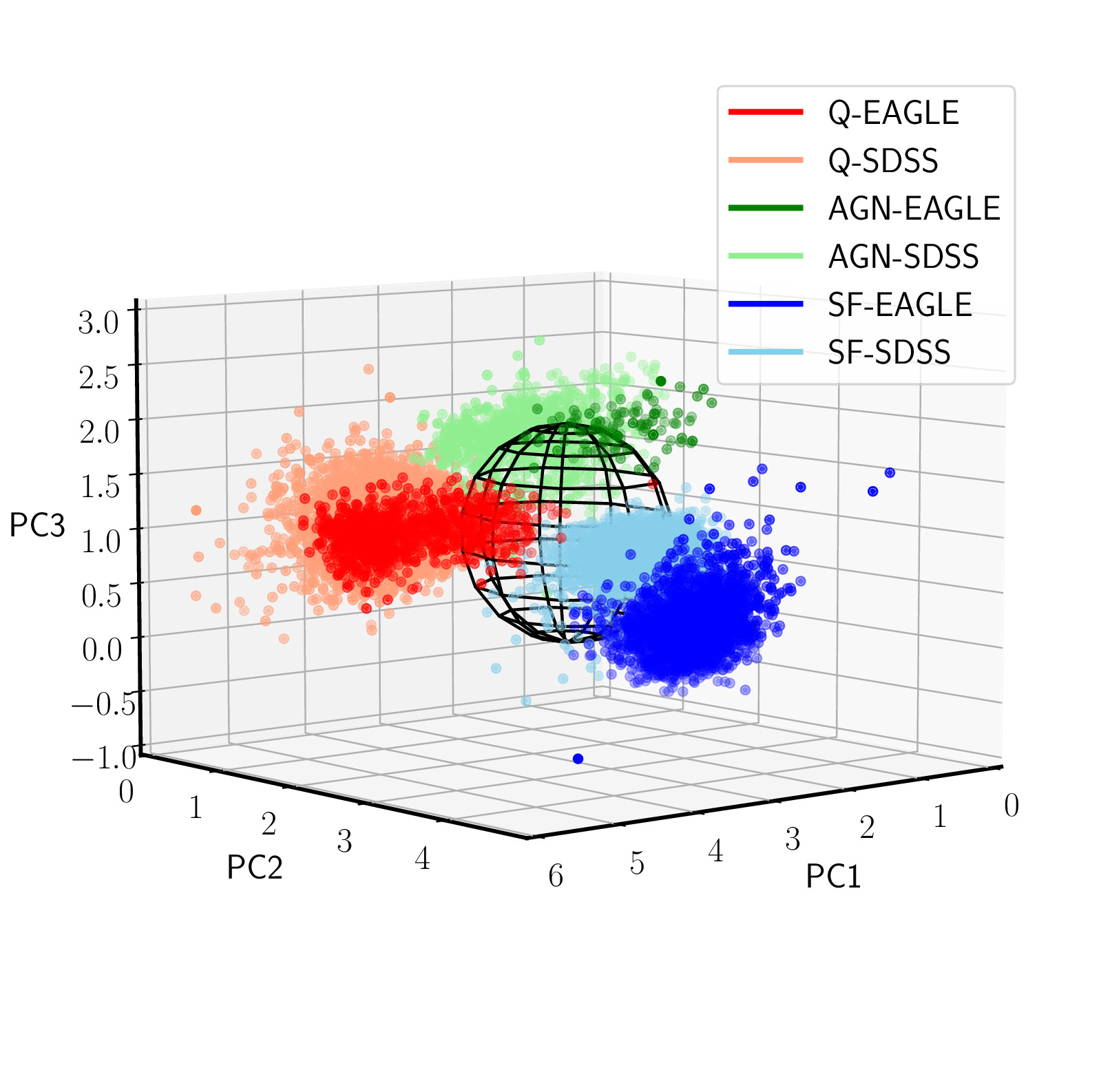}
    \includegraphics[width=60mm]{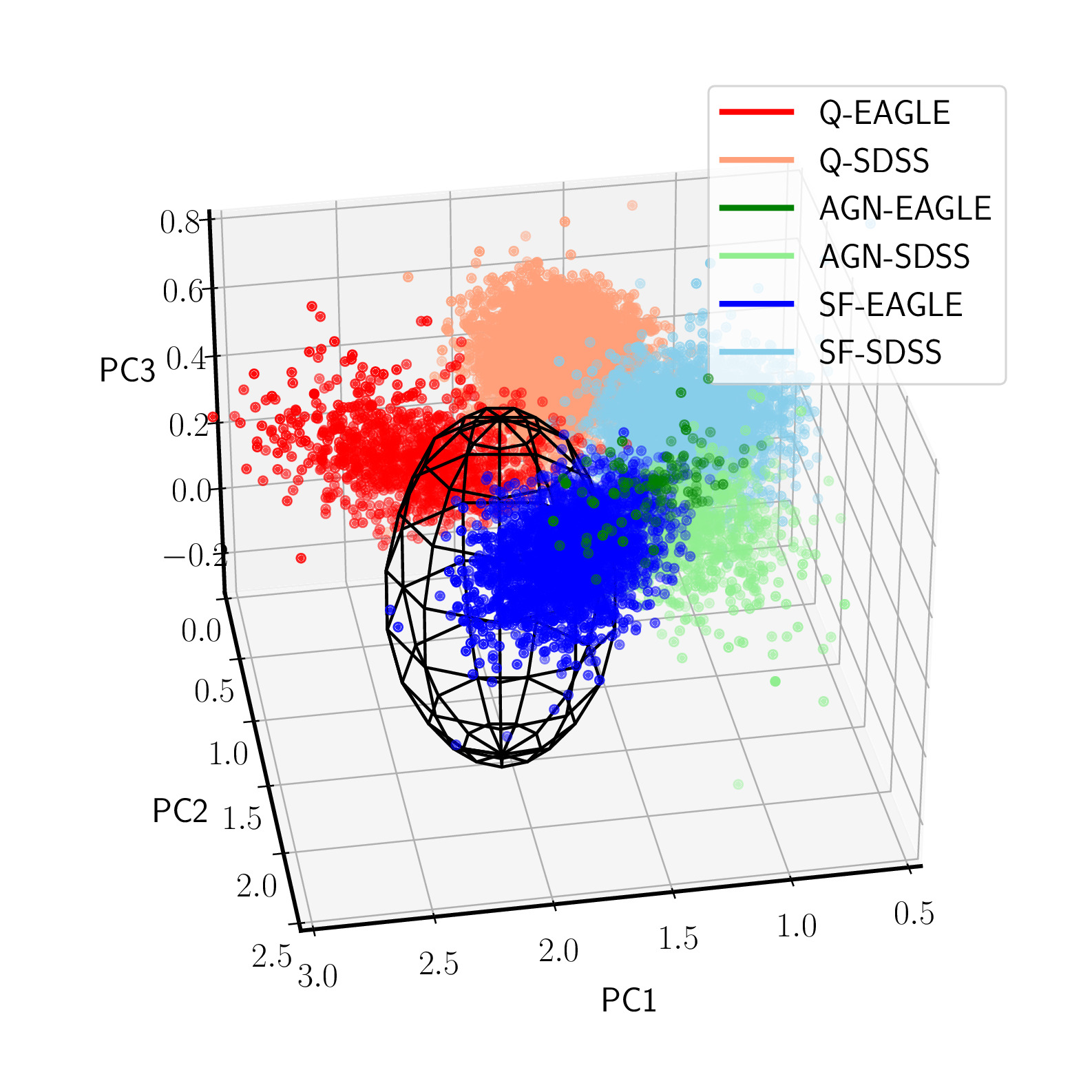}
    \includegraphics[width=60mm]{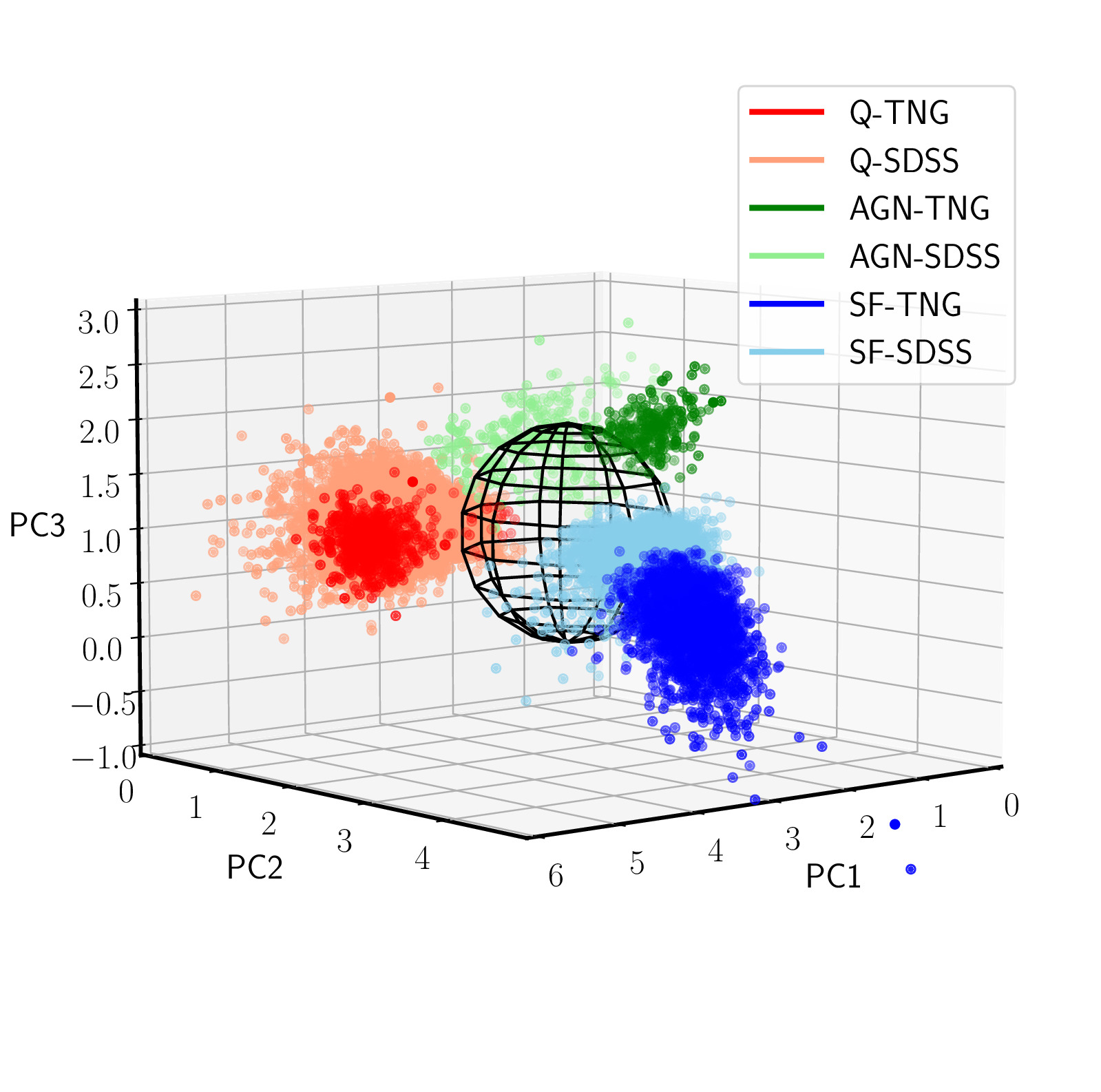}
    \includegraphics[width=60mm]{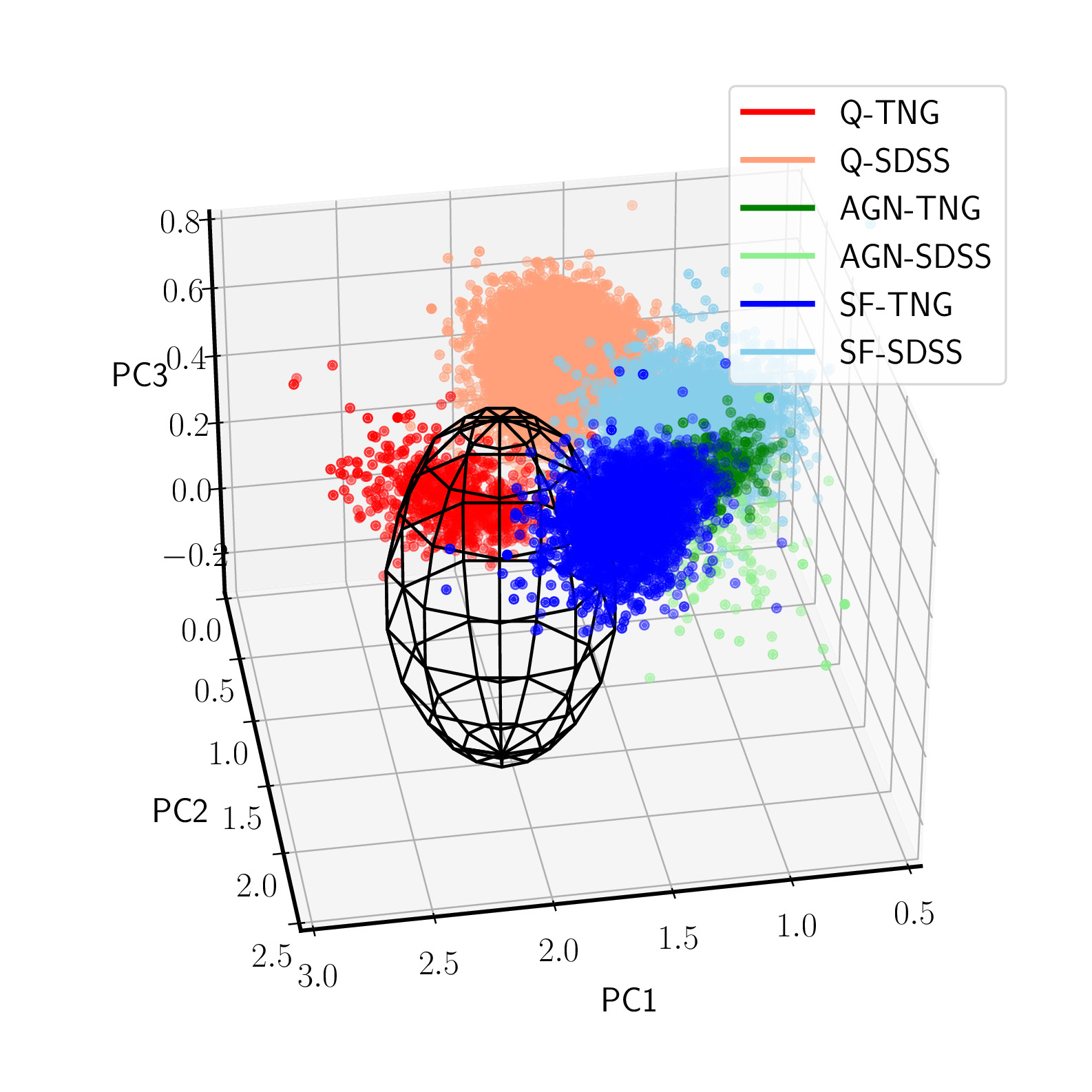}
    \caption{Three dimensional equivalent of Fig.~\ref{fig:corner_sims}, showing the distribution of the projections of a representative sample of EAGLE, TNG100 and SDSS galaxy spectra onto the first three principal components of PCA derived by the SDSS spectra. In each case the pair EAGLE-SDSS and TNG100-SDSS are homogenised, as desccribed in \S\S~\ref{Homog}. The data are colour coded as star-forming (blue), AGN (green), and quiescent (red). The left (right) panels correspond to the blue (red) spectral interval. The framed structure shows, for reference, a sphere with a radius of 1 for the blue interval and a radius of 0.5 for the red interval.}
    \label{fig:3D_sims}
\end{figure*}

This appendix presents a three dimensional rendering of the latent space with the distribution
of the three types of galaxy spectra projected on to the first three principal components, comparing thhe SDSS, EAGLE and TNG100 samples (Fig.~\ref{fig:3D_sims}). An animated version of a similar plot can be found in \citet{variance}.

\section{Latent Space comparision}\label{C}

\begin{figure}
    \centering
    \includegraphics[width=0.47\textwidth]{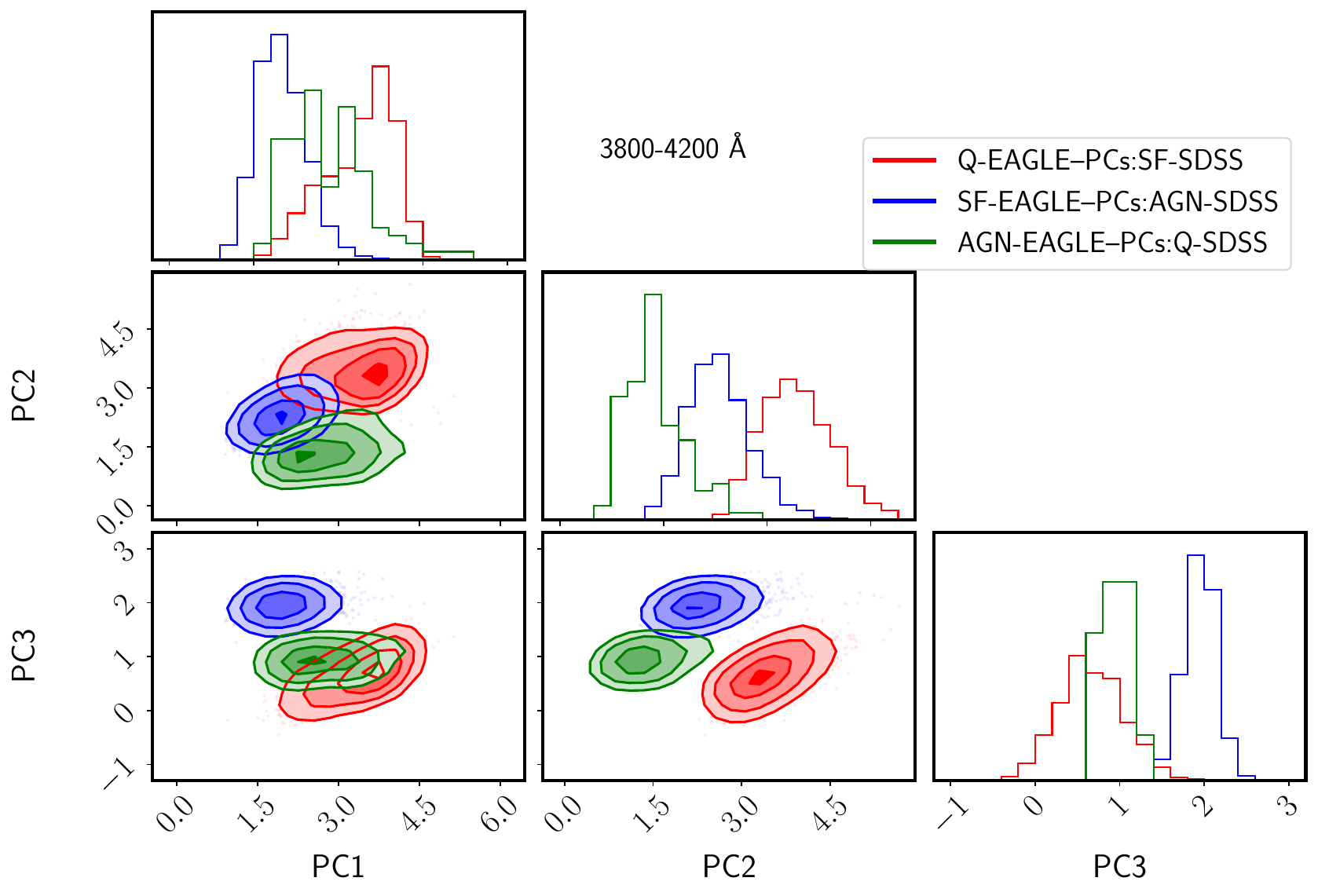}
    \caption{Projection of EAGLE galaxy spectra to non-equivalent galaxy group principal components from the observed SDSS spectra (blue interval).}
    \label{fig_latent_space}
\end{figure}

As stated in the main body of the paper, synthetic galaxy spectra from the EAGLE and TNG100 simulations were projected onto the first three principal components derived from SDSS spectra. PCA is applied separately in each subgroup (quiescent, star-forming and AGN). It can be argued that this is a disjoint comparison of three different input samples or covariance matrices. We emphasize that although the projections are onto different eigenvectors, depending on the galaxy group, the underlying data concern the same system, namely the stellar populations of the galaxies -- after careful removal of features affected by dust or ionized gas. So this is not a disjoint comparison, and the eigenvectors of the three groups are not completely independent. The three sets of eigenvectors reflect, instead, the typical stellar populations found in each group. To prove this, in Fig.~\ref{fig_latent_space}, we show the projection of EAGLE galaxies from a subgroup onto another group. Despite the different absolute values of the projected components, the separation between different groups of galaxies remains. This means that each part of the latent space has a piece of information regarding the various underlying stellar populations. The difference between this latent space and the latent spaces of Fig.~\ref{fig:corner_sims}, where we project the spectra onto the PCs of the same group, is caused by the differences in the stellar population content of Q/SF/AGN subsets. The highest variance components (PC1, PC2, PC3) will give different weights to specific parts of the spectra, for instance, Q galaxies are overall rather old, so there might be a more significant contribution due to metallicity or abundance ratios, whereas SF eigenvectors would have more variance regarding differences in the age of the populations.

\section{Testing the role of noise modelling}\label{D}

We give in this appendix additional information about the potential systematic caused by our adoption of noise to produce realistic synthetic spectra. Firstly, in addition to the strict cut in redshift and stellar velocity dispersion, the SDSS sample is chosen, on purpose, with a rather high threshold in the S/N of the data (above 15), to minimise the effect of noise in the covariance of the spectra. Moreover, note that the components from PCA are ranked, and only a few of those with the highest variance are chosen. At high S/N we do not expect a large fraction of the variance to be dominated by noise.

\begin{figure}
    \centering
    \includegraphics[width=0.47\textwidth]{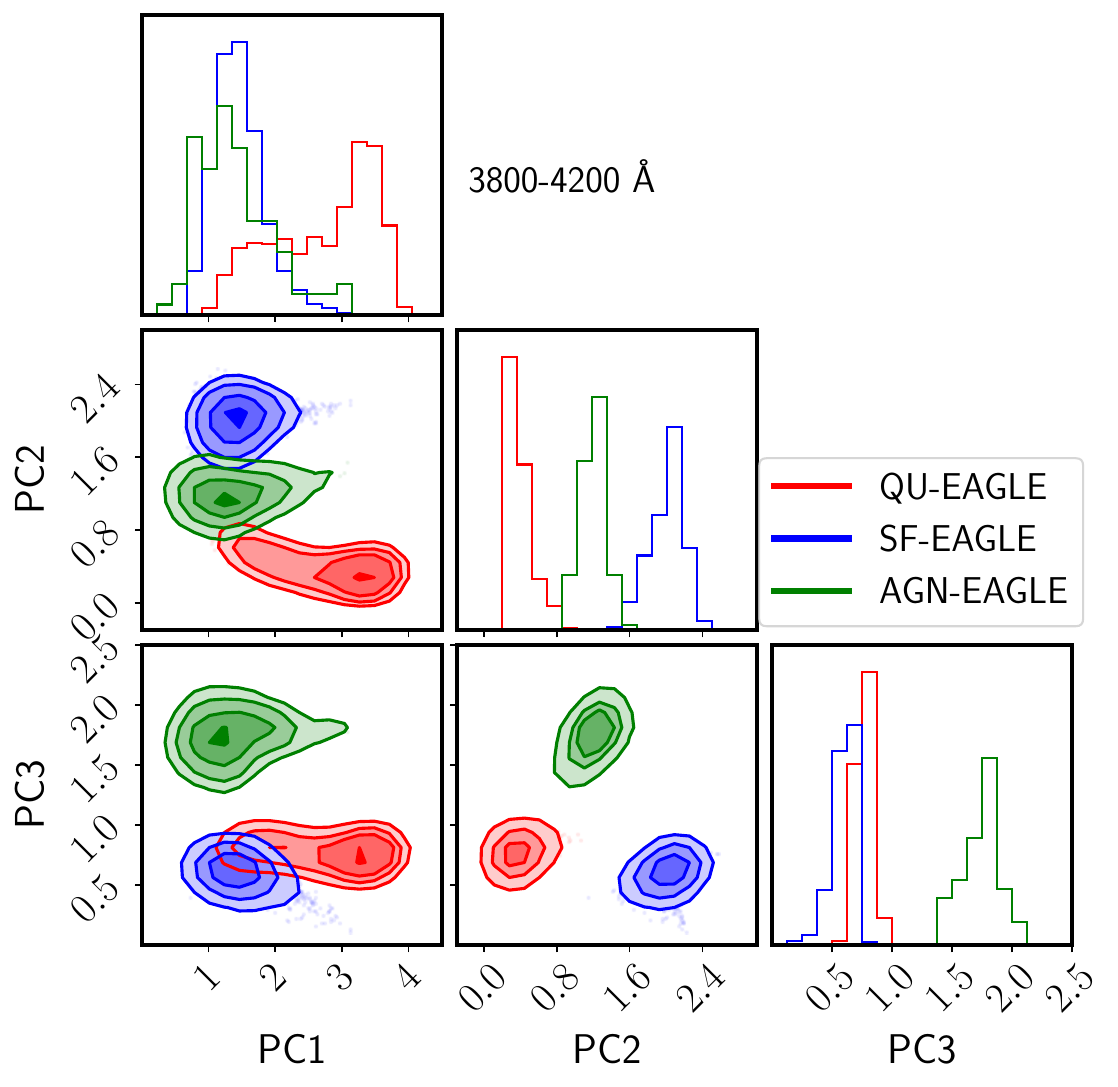}
    \caption{Projections of the Noiseless EAGLE synthetic spectra onto the SDSS eigenvectors (blue interval).}
    \label{fig_noiseless_gal}
\end{figure}

To confirm this, Fig.~\ref{fig_noiseless_gal} shows the projection of noiseless EAGLE synthetic galaxy spectra onto the principal components from the SDSS data, i.e. the identical procedure as in the full study but removing the contribution from noise in the simulation data. The separation between different groups of galaxies in the latent space remains the same with respect to the more realistic, noisy spectra shown in Figure~\ref{fig:corner_sims}. As expected, the absolute values of the projected components have changed as well as the overlap between the different groups of the galaxy. However, we conclude that the actual modelling of the noise is less important with respect to the actual variations found in the spectra. Nevertheless, note that our analysis enforces the same noise characteristics as the observational SDSS spectra.

\clearpage
\onecolumn

\begin{center}
  {\Large Supplementary material for:}

  {\Large \bf Evaluating quenching in cosmological simulations of galaxy formation with spectral covariance in the optical window}

  {\Large Sharbaf, Ferreras, Negri, Angthopo, Dalla Vechia, Lahav \& Somerville, MNRAS, 2024}
\end{center}

\section*{SPECTRAL FITTING FOR PCA and SFH IN THE RED INTERVAL} \label{D}
Concerning the result of the spectral fitting of stacked data according to extreme values of the principal component projections, we show in Figs.~1, 2, 3 of this Supplementary Material the equivalent of Figs.~7 ,8 and 9 for the analysis performed in the red spectral interval ($\lambda\in$[5000,5400]\AA), using the same fitting models and parameter ranges as described in the text. Figs.~4, 5, 6 correspond to the equivalent of Figs.~10, 11, 12 of the paper for the red spectral interval.

\begin{figure}
\includegraphics[width=.05\linewidth]{Figs/SSPFits/Row0_PC1.pdf}
\includegraphics[width=.3\linewidth]{Figs/SSPFits/Col1.pdf}
\includegraphics[width=.3\linewidth]{Figs/SSPFits/Col2.pdf}
\includegraphics[width=.3\linewidth]{Figs/SSPFits/Col3.pdf}
\includegraphics[width=.05\linewidth]{Figs/SSPFits/Row1.pdf}
\includegraphics[width=.3\linewidth]{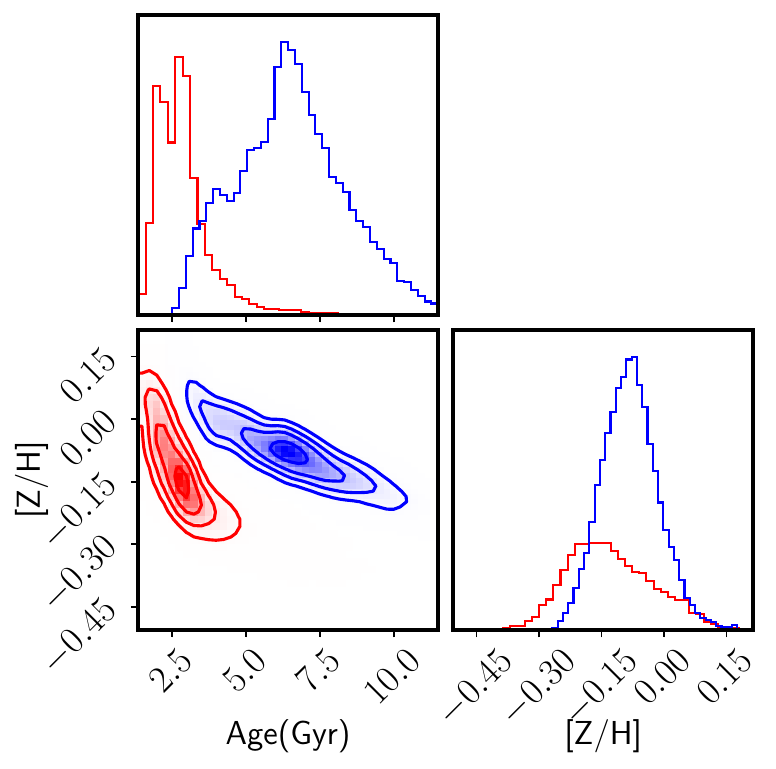}
\includegraphics[width=.3\linewidth]{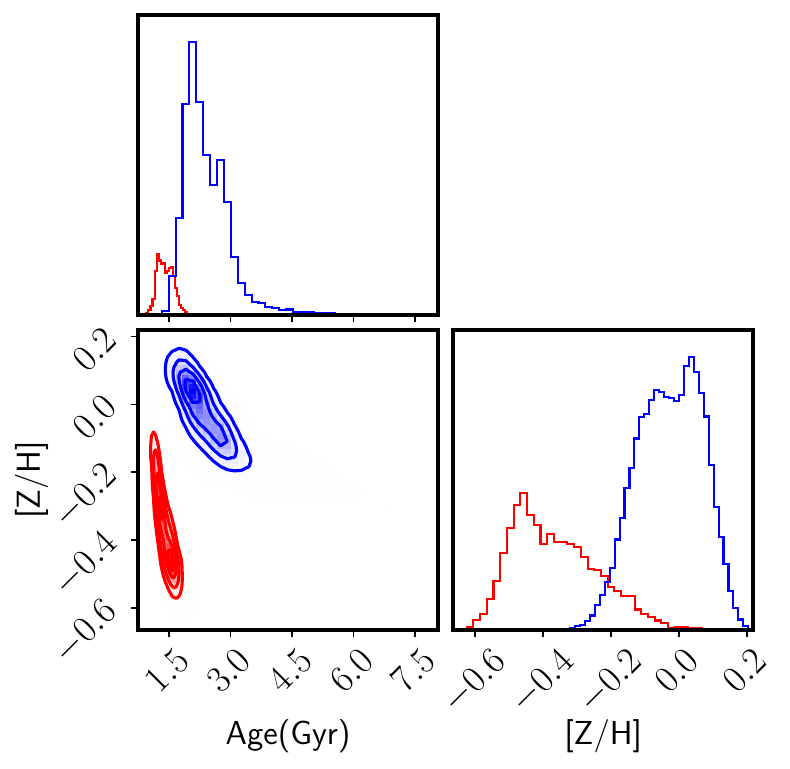}
\includegraphics[width=.3\linewidth]{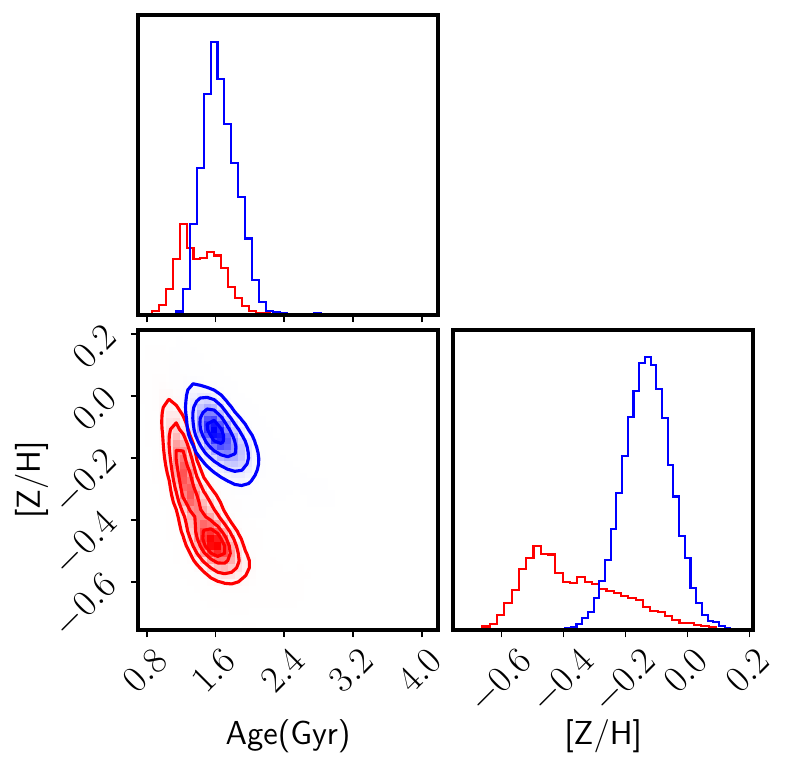}
\includegraphics[width=.05\linewidth]{Figs/SSPFits/Row2.pdf}
\includegraphics[width=.3\linewidth]{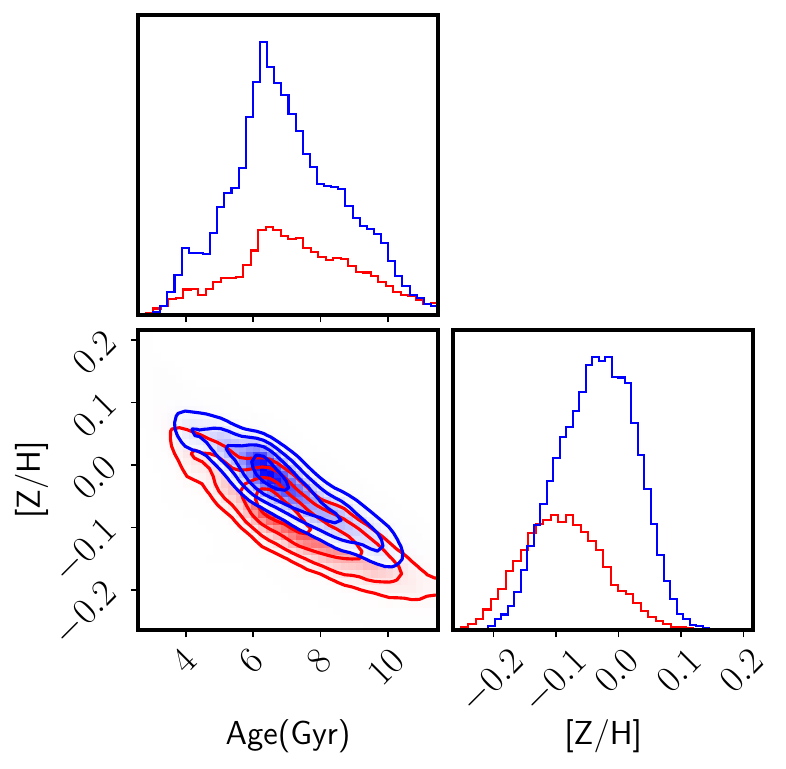}
\includegraphics[width=.3\linewidth]{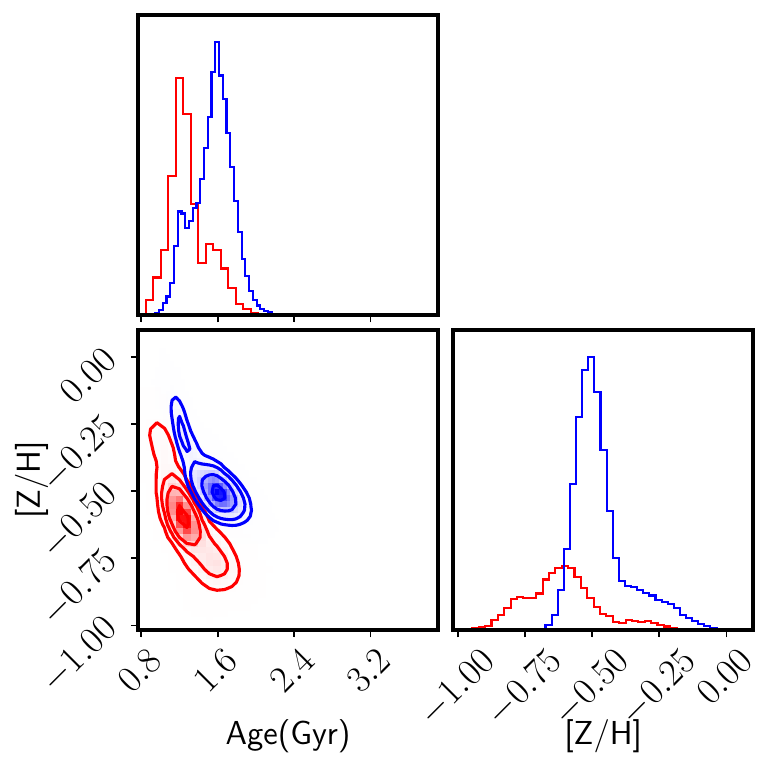}
\includegraphics[width=.3\linewidth]{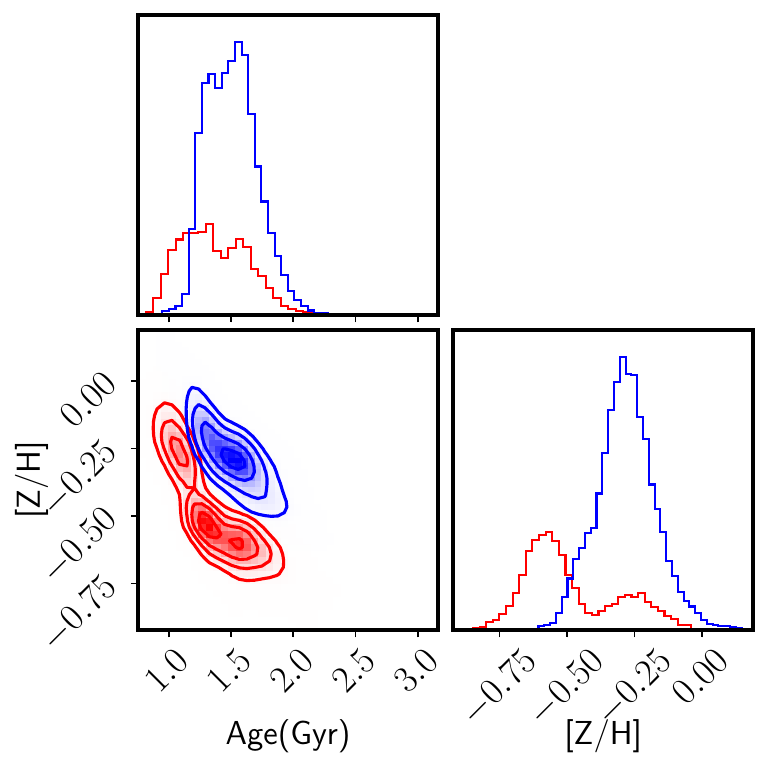}
\includegraphics[width=.05\linewidth]{Figs/SSPFits/Row3.pdf}
\includegraphics[width=.3\linewidth]{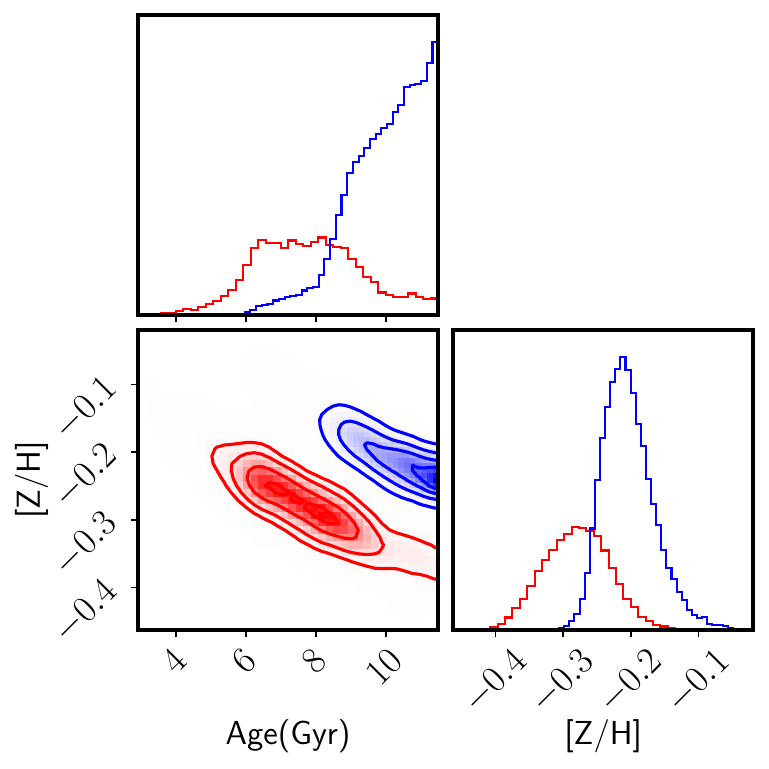}
\includegraphics[width=.3\linewidth]{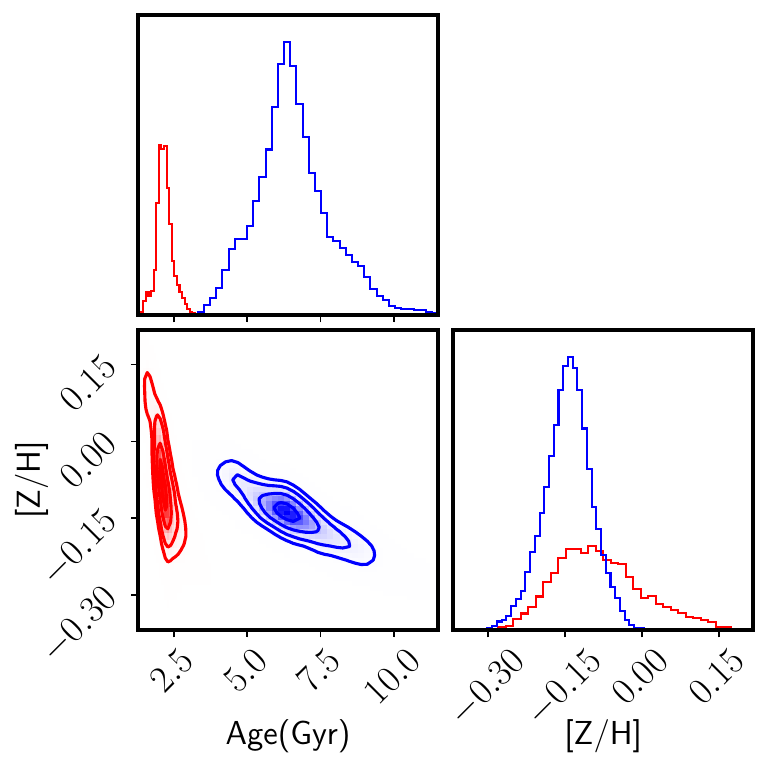}
\includegraphics[width=.3\linewidth]{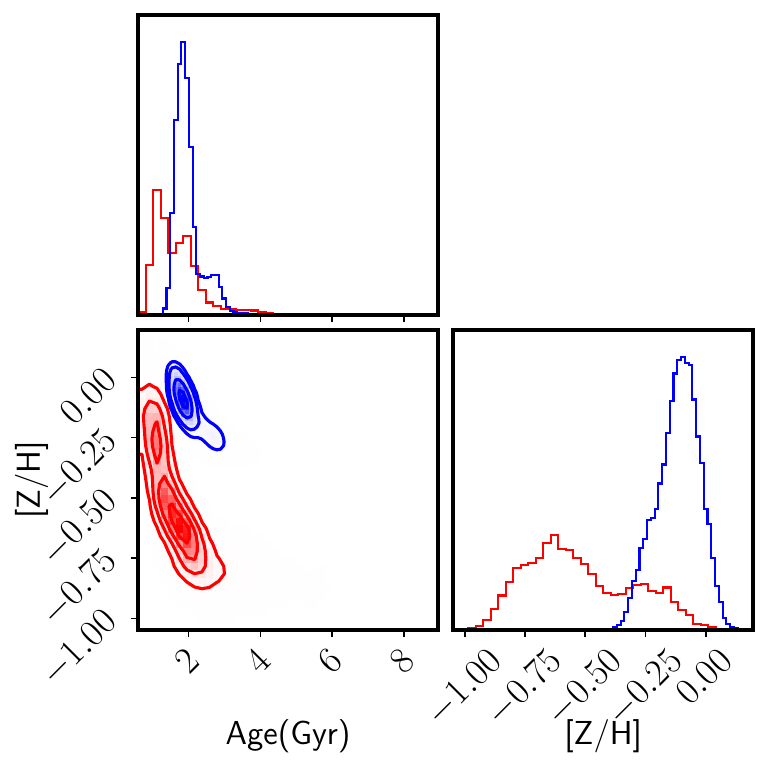}
\caption{Confidence levels of the SSP-equivalent age (in Gyr) and metallicity ([Z/H]) obtained from fitting the spectra (continuum subtracted) with the E-MILES population models \citep{EMILES}. We fit the stacked spectra with the lowest (33 percentile, red) and highest (67 percentile, blue) value of PC1, for EAGLE, TNG100, and SDSS samples  (from top to bottom), for Q (left column), AGN (middle column), and SF galaxies (right column). The principal components obtained correspond to the SDSS covariance defined in the red interval.}
\label{fig:MCMC-pc1-red}
\end{figure}

\begin{figure}
\includegraphics[width=.05\linewidth]{Figs/SSPFits/Row0_PC2.pdf}
\includegraphics[width=.3\linewidth]{Figs/SSPFits/Col1.pdf}
\includegraphics[width=.3\linewidth]{Figs/SSPFits/Col2.pdf}
\includegraphics[width=.3\linewidth]{Figs/SSPFits/Col3.pdf}
\includegraphics[width=.05\linewidth]{Figs/SSPFits/Row1.pdf}
\includegraphics[width=.3\linewidth]{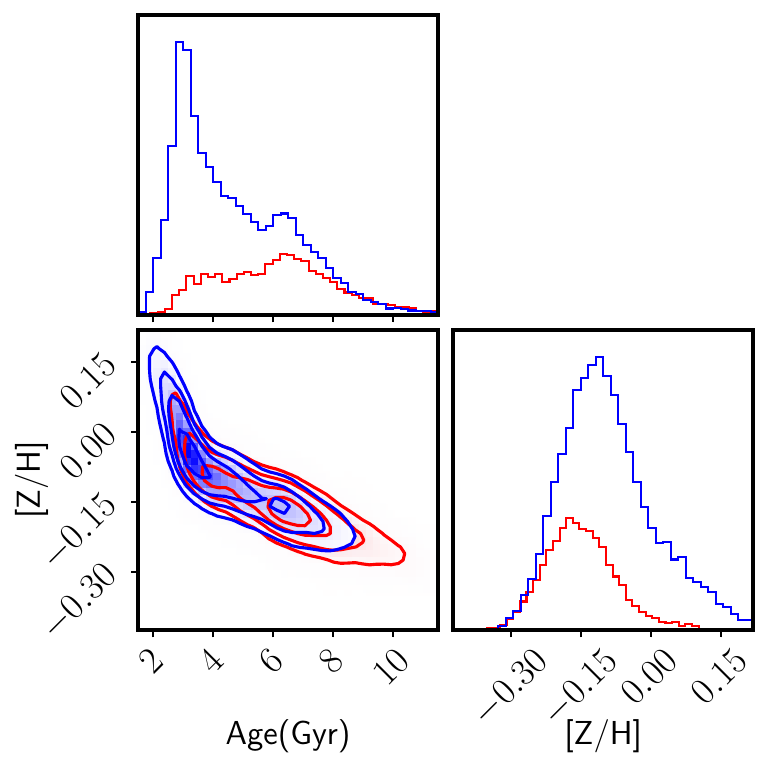}
\includegraphics[width=.3\linewidth]{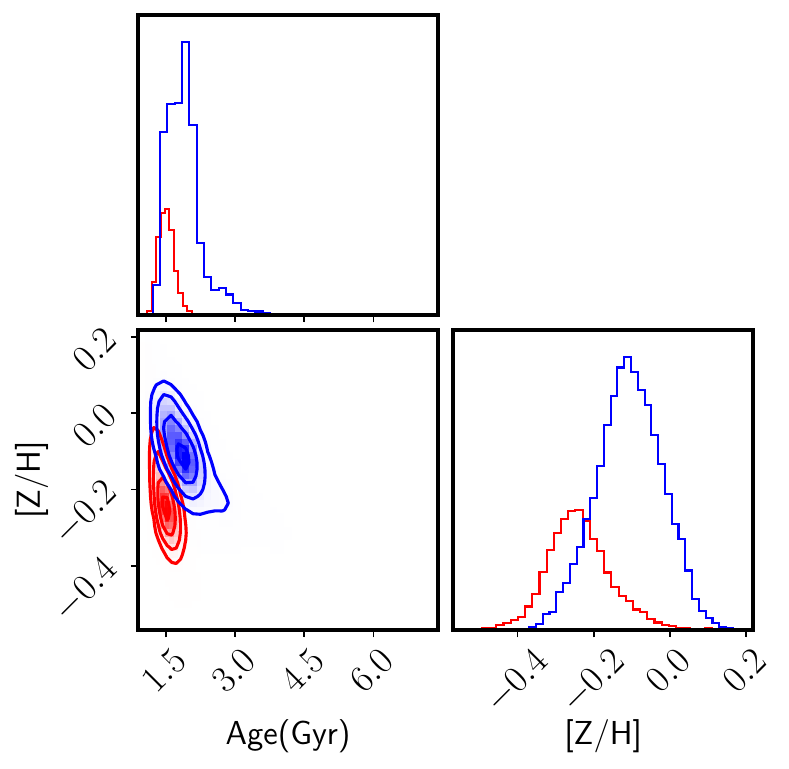}
\includegraphics[width=.3\linewidth]{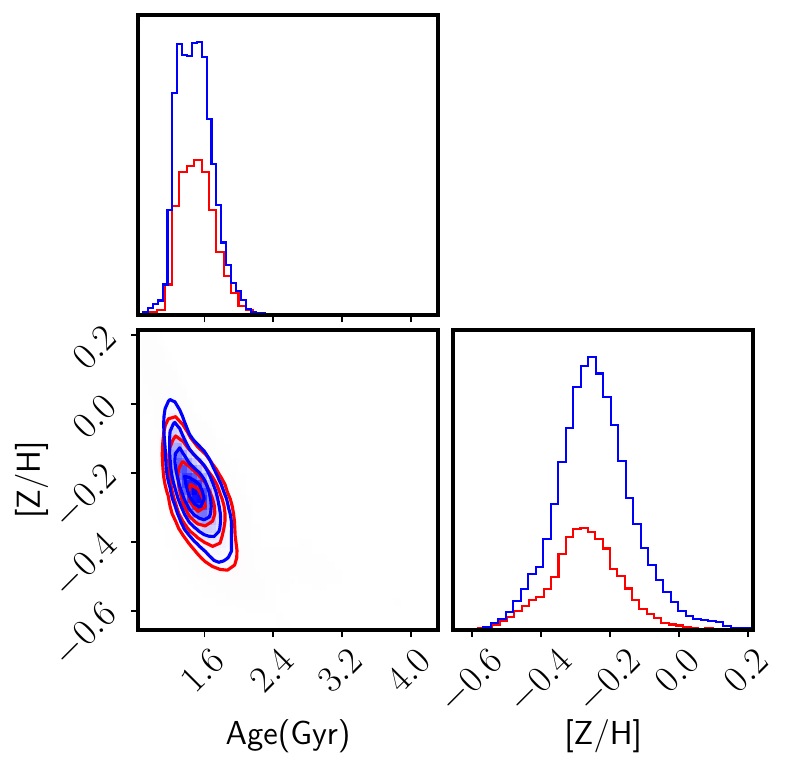}
\includegraphics[width=.05\linewidth]{Figs/SSPFits/Row2.pdf}
\includegraphics[width=.3\linewidth]{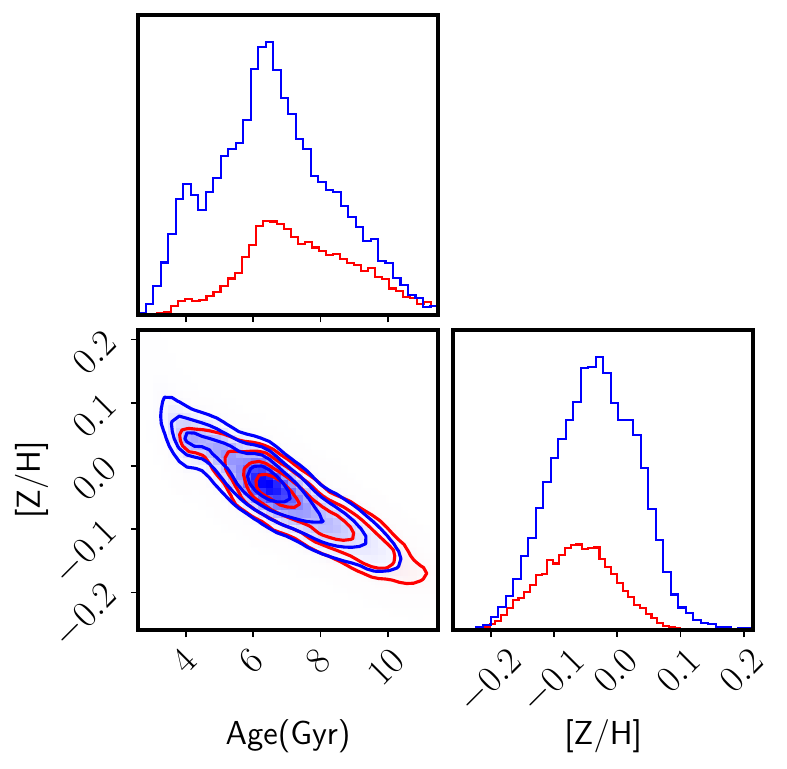}
\includegraphics[width=.3\linewidth]{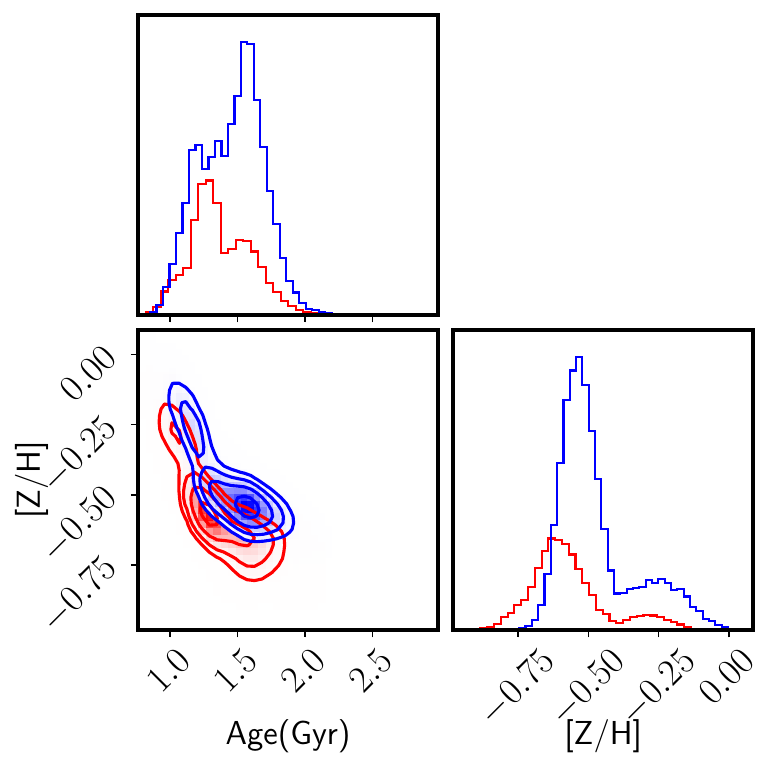}
\includegraphics[width=.3\linewidth]{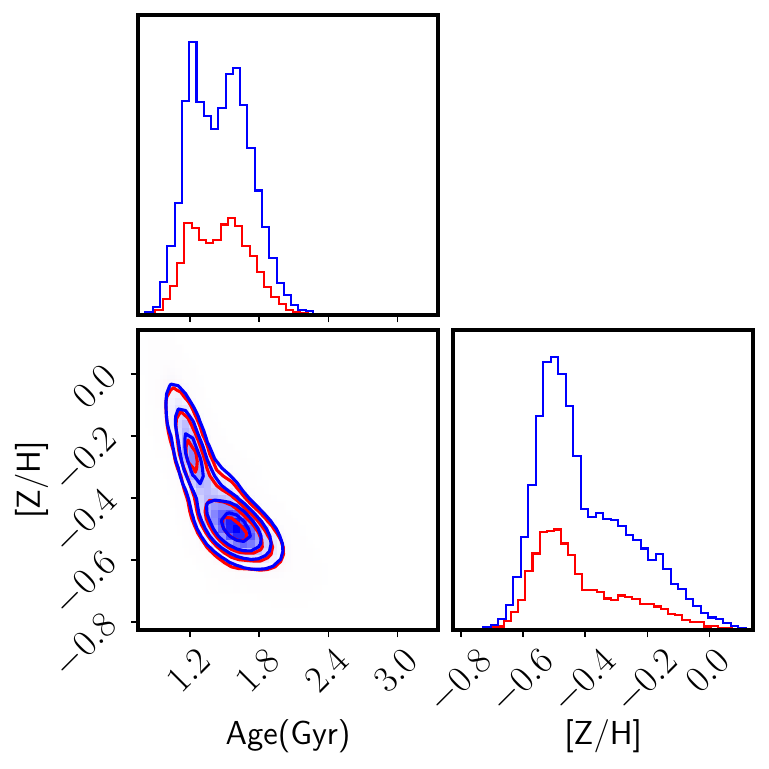}
\includegraphics[width=.05\linewidth]{Figs/SSPFits/Row3.pdf}
\includegraphics[width=.3\linewidth]{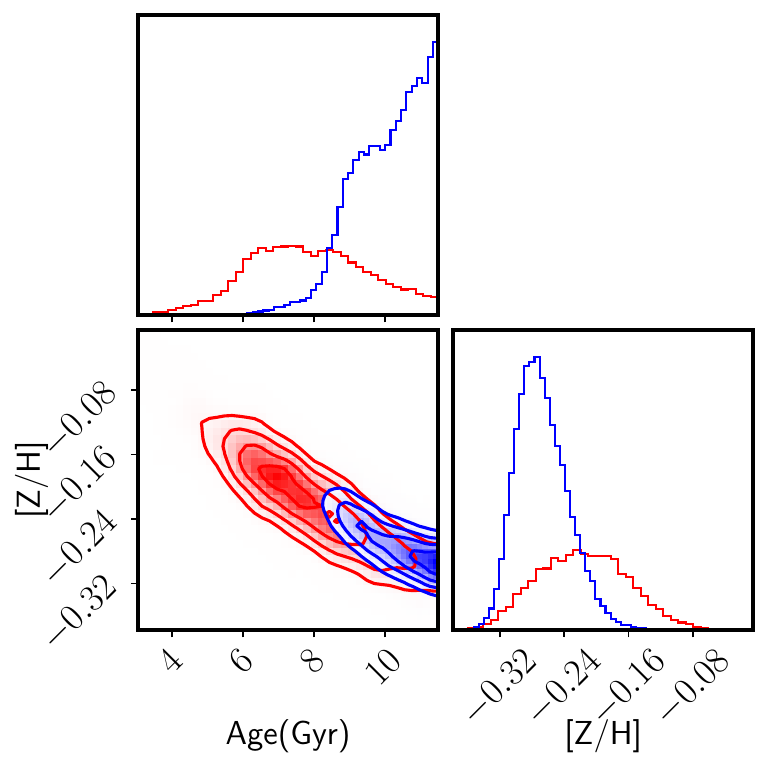}
\includegraphics[width=.3\linewidth]{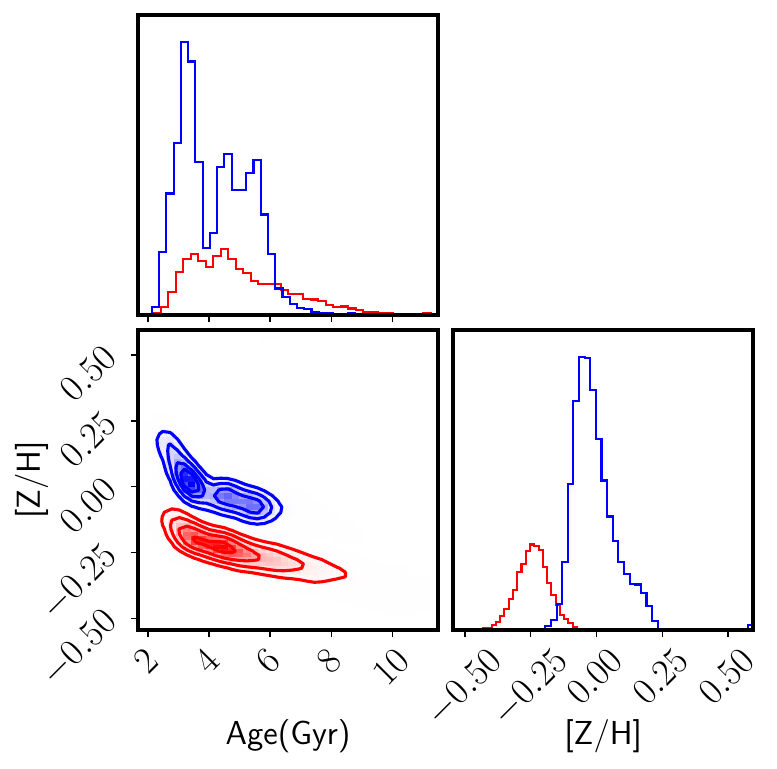}
\includegraphics[width=.3\linewidth]{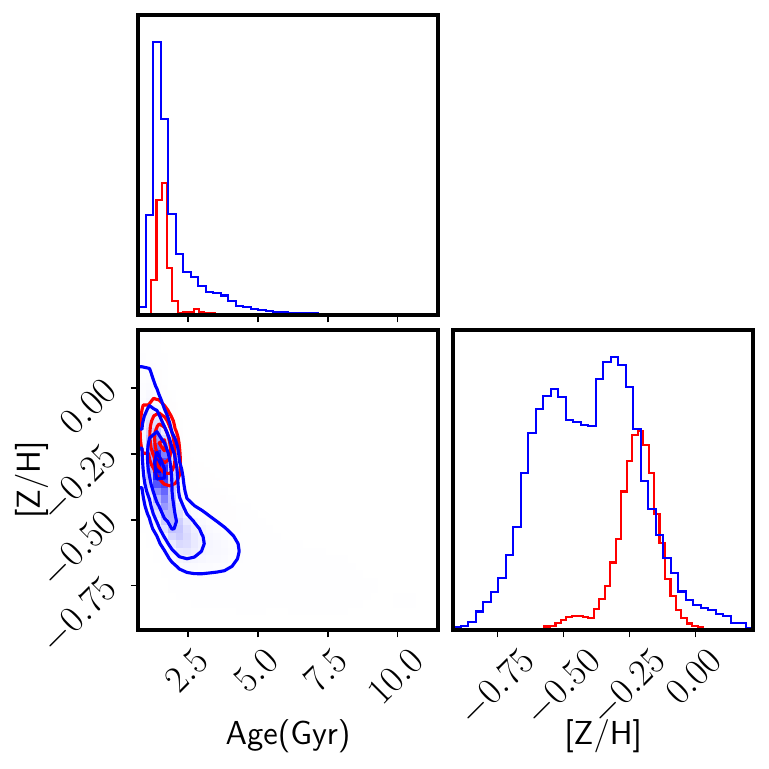}
\caption{Confidence levels of the SSP-equivalent age (in Gyr) and metallicity ([Z/H]) obtained from fitting the spectra (continuum subtracted) with the E-MILES population models \citep{EMILES}. We fit the stacked spectra with the lowest (33 percentile, red) and highest (67 percentile, blue) value of PC2, for EAGLE, TNG100, and SDSS samples  (from top to bottom), for Q (left column), AGN (middle column), and SF galaxies (right column). The principal components obtained correspond to the SDSS covariance defined in the red interval.}
\label{fig:MCMC-pc2-red}
\end{figure}

\begin{figure}
\includegraphics[width=.05\linewidth]{Figs/SSPFits/Row0_PC3.pdf}
\includegraphics[width=.3\linewidth]{Figs/SSPFits/Col1.pdf}
\includegraphics[width=.3\linewidth]{Figs/SSPFits/Col2.pdf}
\includegraphics[width=.3\linewidth]{Figs/SSPFits/Col3.pdf}
\includegraphics[width=.05\linewidth]{Figs/SSPFits/Row1.pdf}
\includegraphics[width=.3\linewidth]{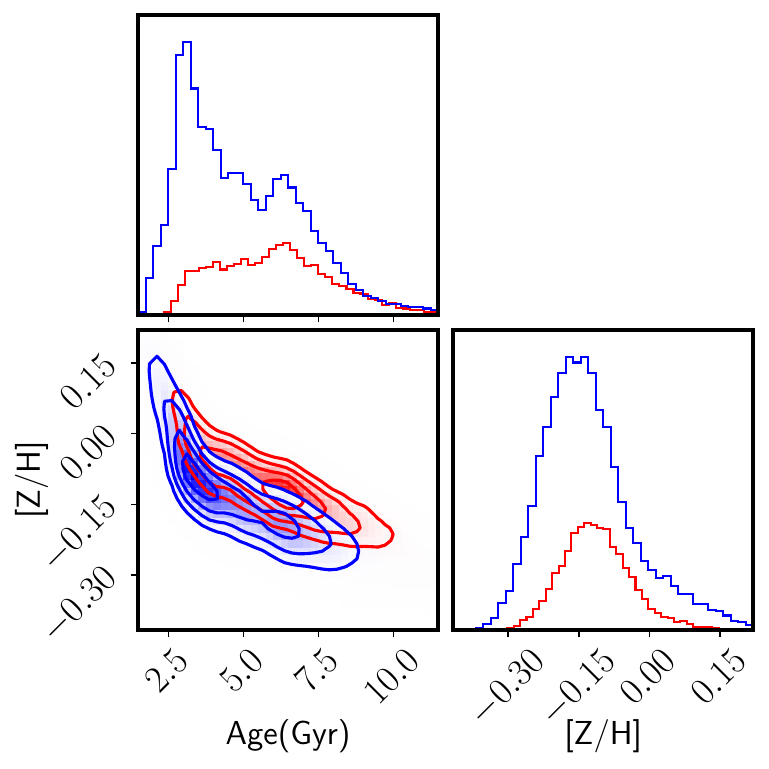}
\includegraphics[width=.3\linewidth]{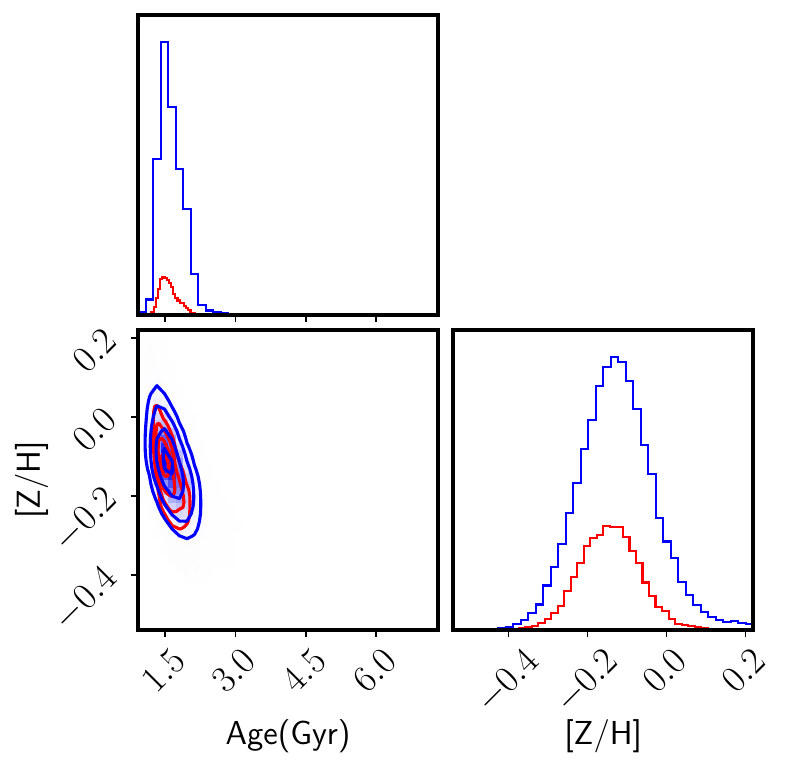}
\includegraphics[width=.3\linewidth]{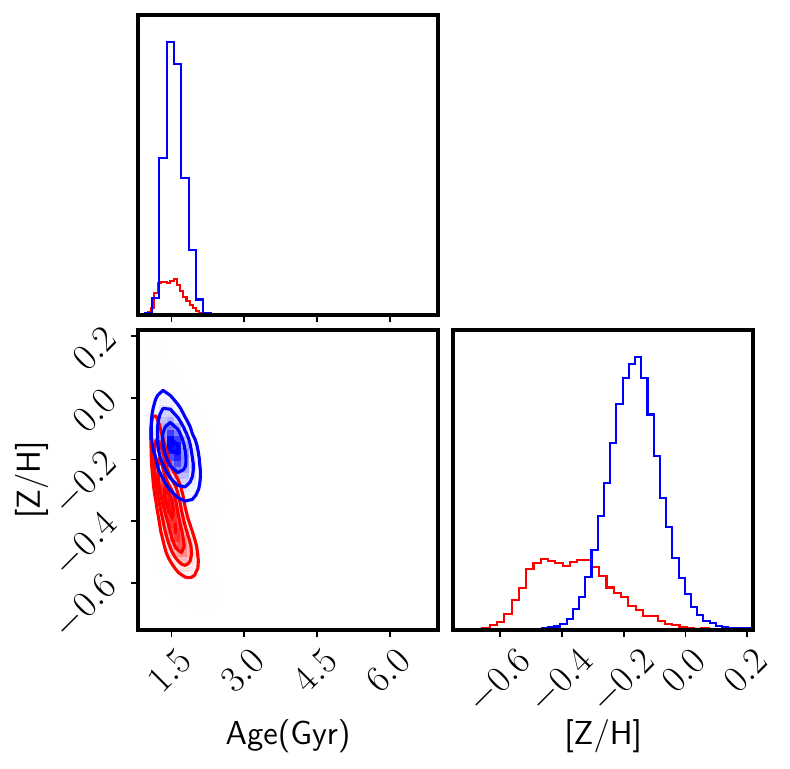}
\includegraphics[width=.05\linewidth]{Figs/SSPFits/Row2.pdf}
\includegraphics[width=.3\linewidth]{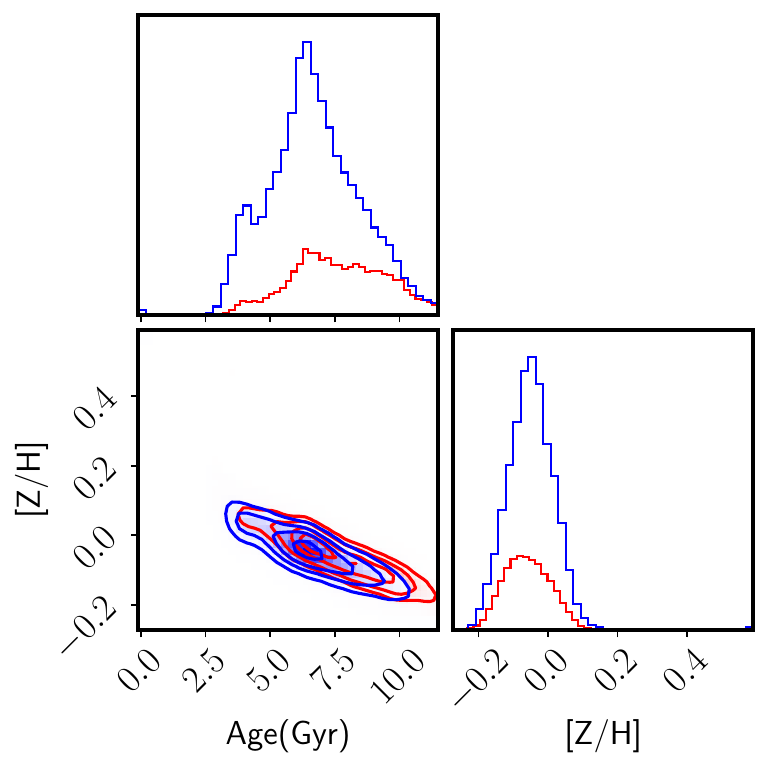}
\includegraphics[width=.3\linewidth]{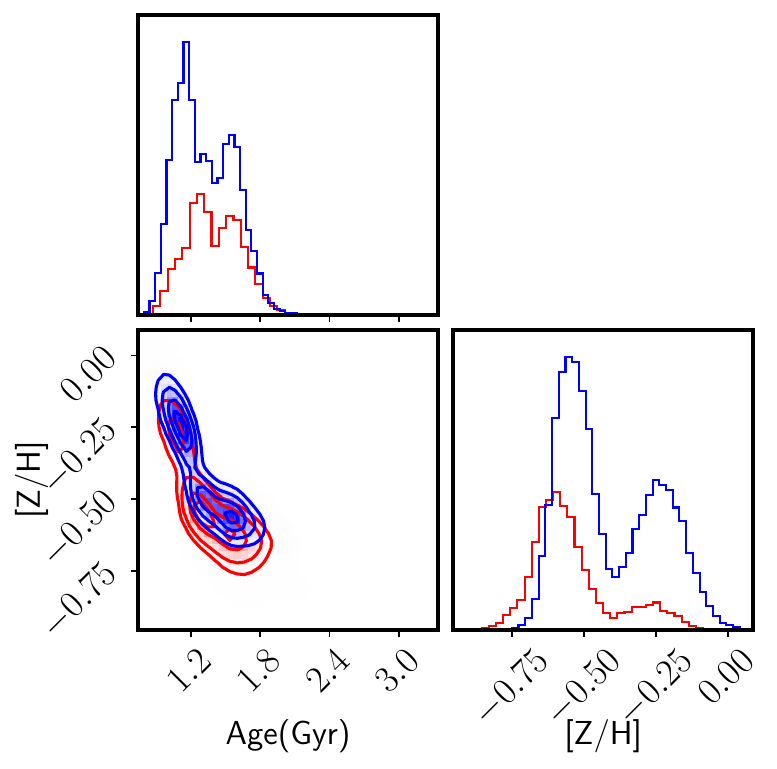}
\includegraphics[width=.3\linewidth]{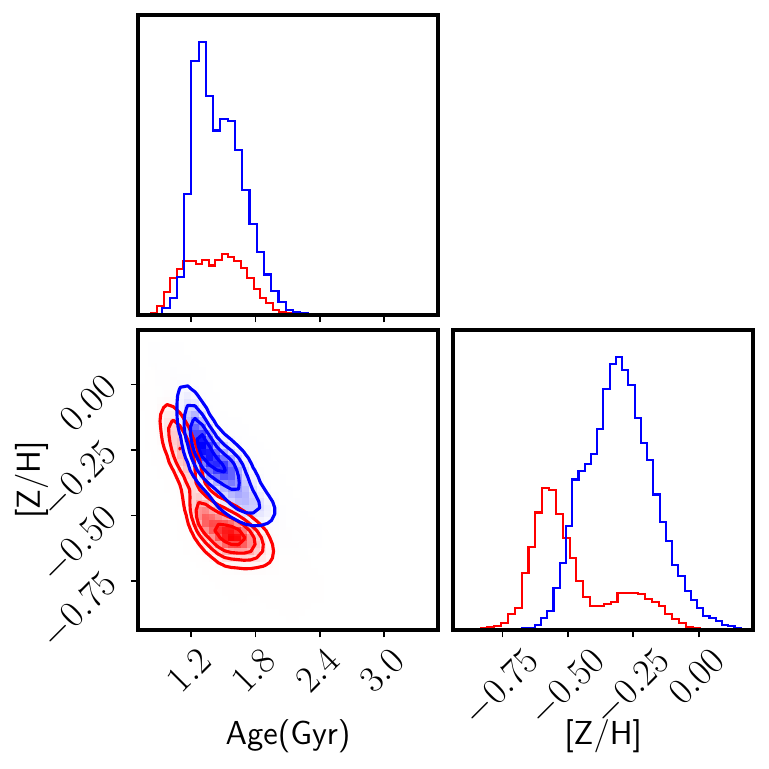}
\includegraphics[width=.05\linewidth]{Figs/SSPFits/Row3.pdf}
\includegraphics[width=.3\linewidth]{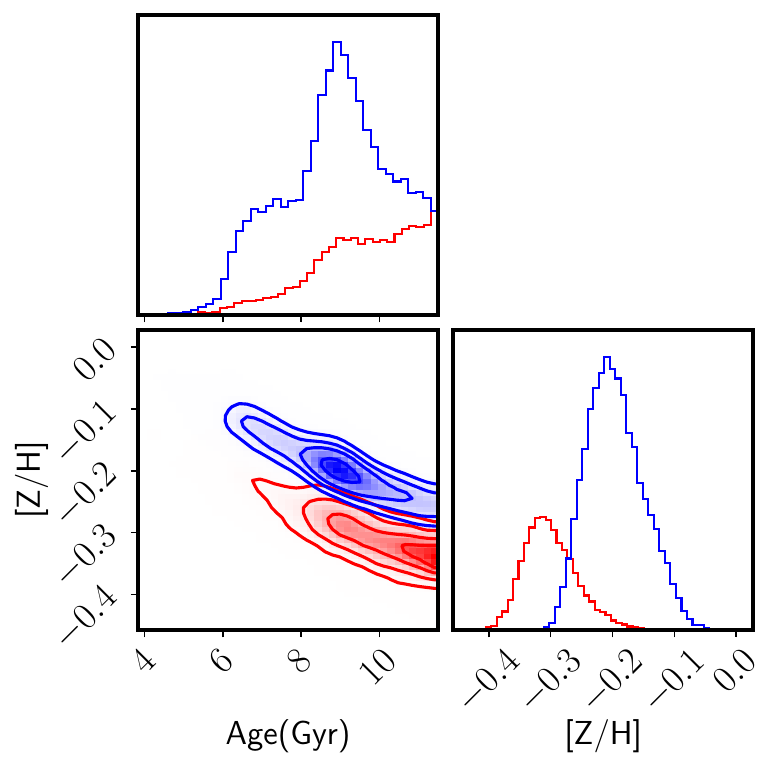}
\includegraphics[width=.3\linewidth]{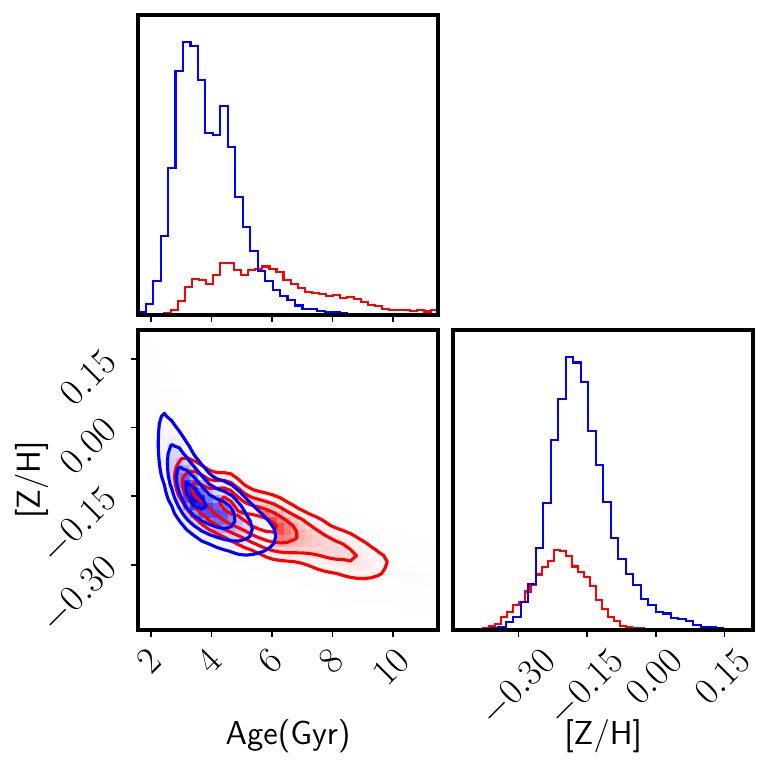}
\includegraphics[width=.3\linewidth]{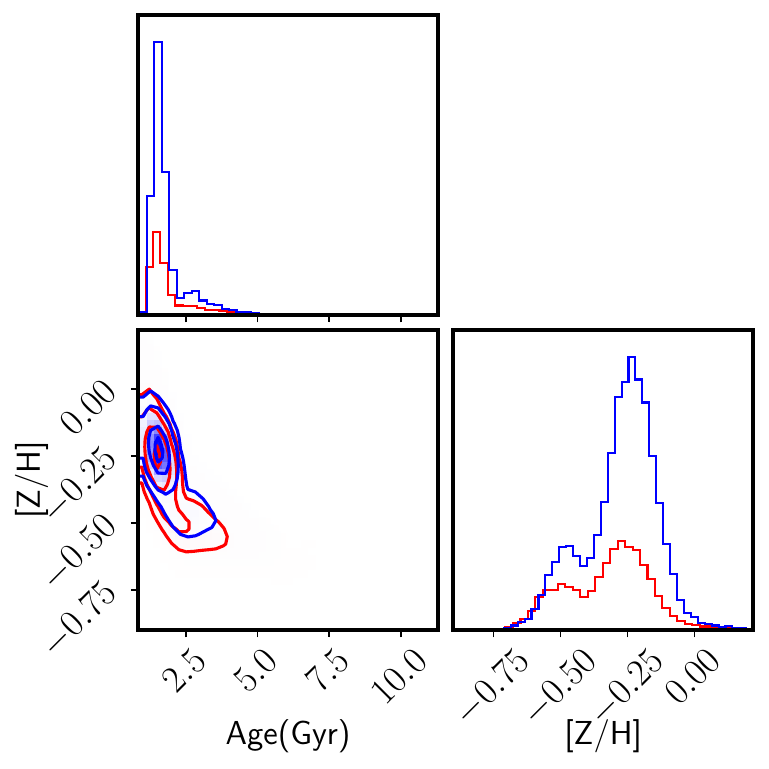}
\caption{Confidence levels of the SSP-equivalent age (in Gyr) and metallicity ([Z/H]) obtained from fitting the spectra (continuum subtracted) with the E-MILES population models \citep{EMILES}. We fit the stacked spectra with the lowest (33 percentile, red) and highest (67 percentile, blue) value of PC3, for EAGLE, TNG100, and SDSS samples  (from top to bottom), for Q (left column), AGN (middle column), and SF galaxies (right column). The principal components obtained correspond to the SDSS covariance defined in the red interval.}
\label{fig:MCMC-pc3-red}
\end{figure}


\begin{figure}
    \centering
    \includegraphics[width=0.3\textwidth]{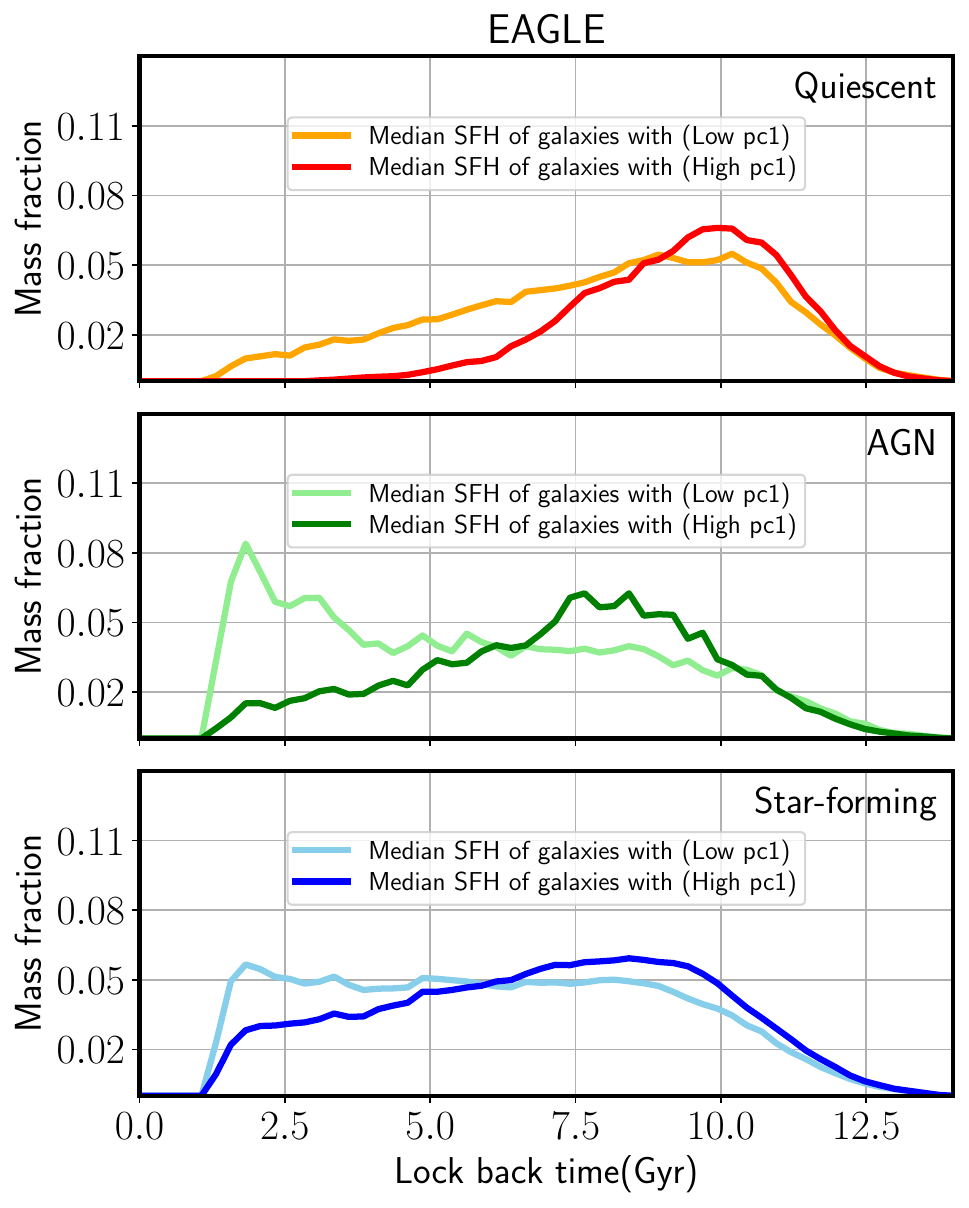}
    \includegraphics[width=0.3\textwidth]{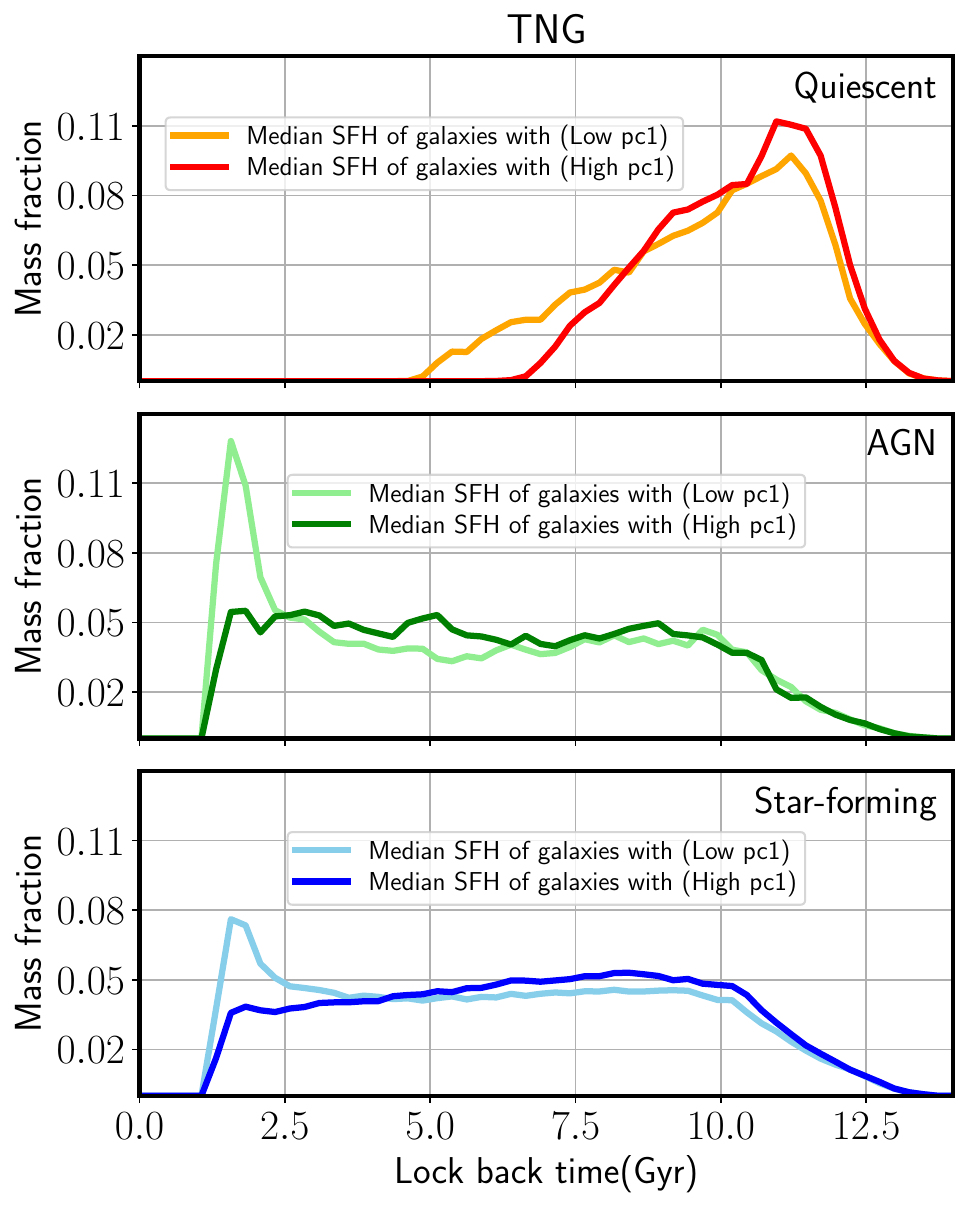}
    \caption{Median star formation history of the galaxies with the lowest (33 percentile) and highest (67 percentile) value of PC1 in the red interval, for Q, AGN, and SF sub-classes (from top to bottom), and the EAGLE sample in the left column and the TNG100 sample in the right column.} 
    \label{figsfh-eag-pc1-red}
\end{figure}

\begin{figure}
    \centering
    \includegraphics[width=0.3\textwidth]{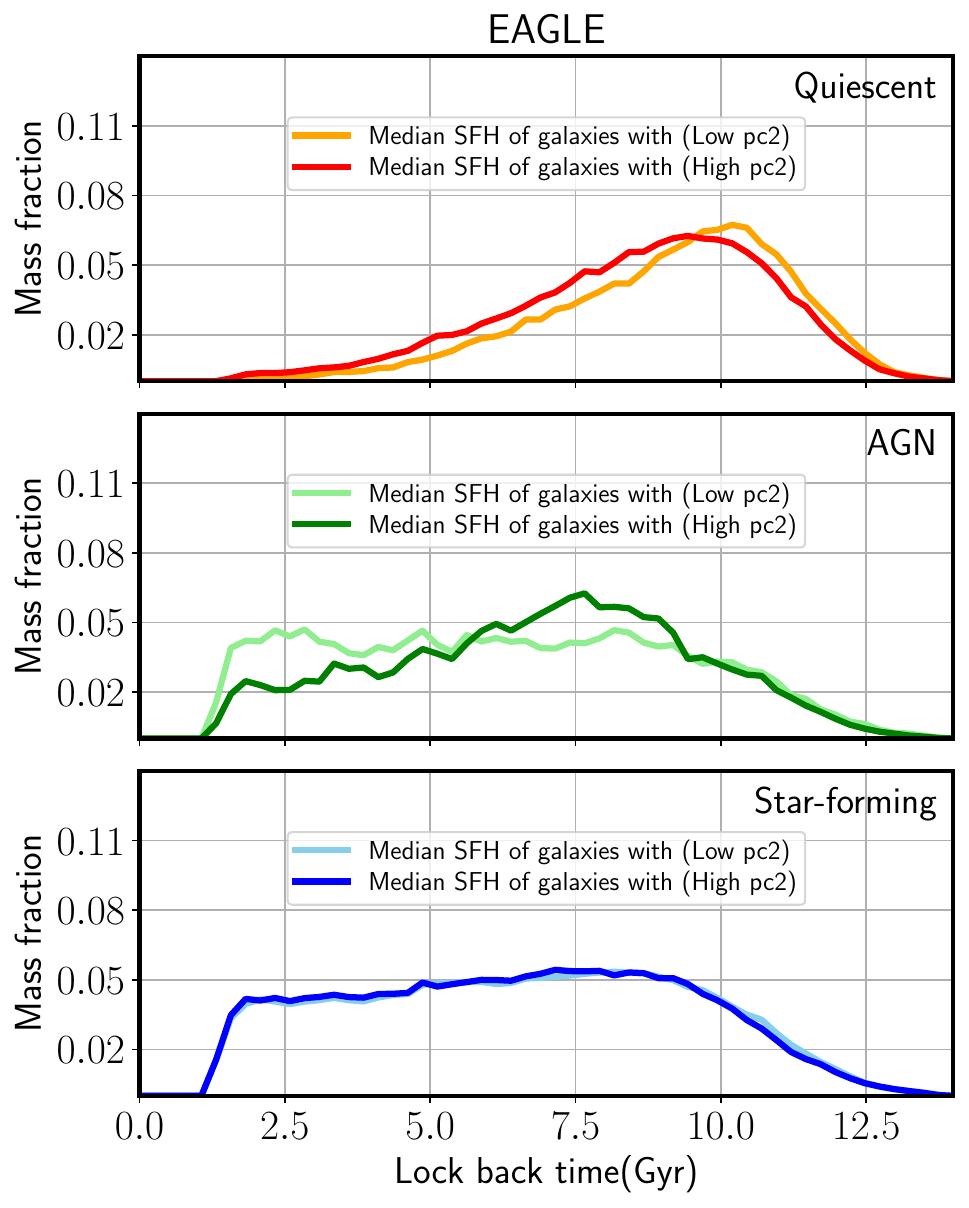}
    \includegraphics[width=0.3\textwidth]{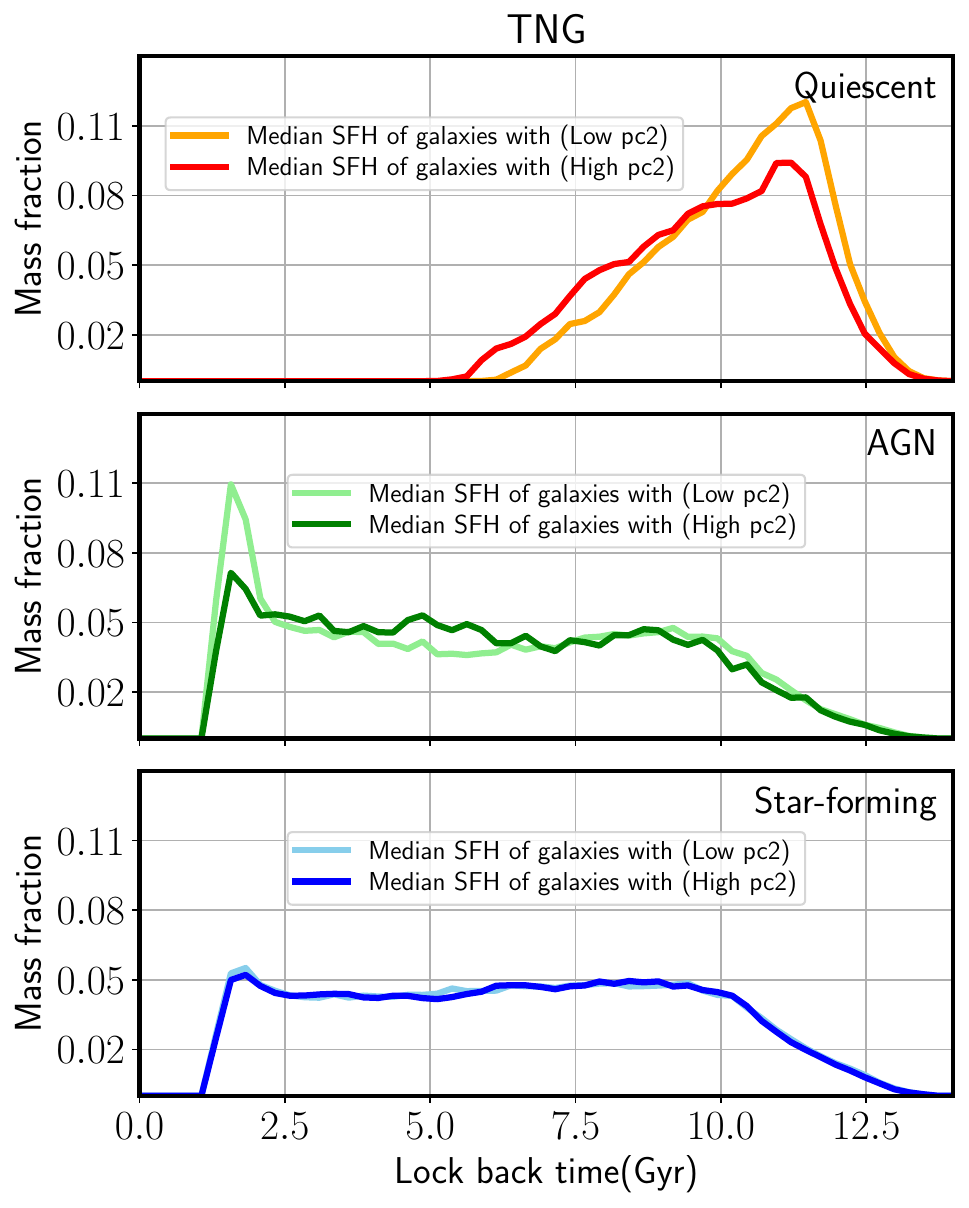}
    \caption{Median star formation history of the galaxies with the lowest (33 percentile) and highest (67 percentile) value of PC2 in the red interval, for Q, AGN, and SF sub-classes (from top to bottom), and the EAGLE sample in the left column and the TNG100 sample in the right column.} 
    \label{figsfh-eag-pc2-red}
\end{figure}

\begin{figure}
    \centering
    \includegraphics[width=0.3\textwidth]{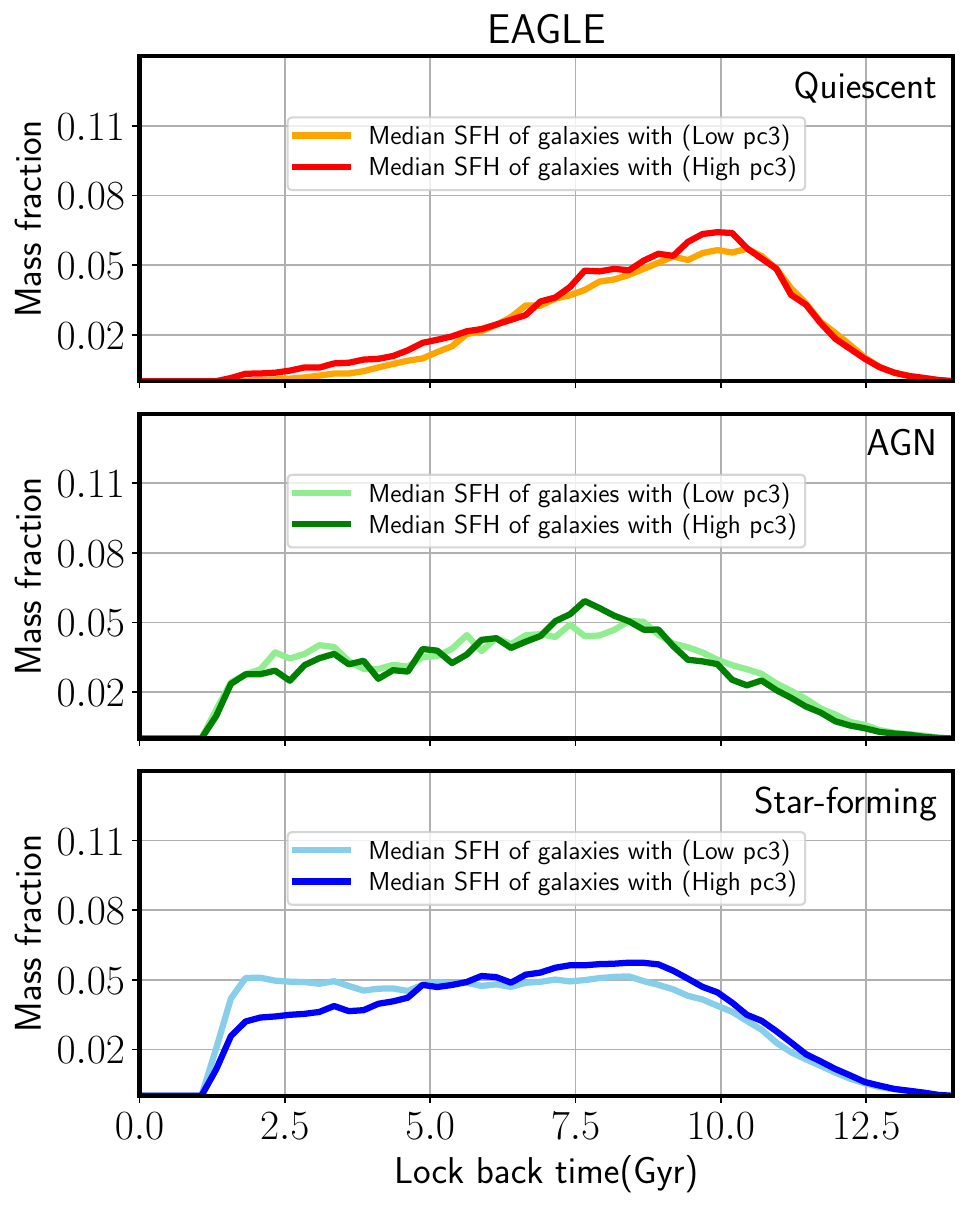}
    \includegraphics[width=0.3\textwidth]{Figs/Supp_PC3_EAGRed.pdf}
    \caption{Median star formation history of the galaxies with the lowest (33 percentile) and highest (67 percentile) value of PC3 in the red interval, for Q, AGN, and SF sub-classes (from top to bottom), and the EAGLE sample in the left column and the TNG100 sample in the right column.} 
    \label{figsfh-eag-pc3-red}
\end{figure}

\bsp	
\label{lastpage}
\end{document}